\documentclass[%
 reprint,
 showkeys,
nofootinbib,
 amsmath,amssymb,
 aps,
]{revtex4-2}

\usepackage{placeins} 
\usepackage{graphicx}
\usepackage{dcolumn}
\usepackage{bm}
\usepackage[
  breaklinks=true
]{hyperref}\usepackage{graphicx}	
\usepackage{color}
\usepackage{amsmath}
\usepackage{xcolor}
\usepackage{changepage}
\usepackage[normalem]{ulem}

\newcommand{\dd}{\mathrm{d}}
\newcommand{\Rc}{\mathcal{R}_c}
\newcommand{\Rci}{\mathcal{R}_{c,i}}
\newcommand{\Rsp}{^{(3)}\!R}

\newcommand{\deltaM}{\delta M}

\newcommand{\eqone}{\overset{(1)}{=}}

\begin{document}

\preprint{APS/123-QED}

\title{Supermassive black hole seeds from direct collapse of CDM-curvature peaks}
\author{Marco Galoppo$^{a}$}
\email{Marco.Galoppo@pg.canterbury.ac.nz}
\author{Marco Bruni$^{b,c}$}
\email{Marco.Bruni@port.ac.uk}
\author{Tomohiro Harada$^{d}$}
\email{Harada@rykkio.ac.jp}

\affiliation{$^a$School of Physical \& Chemical Sciences, University of Canterbury, Private Bag 4800, Christchurch 8140, New Zealand}
\affiliation{$^b$Institute of Cosmology \& Gravitation, University of Portsmouth, Dennis Sciama Building, Burnaby Road,
Portsmouth, PO1 3FX, United Kingdom
\\
$^c$INFN Sezione di Trieste, Via Valerio 2, 34127 Trieste, Italy}
\affiliation{$^d$Department of Physics, Rikkyo University, Toshima, Tokyo 171-8501, Japan}
\affiliation{$^e$iCLA, Yamanashi Gakuin University, Kofu, Yamanashi, 400-8575, Japan\\
 $^f$Centre for Theoretical Physics, Jamia Millia Islamia, New Delhi, 110025, India}

\date{\today}

\begin{abstract}
We study black hole (BH) formation from the nonlinear growth and collapse of primordial perturbations during the matter-dominated era. Modelling cold dark matter (CDM) as pressureless dust, we describe the collapse in a fully nonlinear general-relativistic setting using the Lema\^{i}tre–Tolman–Bondi (LTB) and quasi-spherical Szekeres solutions, interpreted as exact nonlinear perturbations of a global and spatially-flat Friedmann–Lema\^{i}tre–Robertson–Walker (FLRW) $\Lambda$CDM background. In this context, at first-order in the relativistic theory of scalar perturbations the growing mode of any function of interest can be expressed in terms of a single conserved quantity, the gauge-invariant curvature perturbation variable $\mathcal{R}_c$. This acts as a potential for the 3-curvature of spatial slices orthogonal to the matter 4-velocity. We exploit this result to express the active gravitational mass and curvature functions of the LTB and Szekeres solutions in terms of the local initial values of $\mathcal{R}_c$ and its spatial derivatives. In this framework we derive, directly from the initial curvature data: (i) the turn-around, collapse, and apparent-horizon formation times, and (ii) the regularity conditions required for genuine BH formation. We further show that sinusoidal and Gaussian profiles do not constitute viable BH–forming channels in this setting, whereas broad, compensated curvature peaks --- naturally emerging from peak theory --- provide a possible, viable channel. We then provide first estimates for the formation times of $10^{3}\!-\!10^{6}~\mathrm{M}_\odot$ massive BH seeds produced by the direct collapse of primordial CDM-curvature peaks, finding possible full BH formation at redshifts $z > 5$, with their cores starting to collapse at redshifts $10 \lesssim z \lesssim 16$. Finally, we fully characterize the local dynamical nature of the gravitational collapse and the type of singularity that follows --- point-like, cigar-like, or pancake-like --- directly in terms of the initial comoving curvature data, also clarifying the role of the local initial shear in selecting the collapse end-state. 
\end{abstract}


\maketitle

\section{Introduction}\label{sec:Intro}
The formation of Black Holes (BHs) from cosmological initial perturbations is a central problem in relativistic structure formation. In the early Universe, collapse of large-amplitude primordial fluctuations is the basic mechanism underlying Primordial Black Hole (PBH) formation scenarios~\cite{Y.B.Zeldovich_I.D.Novikov_1967,S.W.Hawking_1971,B.J.Carr_S.W.Hawking_1974,B.J.Carr_1975}, thereby linking BH abundances and mass functions to the statistics of the primordial spectrum and to the microphysics of the medium at formation~\cite{T.Harada_etal_2016,C.M.Yoo_etal_2018,T.Harada_etal_2023,K.Uehara_etal_2025,W.Ye_etal_2025,C.Byrnes_etal_2025}. A similar scenario then emerges within the late-time cosmology regime: sufficiently large-amplitude, large-scale perturbations in the Cold Dark Matter (CDM) distribution may undergo relativistic collapse in the matter era. In this setting, however, BH formation is not determined solely by the overdensity amplitude: it depends sensitively on the spatial structure of the initial perturbation and on the tidal anisotropies it induces.

This problem has acquired renewed observational relevance owing to recent high-redshift detections of unexpectedly massive compact objects. In particular, \textit{James Webb Space Telescope} observations have identified candidate active galactic nuclei and massive BH populations already in place at redshifts $z \gtrsim 7-10$, with inferred masses up to $10^{6}-10^{9}\,M_\odot$ in some analyses~\cite{F.Pacucci_etal_2023,R.Larson_etal_2023,R.Maiolino_etal_2024}. Although the interpretation of these sources remains subject to substantial systematic uncertainties, such observations have sharpened the question of whether massive BH seeds can form sufficiently early within standard cosmological physics, thereby motivating a careful re-examination of relativistic collapse in pressureless CDM.

A key qualitative distinction for BH formation is whether the collapse occurs in a radiation-dominated environment, where pressure gradients introduce a non-trivial threshold for collapse, or in a pressureless environment~\cite{T.Harada_etal_2016,K.Uehara_etal_2025,M.Y.Khlopov_A.G.Polnarev_1980,T.Harada_etal_2023}, where  matter effectively behaves as ``dust", i.e.\ a pressurless fluid, thus facilitating the collapse. In the latter case --- relevant to collapse during the matter-dominated era and, locally, to CDM dynamics --- the usual pressure-supported threshold is absent. Nevertheless, anisotropic tidal fields, angular momentum acquisition, and inhomogeneity can strongly suppress (or qualitatively alter) the route to horizon formation, often favouring pancake-like configurations and/or the appearance of shell-crossing singularities before any trapped surface can form~\cite{M.Y.Khlopov_A.G.Polnarev_1980,T.Harada_etal_2016,T.Harada_S.Jhingan_2015,W.Ye_etal_2025}. A consistent relativistic treatment must therefore track not only the growth of the overdensity, but also the development of shear and Weyl curvature that drive departures from spherical, Top-Hat (TH)~\cite{J.E.Gunn_J.R.Gott_1972} intuition on BH formation.

In this work we study BH formation from the nonlinear growth and collapse of primordial perturbations during the matter-dominated era, modelling CDM as irrotational pressureless dust. We employ exact inhomogeneous dust solutions as fully nonlinear realisations of perturbations about a homogeneous background; namely, we interpret the Lema\^{i}tre--Tolman--Bondi (LTB) ~\cite{G.Lemaitre_1933,R.C.Tolman_1934,H.Bondi_1947} and quasi-spherical Szekeres families~\cite{P.Szekeres_1975a,P.Szekeres_1975b} as exact nonlinear perturbations of a global flat Friedmann--Lema\^{i}tre--Robertson--Walker (FLRW) background. These spacetimes are particularly well suited to isolating the interplay between inhomogeneity, shear and horizon formation as their dynamics reduces to local evolution along each dust flow line~\cite{M.Bruni_etal_1995_Jun,J.Wainwright_G.F.R.Ellis_1997,G.F.R.Ellis_etal_2012}.

In this context, at first order in relativistic perturbation theory, the growing scalar mode of irrotational dust is fully specified by a single gauge-invariant linearly-conserved quantity, i.e., the comoving curvature perturbation $\Rc$~\cite{K.A.Malik_D.Wands_2008,M.Bruni_etal_2014_Mar,R.L.Munoz_M.Bruni_2023}. This is the potential for the 3-curvature perturbation of the spatial slices orthogonal to the matter 4-velocity. We therefore formulate the initial conditions in terms of $\Rc$ and provide an explicit map between a comoving curvature profile and the exact LTB/Szekeres free initial data, allowing the subsequent collapse dynamics to be expressed directly in terms of $\Rc$ and its derivatives.

We then derive analytic expressions for the turn-around, collapse, and apparent-horizon (future trapping horizon) formation times of each comoving shell. Additionally, we rewrite the regularity requirements for BH formation~\cite{P.Szekeres_1975b,C.Hellaby_A.Krasinski_2002,C.Hellaby_A.Krasinski_2008,T.Harada_S.Jhingan_2015} --- in particular shell-crossing avoidance and conditions controlling the causal nature of the central focusing singularity --- as constraints on $\Rc$ and its derivatives. 

 Applying the resulting criteria to representative profiles, we find that single-mode and simple Gaussian choices must be excluded as BH formation channels: beside being unrealistic,  these undergo collapses producing a central naked singularity. Instead, we find that broad compensated peaks of the type naturally emerging from peak theory (see e.g.,~\cite{A.G.Doroshkevich_1970,J.M.Bardeen_J.R.Bond_N.Kaiser_A.S.Szalay_1986,R.J.Adler_1981,C.Germani_T.Prokopec_2017,C.Germani_2025}) provide a viable channel: an almost homogeneous core covers the central singularity, whilst a steepened compensated tail delays shell-crossing. We thus provide mass and redshift formation estimates for the massive BH seed in the matter era, finding that BHs of masses in the range $10^{3}-10^{6}~\rm{M}_\odot$ can fully form, within the limitations of our conservative choice of realistic peak-perturbation profiles,  in the  $5 \lesssim z \lesssim 7$.

Finally, we also characterise the local dynamical nature of the collapse end-state using the covariant $1+3$ formalism~\cite{J.Ehlers_1961,S.W.Hawking_1966,G.F.R._Ellis1971, S.W.Hawking_G.F.R.Ellis_1973} and its dynamical-systems representation~\cite{J.Wainwright_L.Hsu_1989,M.Bruni_etal_1995_Jun,M.Bruni_etal_1995_Jul,J.Wainwright_G.F.R.Ellis_1997}. Interestingly, we find that even when a BH forms, the approach to the shell-focusing singularity is generically locally anisotropic, being typically represented by  a Kasner-like cigar (AKA as spindle) state \cite{J.Wainwright_G.F.R.Ellis_1997} rather than an isotropic, point-like focusing. Additionally, we directly link shell-crossing to pancake-like collapse and show that the dynamical end-state is completely determined by the local initial shear. The bottom line is that, unlike in Newtonian gravity where gravity is always linearly related to matter density via the Poisson equation and pancakes always prevail as first caustics in the collapse of pressureless matter, in GR a horizon can form even in the presence of anisotropy in the CDM distribution, and actually this leads to spaghettification and, ultimately, to cigar-like Weyl-dominated singularities and the irrelevance of matter in the final stages of collapse, when {\it gravity gravitates}.

The paper is organised as follows.  In Sec.~\ref{sec:1+3} we summarise the covariant $1+3$ description of irrotational dust and introduce the kinematic classification of collapse end-states.  In Sec.~\ref{sec:3+1} we present the complementary $3+1$ viewpoint and the linear perturbation framework in terms of initial comoving curvature perturbations used to seed initial data.  In Sec.~\ref{sec:LTB} we review the quasi-spherical Szekeres/LTB solutions and introduce the formalisms employed in this work.  In Sec.~\ref{sec:Collapse} we derive the relevant collapse times and the regularity conditions (shell-crossing avoidance and horizon-formation criteria). In Sec.~\ref{sec:BH} we construct the comoving curvature perturbations-based initial conditions, analyse explicit curvature profiles, and identify the peak-theory motivated channel leading to black-hole formation, as well as provide estimates on massive BH formation in the matter-dominated era. Sec.~\ref{sec:conc} is left for a summary and discussion of future works. Further technical derivations are collected in Apps.~\ref{app:mapping} and~\ref{app:param}. 

Throughout this work, we set the vacuum speed of light $c$ to unity and use the metric signature $(-,+,+,+)$.

\section{1+3 approach for CDM}\label{sec:1+3}

The $1 + 3$ approach allows for a covariant decomposition of the field equations, by introducing a threading timelike congruence $u^{\mu}$, representing the worldlines of a family of observers or fluid elements. The geometry and dynamics of spacetime are then encoded in covariantly defined kinematic, dynamical, and curvature quantities obtained by projecting all tensors either parallel or orthogonally to $u^{\mu}$, thereby splitting the field equations directly into evolution equations along the flow and constraint equations on the instantaneous local rest spaces.

The \textit{manifestly} covariant $1 + 3$ approach --- originally developed by Ehlers, Hawking, and Ellis~\cite{J.Ehlers_1961,S.W.Hawking_1966,G.F.R._Ellis1971, S.W.Hawking_G.F.R.Ellis_1973} --- is a powerful analytical framework for studying both relativistic cosmology and gravitational collapse, e.g.,~\cite{S.W.Goode_J.Wainwright_1982,G.F.R.Ellis_M.Bruni_1989,M.Bruni_etal_1992,P.K.S.Dunsby_etal_1992,M.Bruni_etal_1995_Jul,J.Wainwright_G.F.R.Ellis_1997,G.F.R.Ellis_etal_2012}. We will employ this formalism within a CDM framework --- where the matter content of the spacetime is modelled as an irrotational, pressureless fluid (dust) --- to investigate the role of dynamical variables in LTB and Szekeres gravitational collapse. 

\subsection{Defining the covariant variables}
Let us consider a $4-$dimensional, Lorentzian spacetime manifold which is assumed to be globally hyperbolic, and to obey the Einstein equations sourced by an irrotational dust source with energy density $\rho$ and timelike $4-$velocity $u^\mu$, i.e., with energy-momentum tensor $T_{\mu\nu} = \rho \;\! u_\mu u_\nu$. The metric $\bm{g} = g_{\mu\nu} \, \mathrm{d}x^\mu \mathrm{d}x^\nu$ defines a local projector orthogonal to the $4-$velocity, with components $h_{\mu \nu} := g_{\mu \nu} + u_\mu u_\nu$. The $4$-velocity field $\bm u$ is further characterized by its expansion scalar $\Theta$ and by its shear tensor $\sigma_{\mu \nu}$ and scalar $\sigma^2 := (1/2) \, \sigma^{\mu \nu} \sigma_{\mu\nu}$. These quantities arise from the kinematic decomposition of the fluid's expansion tensor, namely~\cite{J.Wainwright_G.F.R.Ellis_1997,G.F.R.Ellis_etal_2012}
\begin{align}
    &\Theta_{\mu \nu} := h^\kappa_{\,(\mu} h^\lambda_{\,\nu)}\nabla_\kappa u_\lambda = \nabla_\mu u_\nu = \frac{1}{3} \Theta \, h_{\mu \nu} + \sigma_{\mu\nu} \, \\
    &\Theta := \Theta^\mu_{\,\mu} = h^{\mu \nu} \nabla_\mu u_\nu \, , \\
    & \sigma_{\mu \nu} := \Theta_{\mu \nu} - \frac{1}{3} \Theta \, h_{\mu \nu} \, ,
\end{align}
with the $4-$acceleration $a^\mu :=u^\nu \nabla_\nu u^\mu$ and vorticity $\omega_{\mu\nu}:=h^\kappa_{\,[\mu} h^\lambda_{\,\nu]} \nabla_\kappa u_\lambda$ of the fluid vanishing by the irrotational dust assumption. In addition, the shear tensor is symmetric, traceless and fluid-orthogonal by construction, i.e., $\sigma_{[\mu \nu]} = 0$, $\sigma^{\mu}_{\,\mu} = 0$, $u^\mu \sigma_{\mu \nu} = 0$. Furthermore, let us consider the decomposition of the Weyl tensor, $C_{\mu\rho\nu\sigma}$, in terms of its electric and magnetic parts~\cite{J.Wainwright_G.F.R.Ellis_1997, G.F.R.Ellis_etal_2012}, i.e.,
\begin{equation}
    E_{\mu\nu} = C_{\mu\rho\nu\sigma}u^{\rho}u^{\sigma} \, , \; \; \; H_{\mu\nu} = \frac{1}{2}\eta_{\alpha\beta\gamma\mu}C^{\alpha\beta}{}{}_{\nu\delta}u^{\gamma}u^{\delta} \, ,
\end{equation}
with $\eta = \sqrt{-g}\,\epsilon_{\alpha\beta\gamma\mu}$, where $\epsilon_{\alpha\beta\gamma\mu}$ is the alternating Levi-Civita symbol, and both the resulting electric and magnetic parts are traceless and symmetric.

To model the dynamical collapse of structures in GR, it is often meaningful to assume that the magnetic part of the Weyl tensor vanishes, i.e., to impose $H_{\mu\nu} = 0$. While this condition cannot be imposed in general --- even within a Newtonian setting~\cite{L.Kofman_D.Pogosian_1995} --- if coupled to the irrotational dust assumption, it simplifies the resulting evolution equations into a set of \textit{local} Ordinary Differential Equations (ODEs)~\cite{A.Barnes_R.R.Rowlingson_1989,S.Matarrese_etal_1993,M.Bruni_etal_1995_Jun,J.Wainwright_G.F.R.Ellis_1997}. Spacetimes in which these conditions are met are referred to as \textit{silent universes}~\cite{S.Matarrese_etal_1993,M.Bruni_etal_1995_Jun,M.Bruni_etal_1995_Jul,J.Wainwright_G.F.R.Ellis_1997}, since each fluid worldline evolves independently of its neighbours. In such models, the evolution of the kinematic and curvature variables along a given worldline is determined entirely by local quantities, reproducing a key feature of Newtonian-Lagrangian fluid dynamics in the Zel’dovich approximation~\cite{Y.B.Zeldovich_1970}, widely used in structure formation studies, see e.g.~\cite{Bruni:2002xk}.

Then, within a silent universe framework, the set of kinematic, matter and curvature variables given by $\{\Theta,\, \sigma_{\mu\nu},\, \rho,\, E_{\mu\nu}\}$ fully characterises the dynamics of the spacetime. In particular, we emphasise that the LTB and Szekeres models are exact silent dust solutions, so that within the framework of this work the silent-universe condition is not introduced as an approximation, but follows identically from the underlying geometry and matter content of the employed exact solutions.
\subsection{Evolution and spatial constraint equations}
The evolution equations for the variables of interest can be directly derived by applying appropriate projections with respect to $u^\mu$ and $h^{\mu\nu}$ of the Ricci and  Bianchi identities, using Einstein Field Equations (EFEs) to replace the Ricci tensor with the energy-momentum tensor, obtaining~\cite{S.Matarrese_etal_1993,M.Bruni_etal_1995_Jun,J.Wainwright_G.F.R.Ellis_1997, G.F.R.Ellis_etal_2012} 
\begin{align}
    &\dot{\Theta} = -\frac{1}{3}\Theta^2 - 2\sigma^2 - 4\pi G\rho \, ;\label{eq:Raychaudhuri_silent}\\
    &\dot{\rho} = -\Theta\rho \, ;\label{eq:continuity_silent}\\
    & \dot{\sigma}_{\langle \mu\nu \rangle} = -\frac{2}{3}\Theta\sigma_{\mu\nu} - \sigma_{\alpha \langle \mu}\sigma_{\nu \rangle}{}^{\alpha} - E_{\mu\nu}\, ;\label{eq:ev_sigmamunu_silent}\\
    & \dot{E}_{\langle \mu\nu \rangle} = -\Theta E_{\mu\nu}-3\sigma_{\alpha \langle \mu}E_{\nu \rangle}{}^{\alpha} - 4\pi G\rho\sigma_{\mu\nu} \, ; \label{eq:ev_Emunu_silent}
\end{align}
where we have defined the covariant derivative along $u^\mu$ as $\dot{}:=u_\mu\nabla^{\mu}$, the spatially projected derivative $D_\mu$ of a tensor via $D_\mu M_{\alpha}{}^{\beta\gamma}:= h_{\mu}{}^{\rho} h_{\alpha}{}^{\sigma} h_{\xi}{}^{\beta} h_{\zeta}{}^{\gamma}\nabla_\rho M_{\sigma}{}^{\xi\zeta}$, and the symmetric, traceless projection of a tensor through $M_{\langle \mu\nu\rangle}:= \left( h_{(\mu}{}^{\alpha}h_{\nu)}{}^{\beta} - \frac{1}{3}h_{\mu\nu}h^{\alpha\beta}\right)M_{\alpha\beta}$. The system of equations~\eqref{eq:Raychaudhuri_silent}--\eqref{eq:ev_Emunu_silent} fully determines the evolution of the spacetimes with, in particular, Eqs.~\eqref{eq:Raychaudhuri_silent} and \eqref{eq:continuity_silent} representing the Raychaudhuri and continuity equations, respectively. The evolution equations are then completed by a set of spatial constraint, namely~\cite{J.Wainwright_G.F.R.Ellis_1997, G.F.R.Ellis_etal_2012} 
\begin{align}
    & D^\nu\sigma_{\mu\nu} = \frac{2}{3}D_\mu\Theta \, ; \\
    & D^\nu E_{\mu\nu} = \frac{4\pi G}{3}D_\mu\rho \, ; \\
    & \eta_{\mu\alpha\beta}\sigma^{\alpha\gamma}E^{\beta}{}_{\gamma} = 0\, ; \\
    & \eta_{\alpha\beta\langle\mu}D^\alpha \sigma_{\nu\rangle}{}^{\beta} = 0 \, ;  \label{eq:const_interesting}\\
    & \eta_{\alpha\beta\langle\mu}D^\alpha E_{\nu\rangle}{}^{\beta} = 0 \, ;\label{eq:const_weird}\\
    & \Rsp = -\frac{2}{3}\Theta^2 + 16\pi G\rho + 2\sigma^2 \, ; \label{eq:Hamiltonian_constraint_silent}
\end{align}
where Eq.~\eqref{eq:Hamiltonian_constraint_silent} is the Hamiltonian constraint relating the spatial curvature, i.e., the 3--Ricci scalar, to the matter and kinematic variables. Interestingly, Eq.~\eqref{eq:const_interesting} is equivalent to a commutation relation between the shear tensor and the electric Weyl tensor~\cite{A.Barnes_R.R.Rowlingson_1989}. Consequently, one can introduce a triad of 4--vectors $\{e_\alpha^{(i)}\}_{i =1}^3$, alongside $u^\alpha$, that simultaneously diagonalises both tensors. We can thus define the eigenvalues of $\sigma_{\mu\nu}$ and $E_{\mu\nu}$ with respect to this triad,  namely $\{\sigma_{ii}\}_{i =1}^3$ and $\{E_{ii}\}_{i =1}^3$, and, using the trace-free property of the shear and the electric Weyl tensor, construct the independent sets $\{\sigma_\pm\}$ and $\{E_\pm\}$, defined by~\cite{J.Wainwright_G.F.R.Ellis_1997}
\begin{align}
   & \sigma_+ := \frac{1}{2}(\sigma_{22}+\sigma_{33}) = -  \frac{1}{2}\sigma_{11}\, , \\
    & \sigma_- := \frac{1}{2\sqrt{3}}(\sigma_{22}-\sigma_{33}) \, ,
\end{align}
and analogously for $\{E_\pm\}$. Then, the evolution equations for both the shear and electric Weyl tensors( see Eqs.~\eqref{eq:ev_sigmamunu_silent} and~\eqref{eq:ev_Emunu_silent}) can be fully written in terms of the newly introduced scalar variables. These reduce to the evolution equations for LTB and Szekeres models upon setting $\sigma_- = E_- = 0$, following their symmetry restrictions as locally rotational invariant spacetimes~\cite{J.Wainwright_G.F.R.Ellis_1997, G.F.R.Ellis_etal_2012}. Note then, that within the Szekeres and LTB framework $e_1^{(i)}$ conventionally indicates the principal radial direction, whilst $e_2^{(i)}$ and $e_3^{(i)}$ the transverse angular ones. Finally, we obtain as complete set of evolution equations for LTB and Szekeres models \cite{S.Matarrese_etal_1993,M.Bruni_etal_1995_Jun}
\begin{align}
     & \dot{\rho} = -\Theta\rho \, , \label{eq:Sk1}\\
     & \dot{\Theta} = -\frac{1}{3}\Theta^2 -6\sigma^2_+ -4\pi G\rho\, , \\
     & \dot{\sigma}_+ = -\frac{2}{3}\Theta\sigma_+ + \sigma_+^2  - E_+ \, ,\label{eq:sigma++}\\
    & \dot{E}_+ = -\Theta E_+ -3E_+\sigma_+- 4\pi G\rho\sigma_+ \label{eq:E++}\, .
\end{align}
\subsection{Kinematic classification of singularities}

Within the $1+3$ formalism, spacetime singularities arising in dust collapse admit a physically meaningful and coordinate-independent characterisation in terms of the expansion tensor $\Theta_{ij}$, which describes the local deformation of comoving fluid elements~\cite{G.F.R._Ellis1971,M.A.H.MacCallum_1973,J.Wainwright_G.F.R.Ellis_1997,G.F.R.Ellis_etal_2012}. 

To do so, we begin by noting that it is always possible to introduce a local orthonormal triad $\{e_\alpha^{(i)}\}_{i =1}^3$, Fermi--propagated along the dust flow, which diagonalises $\Theta_{ij}$ at each spacetime point. In this principal frame, it is then meaningful to introduce the eigenvalues of the expansion tensor, i.e., $\{\Theta_{(i)}\}_{i =1}^3$, which correspond to the local expansion or contraction rates along the principal axes of deformation of an infinitesimal fluid element. Furthermore, we can consider the associated principal scale factors $\ell_i$ defined via the evolution equation~\cite{G.F.R._Ellis1971,M.A.H.MacCallum_1973,J.Wainwright_G.F.R.Ellis_1997,G.F.R.Ellis_etal_2012}
\begin{equation}
\frac{\dot{\ell}_i}{\ell_i} = \Theta_{(i)}  = \frac{1}{3}\Theta + \sigma_{ii}\, ,\label{eq:define_singularities}
\end{equation}
from which it follows that the local volume element evolves as $\dd V = \ell_1 \ell_2 \ell_3 \propto \exp\!\left(\int \Theta \, dt\right)$, with $\Theta = \sum_i \Theta_{(i)}$.

A spacetime singularity is therefore signalled by the vanishing of at least one principal scale factor $\ell_i$ within finite proper time, equivalently by the divergence of at least one eigenvalue $\Theta_{(i)}$~\cite{G.F.R._Ellis1971,M.A.H.MacCallum_1973,J.Wainwright_G.F.R.Ellis_1997,G.F.R.Ellis_etal_2012}. The relative behaviour of the three $\Theta_{(i)}$, equivalently $\ell_i$, determines the local dynamical character of the singularity, providing a natural kinematic classification of collapse end states. These are reported in Tab.~\ref{tab:singularity_types}. 
\begin{table*}[t]
\centering
\setlength{\tabcolsep}{10pt}        
\renewcommand{\arraystretch}{1.15}  
\begin{tabular}{lll}
\hline\hline
Singularity type 
& Behaviour of $\ell_i$ 
& Asymptotic behaviour of $\Theta_{(i)}$ \\[6pt]
\hline
Point-like 
& $\ell_1,\ell_2,\ell_3 \to 0$  
& $\Theta_{(1)},\,\Theta_{(2)},\,\Theta_{(3)} \to -\infty$  \\[6pt]
Cigar-like  
& $\ell_1,\ell_2 \to 0,\, \ell_3 \to +\infty$  
& $\Theta_{(1)},\,\Theta_{(2)} \to -\infty,\,\Theta_{(3)} \to \mathrm{const} > 0$ \\[6pt]
Filament-like   
& $\ell_1,\ell_2 \to 0,\,\ell_3 \to \mathrm{const} > 0$ 
& $\Theta_{(1)},\,\Theta_{(2)} \to -\infty,\,\Theta_{(3)} \to 0$ \\[6pt]
Pancake-like   
& $\ell_1 \to 0,\,\ell_2,\ell_3 \to \mathrm{const} > 0$ 
& $\Theta_{(1)} \to -\infty,\,\Theta_{(2)},\,\Theta_{(3)} \to 0$ \\
\hline\hline
\end{tabular}
\caption{Kinematic classification of singularities.}
\label{tab:singularity_types}
\end{table*}

This classification is invariant under coordinate transformations and provides a local, kinematic characterisation of the end state of collapse. In particular, the nature of the singularity is controlled by the relative magnitude of the shear with respect to the isotropic expansion, as encoded in the eigenvalue structure of \(\Theta_{ij}\). 

Point-like singularities correspond to an isotropic, shear-free collapse, in which all principal directions contract at the same rate. By contrast, cigar-like\footnote{
These may also be referred to as spindle-like kinematic singularities. However, note that this terminology differs from that traditional employed in numerical relativity, see e.g., Shapiro and Teukolsky~\cite{S.L.Shapiro_S.A.Teukolsky_1991}.
}, filament-like, and pancake-like singularities arise from increasingly anisotropic, shear-dominated dynamics, characterised by a decreasing number of collapsing principal directions. In this hierarchy, cigar-like singularities exhibit contraction along two directions with expansion along the third, filament-like singularities involve two collapsing directions and one asymptotically static axis, while pancake-like singularities correspond to collapse along a single direction. In the Newtonian case the pancake is the generic attractor of the dynamics, as well illustrated by the Zel'dovich approximation~\cite{Bruni:2002xk}. Remarkably, this is not the case in GR, at least in the LTB/Szekeres class of models\footnote{
Although  no one knows the general properties of singularities in dust collapse in GR, one of the authors, MB, considers  the Belinski–Khalatnikov–Lifshitz (BKL) conjecture~\cite{Landau1975,J.Wainwright_G.F.R.Ellis_1997} as totally plausible, so that in the first Kasner-like phase the dust fluid locally evolves \lq\lq silently"~\cite{S.Matarrese_etal_1993,Matarrese:1993zf,M.Bruni_etal_1995_Jun,M.Bruni_etal_1995_Jul,J.Wainwright_G.F.R.Ellis_1997}, with negligible contribution from the magnetic Weyl tensor, until the jump to the next Kasner phase~\cite{Bruni:2003hm}.
}: as we are going to see, in LTB and Szekeres models there are rather general initial conditions that lead to a cigar-like singularity~\cite{S.Matarrese_etal_1993,M.Bruni_etal_1995_Jun}. From a GR perspective, pancakes are Ricci-dominated (and therefore, matter dominated), and can be interepreted more like the break-down of the fluid description rather than true physical singularities, while in the late stage of the collapse leading to a cigar singularity spaghettification occurs,  matter becomes negligible,  and the singularity is Weyl-dominated but, crucially, covered by the forming BH horizon.

\section{3+1 approach for CDM}\label{sec:3+1}
The $3 + 1$ decomposition considers the opposite perspective with respect to the $1 + 3$ approach, where the fundamental object is a timelike 4--vector $u^{\mu}$ threading the spacetime. Rather, in $3 + 1$ framework, the spacetime is studied by determining the  properties of the set of ``slices" $\Sigma_t$, i.e., the level surfaces  of a globally defined time function $t(x^{\mu})$, and how those surfaces are related to one another as the function $t$ changes. In practice, for irrotational dust the 4-velocity $u^\mu$ is hypersurface-orthogonal, and the two approaches are equivalent, even if in 1+3 the formalism is anchored to the physical 4-vector $u^\mu$, while $3 + 1$ is more general, being based on a generic time-like vector $n^\mu$ normal to the slices.

This decomposition, originally developed in a seminal series of papers by Arnowitt, Deser, and Misner~\cite{R.Arnowitt_S.Deser_C.W.Misner_1959a,R.Arnowitt_S.Deser_C.W.Misner_1959b, R.Arnowitt_S.Deser_C.W.Misner_1960,R.Arnowitt_S.Deser_C.W.Misner_1961, R.Arnowitt_S.Deser_C.W.Misner_1962}, has played a central role in mathematical GR~\cite{Y.ChoquetBruhat_1952,Y.ChoquetBruhat_R.Geroch_1969,York1973_ConformalDecomposition}, provided the foundation for the Hamiltonian formulation of relativistic theories~\cite{P.A.M.Dirac_1958,P.A.M.Dirac_1959,S.A.Hojman_K.Kuchar_C.Teitelboim_1976}, and --- perhaps most prominently --- represents the cornerstone of modern numerical relativity~\cite{M.Alcubierre_2008,T.W.Baumgarte_S.L.Shapiro_2010,M.Shibata_2015}. In this work, we employ the $3+1$ decomposition to specify well-posed initial conditions for BH formation in Szekeres models via first-order perturbation theory on a flat FLRW background in synchronous-comoving gauge.

\subsection{Theoretical framework}

Let us consider again a $4-$dimensional, Lorentzian spacetime manifold which is assumed to be globally hyperbolic, and to obey the Einstein equations. Then, the starting point of the $3+1$ formulation is the choice of a foliation of spacetime, whose leaves are represented by the level surfaces $\Sigma_t$\footnote{Note that, although the introduction of a preferred slicing in the spacetime break \textit{explicit} covariance, the $3 + 1$ decomposition remains a \textit{non-manifestly} covariant formulation of GR.}.
Each leaf admits a unit normal vector field $n_a$, given by $n^\mu = -N \nabla^\mu t$, where $N(x^\mu)$ is the lapse function, encoding the proper–time separation between neighbouring leaves. Namely, for observers whose worldlines are orthogonal to the level surfaces (i.e., tangent to $n^\mu$), the proper time interval between $\Sigma_t$ and $\Sigma_{t+\dd t}$ is $\dd \tau = N\dd t$. In addition, the leaves of the foliation are connected by a time–evolution vector $t^\mu = N n^\mu + N^\mu$, where $N^\mu$ is the shift vector, which specifies how the spatial coordinates $x^i$ change from $\Sigma_t$ to $\Sigma_{t+\mathrm{d}t}$, such that $\mathrm{d}x^i = N^i\mathrm{d}t$. 
Here, we emphasize that the choice of lapse and shift --- namely, the choice of the foliation $\Sigma_t$ and of the threading of its leafs by constant spatial coordinates worldlines, respectively --- is arbitrary, since general covariance ensures complete coordinate freedom. Hence, $N(x^{\mu})$ and $N^{i}(x^{\mu})$ are gauge variables associated with the freedom to choose spacetime coordinates (see~\cite{M.Alcubierre_2008,T.W.Baumgarte_S.L.Shapiro_2010,M.Shibata_2015} for a complete discussion). 

In this work, we consider spacetimes sourced by irrotational, pressureless dust, which allow us to adopt the synchronous–comoving gauge, i.e., $N = 1$, $N^i = 0$, and $n^\mu = u^\mu$, so that $t$ measures proper time along the dust flow and worldlines of constant $x^i$ follow the dust trajectories. Thus, we can write the spacetime line element
\begin{equation}\label{eq:3+1_line_element}
    \dd s^2 = -\dd t^2 + \gamma_{ij}\dd x^i \dd x^j \, ,
\end{equation}
where $\gamma_{ij}$ represents the induced spatial metric on the leaves $\Sigma_t$, i.e., the first fundamental form of these surfaces. Interestingly, note that within the silent CDM framework, we can directly map $\gamma_{ij}$ to the $h_{ij}$ introduced in Sec.~\ref{sec:1+3}. Ultimately, the entire spacetime geometry can be described via $\gamma_{ij}$ --- encoding all the information about the intrinsic curvature of the surface --- and the extrinsic curvature tensor $K_{ij} = -\frac{1}{2}\dot{\gamma}_{ij} = - \Theta_{ij}$, which gives all the information about how the surface is embedded into the 4-dimensional spacetime manifold, and where the latter equality is valid within the silent CDM framework. 

\subsection{FLRW flat dust models}

We first consider the case of a flat FLRW dust universe, our background spacetime, whose quantities we indicate with an overhead bar. In the synchronous-comoving gauge the spatial metric is then $\bar{\gamma}_{ij} = a(t)^2 \delta_{ij}$, where $a(t)$ is the scale factor and $\delta_{ij}$ is the Kronecker delta. The Hubble expansion is then $\bar{\Theta} = 3 H$ with $H = \dot{a}/a$, and the dust density is $\bar{\rho} = 3 H^2 \Omega_m$, where $\Omega_m$ is the dimensionless matter density parameter. Analogously, we can define  $\Omega_\Lambda := \Lambda/3H^2$ for the cosmological constant. 

For a spatially flat FLRW universe sourced by dust and a cosmological constant, the direct integration of Eqs.~\eqref{eq:Raychaudhuri_silent},~\eqref{eq:continuity_silent}, and~\eqref{eq:Hamiltonian_constraint_silent} --- which reduce to the Friedmann equations --- gives
\begin{align}
    &a(t) = \left(\frac{\Omega_{m,0}}{\Omega_{\Lambda,0}}\right)^{1/3}\sinh^{2/3}\left(\frac{3}{2}H_0\Omega_{\Lambda,0}^{1/2}t\right) \, \\
    &H(t) = H_0\left(\Omega_{m,0}a(t)^{-3}+\Omega_{\Lambda,0}\right)^{1/2}\, , \\
    & \Omega_m (t) = \frac{\Omega_{m,0}}{\Omega_{m,0}+\Omega_{\Lambda,0}a(t)^{3}} \, ,
\end{align}
where we $H_0$, $\Omega_{m,0}$, and $\Omega_{\Lambda,0}$ indicate the present-day value of the respective physical parameters, and we have normalised the scale factor such that $a_0 = 1$. In addition, we note that for an Einstein--de Sitter (EdS) spacetime, i.e., such that $\Lambda = 0$, we have 
\begin{equation}
   a(t) = \left(\frac{3}{2}H_0 t\right)^{2/3}\,, \;\; H(t) = \frac{2}{3t}\;, \;\;\Omega_m = 1\, ,
\end{equation}
where we have normalised once again $a_0 = 1$. 
\subsection{\label{sec:PT} FLRW first-order perturbations}
Perturbation theory on an FLRW background plays a fundamental role in cosmology, providing a systematic framework to study the origin and evolution of large-scale structure in the Universe. In particular, it serves as a powerful tool for investigating BH formation from the collapse of primordial density fluctuations, by tracing the evolution of the perturbations generated during inflation through the radiation- and matter-dominated eras --- when they re-enter the Hubble horizon.

Here, we follow~\cite{M.Bruni_etal_2014_Mar,M.Bruni_etal_2014_Sep} for the treatment of first-order perturbations in the synchronous-comoving gauge, which is formulated in terms of $\Rc$, the first-order gauge-invariant scalar  potential associated with ${}^{(3)}\!R^{(1)}$, the perturbation of the 3--Ricci scalar (for reviews see~\cite{K.A.Malik_D.Wands_2008, D.Langlois_F.Vernizzi_2010}). This approach will serve as the basis for constructing the nonlinear initial conditions underlying the LTB and Szekeres collapse, following the procedure already implemented within a numerical relativity framework in~\cite{R.L.Munoz_M.Bruni_2023}.

We note that by employing $\Rc$: (i) the initial conditions can be directly seeded from inflationary model predictions, and (ii) within a dust framework, $\Rc$ represents a conserved quantity on all scales, allowing the complete specification of first-order scalar perturbation variables corresponding to the growing mode, as we discuss below.

Let us consider scalar perturbations of a flat FLRW background during the matter-dominated era --- i.e., the only relevant first-order perturbations for structure formation --- in the synchronous-comoving gauge and using Cartesian-like coordinates. The line element can then be written as in the $3+1$ approach (see Eq.~\eqref{eq:3+1_line_element}), where the spatial metric $\gamma_{ij}$ is expressed as
\begin{equation}\label{eq:spatial_metric}
\gamma_{ij} =a^2\tilde{\gamma}_{ij}= a^2(t)\left[\left(1-2\psi\right)\delta_{ij} + \chi_{ij}\right]\,,
\end{equation}
with $a(t)$ representing the background scale-factor, $\psi$ the volume perturbation, and $\chi_{ij}$ the traceless anisotropic distortion, respectively. In particular, as we are considering only first-order scalar perturbations, $\chi_{ij}$ can be expressed in terms of a scalar potential $\chi^{(1)}$ via\footnote{%
Hereafter we employ the $\eqone$ symbol to identify equivalence up to first order in the perturbative scheme.%
}
\begin{equation}
    \chi_{ij} \eqone \left(\partial_i\partial_j - \frac{a^2}{3}\nabla^2 \right)\chi^{(1)} \, ,
\end{equation}
where we define the Laplacian as $\nabla^2 = \gamma_{ij}\nabla^i\nabla^j\eqone a^{-2}\delta^{ij}\partial_{x_i}\partial_{x_j}=a^{-2}\bar{\nabla}^2$, where the middle and last equalities are employed when the Laplacian is acting on first-order quantities. We can then express the comoving curvature perturbations $\Rc$ in terms of the metric perturbations as~\cite{D.H.Lyth_1985}
\begin{equation}\label{eq:Rc_metric}
    \Rc = \psi^{(1)} + \frac{1}{6}\bar{\nabla}^2\chi^{(1)}\, .
\end{equation}
With this, the first-order 3-curvature scalar ${}^{(3)}\!R^{(1)}$ of the spatial metric~\eqref{eq:spatial_metric} is then given by~\cite{D.H.Lyth_1985,M.Bruni_etal_2014_Mar}: 
\begin{equation}\label{eq:3R1_Rc}
   {}^{(3)}\!R^{(1)} = \frac{R^{(1)}}{a^2}=\frac{4}{a^2} \bar{\nabla}^2\Rc \, ,
\end{equation}
where $R^{(1)}$ denotes the 3-curvature scalar of the conformal metric $\tilde{\gamma}_{ij}$. Here, we then remark that since ${}^{(3)}\!R^{(1)}$ vanishes on a flat FLRW background, it follows from the Stewart and Walker lemma~\cite{J.M.Stewart_M.Walker_1974} cf.~\cite{G.F.R.Ellis_M.Bruni_1989,M.Bruni_etal_1992,P.K.S.Dunsby_etal_1992}, that ${}^{(3)}\!R^{(1)}$, and consequently $R^{(1)}$ and $\Rc$, are gauge-invariant first-order quantities. Then, to express in terms of $\Rc$, the first-order metric perturbations, as well as the density contrast, we use  the evolution  of the latter in the form of the following  system of two first-order ODEs\footnote{
This is equivalent to the usual second-order homogeneous ODE for $\delta^{(1)}$.
}~\cite{M.Bruni_etal_2014_Mar}
\begin{align}
\label{eq:evolution_linear_density}
    &4H\dot{\delta}^{(1)} +6H^2\Omega_m\delta^{(1)} = a^{-2}R^{(1)} \, ,
    \\
    &\dot{R}^{(1)}=0\, .
\end{align}
Thus $R^{(1)}$ and $\Rc$ are conserved quantities at linear order. Eq.~\eqref{eq:evolution_linear_density} has two solutions: the homogeneous, or decaying mode, $\delta_-\propto H$, associated with the Hubble expansion, and the growing mode, $\delta_+$, sourced by the 3-Ricci perturbation and thus directly related to $\Rc$. Here, we focus solely on the growing mode --- as the decaying mode can be safely set to zero in the matter-dominated era --- and express it in terms of $\Rc$ from Eq.~\eqref{eq:evolution_linear_density} by introducing the linear growth factor $f_1 := \dd \ln \delta / \dd \ln a \eqone \Omega_m^{6/11}$~\cite{P.J.E.Peebles_1980,L.Wang_P.J.Steinhardt_1998} as~\cite{M.Bruni_etal_2014_Mar,M.Bruni_etal_2014_Sep}
\begin{equation}\label{eq:delta(1)}
\delta^{(1)} = \frac{\nabla^2 \Rc}{ \mathcal{F} H^2}\,,
\end{equation}
where $\mathcal{F} = f_1 + (3/2)\Omega_m$. Note that, within the matter-dominated era we can set $\Omega_m = f_1 = 1$, $\mathcal{F} = 5/2$, and $\delta^{(1)}\propto a$, as the EdS model is a good approximation of the cosmological background. 

To express $\psi^{(1)}$ and $\chi^{(1)}$ in terms of $\Rc$ we now employ the deformation tensor $\Theta_{ij}$ associated with the dust fluid in the synchronous-comoving gauge. Indeed, within this gauge we can write $\Theta_{ij} = \bar{\Theta}_{ij} + \Theta^{(1)}_{ij} = \frac{1}{2}\dot{\gamma}_{ij}$, where the background part is given by  $\bar{\Theta}_{ij} = a^{2}H\delta_{ij}$ --- with trace $\bar{\Theta} = a^{-2}\delta^{ij}\bar{\Theta}_{ij} = 3H$ --- and $\Theta^{(1)}_{ij}$ identifies the deformation tensor. Then, we find the first-order trace to be $ \vartheta = \delta^{ij}\Theta^{(1)}_{ij} = -3\dot{\psi}^{(1)}$. Furthermore, the first-order continuity equations yield $\dot{\delta}^{(1)} = - \vartheta$, so that $\dot{\delta}^{(1)} = 3\dot{\psi}^{(1)}$. Then, from Eqs.~\eqref{eq:evolution_linear_density} and \eqref{eq:delta(1)} we find
\begin{equation}\label{eq:psi(1)}
    \psi^{(1)} = \frac{\nabla^2\Rc}{3\mathcal{F}H^2} + \Rc \, .
\end{equation}
Thus, using  Eqs.~\eqref{eq:Rc_metric} and \eqref{eq:psi(1)} we get
\begin{equation}\label{eq:chi(1)}
    \chi^{(1)} = - \frac{2\Rc}{a^2\mathcal{F}H^2} \, .
\end{equation}
Hence, the spatial metric in Eq.~\eqref{eq:spatial_metric} can be expressed in terms of $\Rc$ as 
\begin{equation}\label{eq:gammaij_Rc}
   \gamma_{ij}\eqone a^2\left[(1-2\Rc)\delta_{ij} - \frac{2
   }{a^2\mathcal{F}H^2}\partial_i\partial_j\Rc\right]\, ,
\end{equation}
which, for a general spatial coordinate system, can be cast in the form
\begin{equation}\label{eq:gammaij_Rcg}
   \gamma_{ij}\eqone a^2\left[(1-2\Rc)\hat{\gamma}_{ij} - \frac{2
   }{a^2\mathcal{F}H^2}D_iD_j\Rc\right]\, ,
\end{equation}
where $\hat{\gamma}_{ij}$ is the background metric in the given coordinate system (without the background scale factor), and $D_i$ is the covariant derivative operator associate to the same choice of spatial coordinates.

Finally, within the synchronous-comoving gauge, we also obtain an expression for the expansion tensor in terms of $\Rc$, which thus directly determines the evolution of the perturbed spatial metric, as~\cite{M.Bruni_etal_2014_Mar,M.Bruni_etal_2014_Sep}
\begin{equation}\label{eq:Thetaij_Rc}
    \Theta_{ij} \eqone a^2H\left[(1-2\Rc)\hat{\gamma}_{ij} - \frac{2+f_1}{a^2\mathcal{F}H^2}D_iD_j\Rc\right]\, .
\end{equation}
\section{\label{sec:LTB} The Szekers Solutions}
An exact family of solutions to Einstein’s field equations for a dust-dominated, gravitationally sourced spacetime with no symmetry was first found by Szekeres~\cite{P.Szekeres_1975a}, and later generalised to include a cosmological constant in ~\cite{N.Meures_M.Bruni_2011, I.DelgadoGasparBuchert_T.Buchert_2021}. Further generalisations of the Szekeres solutions to include alternative matter sources --- e.g., fluids with pressure, and non-perfect fluid --- have also been investigated in the literature~\cite{D.A.Szafron_1977,S.Najera_R.A.Sussman_2020,S.Najera_R.A.Sussman_2021,F.A.Pizana_etal_2024}. Owing to their generality, the Szekeres solutions have been widely employed within inhomogeneous cosmology to model both local and large scale structures~\cite{W.B.Bonnor_N.Tomimura_1976,S.W.Goode_J.Wainwright_1982,K.Bolejko_2006,K.Bolejko_2007,K.Bolejko_2008,K.Bolejko_etal_2009,K.Bolejko_2009,K.Bolejko_MN.Celerier_2010,N.Meures_M.Bruni_2011,M.Ishak_A.Peel_2012,A.Peel_etal_2012,N.Meures_M.Bruni_2012,R.A.Sussman_I.DelgadoGaspar_2015,R.A.Sussman_etal_2016,K.Bolejko_etal_2016,MN.Celerier_2024,M.Galoppo_etal_2025,B.Kalbouneh_etal_2025}, as well as to investigate relativistic gravitational collapse and BH formation~\cite{P.Szekeres_1975b, S.W.Goode_J.Wainwright_1982, M.Bruni_etal_1995_Jun, M.Bruni_1996,P.S.Joshi_A.Krolak_1996,S.S.Deshingkar_S.Jhingan_P.S.Joshi_1998,B.C.Nolan_U.Debnath_2007,T.Harada_S.Jhingan_2015,I.DelgadoGaspar_etal_2018}.

The Szekeres solutions naturally split into two inequivalent subclasses, commonly referred to as Class I and Class II, depending on whether the matter density admits an independent radial monopole contribution in synchronous, comoving coordinates~\cite{Krasiński_1997,J.Plebanski_A.Krasinski_2006,K.Bolejko_etal_2009}. In particular, it is Class I models which generalise LTB spacetimes by allowing for fully anisotropic, non-spherical inhomogeneities, while reducing exactly to LTB in the appropriate limit --- see e.g., Kras\'{i}nski~[Sec.~$2.4$]~\cite{Krasiński_1997}. By contrast, Class II models allow only for radially varying dipole terms in the density field, but admit a straightforward interpretation in terms of exact deviations from a homogeneous background (e.g., Goode and Wainwright~\cite{S.W.Goode_J.Wainwright_1982}).

In this work, we restrict our attention to Class I Szekeres models as these more naturally model realistic nonlinear gravitational collapse, and contain LTB spacetimes as their spherical restriction.

\subsection{The Hellaby parametrisation}
Following the prescription of Hellaby~\cite{C.Hellaby_1996}, we can adopt a convenient coordinate system such that the line element in synchronous and comoving coordinates, and in geometric units, can be written as
\begin{equation}\label{eq:Skmetric}
    \dd s^2 = -\dd t^2 +\frac{\left(R'-RD'/D\right)^2}{\epsilon-k}\dd r^2 + \frac{R^2}{D^2}\left(\dd p^2 + \dd q^2\right)\, ,
\end{equation}
where $r$ represents the radial coordinates, whilst $p$ and $q$ are projective angular coordinates. We denote with $'$ the derivative with respect to $r$, $R = R(t,r)$ is the areal radius, $k = k(r)$ is the adimensional curvature parameter, and the dimensionless dipole function, $D = D(r,p,q)$, is defined as
\begin{equation}
    D(r,p,q) = \frac{L}{2}\left[\left(\frac{p-P}{L}\right)^2+\left(\frac{q-Q}{L}\right)^2+\epsilon\right]\, ,
    \label{eq:Drpq}
\end{equation}
where $L$, $P$, and $Q$ are functions only of the radial coordinate and $\epsilon = \pm1,\, 0$. The parameter $\epsilon$ determines whether the $\{p, q\}$ 2-surfaces of constant $\{r, t\}$ are unit 2-spheres ($\epsilon = +1$), unit 2-pseudospheres, i.e., hyperboloids ($\epsilon = -1$), or 2-planes ($\epsilon = 0$). 

In this work, as we are interested in modelling gravitational collapse and BH formation, we consider the quasispherical Szekeres solutions ($\epsilon = +1$). These can be thought of as a set of nonconcentric, evolving spheres, which exhibit a dipole distribution in the energy density variation around each sphere, encoded in the $D'/D$ term in the metric~\cite{Krasiński_1997,J.Plebanski_A.Krasinski_2006}. It is then sometimes advantageous, e.g., when considering regularity conditions, to write~\cite{P.Szekeres_1975b,T.Harada_S.Jhingan_2015} 
\begin{equation}
    \frac{D'}{D} =: -\frac{\vec{n} \cdot \vec{\beta}}{r} \, ,
\end{equation}
where $\vec{n}$ is the unit directional vector in co-moving coordinates, and  $\vec{\beta}$ ultimately characterises the dipole strength and direction. In particular, the latter can be expressed as
\begin{equation}
    \vec{\beta} = r\left(\frac{P'}{L},\, \frac{Q'}{L},\, \frac{L'}{L} \right)\, .
\end{equation}

Here, the reader can find thorough descriptions of the mathematical and physical properties of these solutions in textbooks~\cite{Krasiński_1997, J.Wainwright_G.F.R.Ellis_1997, J.Plebanski_A.Krasinski_2006, K.Bolejko_etal_2009} as well as in a various papers, e.g.,~\cite{P.Szekeres_1975a,P.Szekeres_1975b, S.W.Goode_J.Wainwright_1982, M.M.deSouza_1985, M.Bruni_etal_1995_Jun, C.Hellaby_1996, C.Hellaby_A.Krasinski_2002, C.Hellaby_A.Krasinski_2008, A.Krasinski_2008, A.R.Sussman_K.Bolejko_2012, A.Krasinski_K.Bolejko_2012,G.Ira_C.Hellaby_2017,P.S.Apostolopoulos_2017,R.G.Buckley_E.M.Schlegel_2020}. The Einstein field equations written with a dust source and a cosmological constant reduce to the following two
\begin{align}\label{eq:SkFriedman}
    \left(\frac{\dot{R}}{R}\right)^2 = \frac{2GM}{R^3} -\frac{k}{R^2} + \frac{1}{3}\Lambda \, ,
\end{align}
\begin{equation}\label{eq:SkDensity}
    4\pi \rho = \frac{M' - 3MD'/D}{R^2(R'-RD'/D)} \, ,
\end{equation}
where $M = M(r)$ represents the total active gravitational mass in a shell determined by $r = const$, whilst $k(r)$ can be interpreted as the shell's spatial curvature, and  $\rho$ represents the local matter density. 

We note that the local matter density can diverge only in two circumstances: \textit{shell-focusing} singularities, where $R = 0$, and \textit{shell-crossing} singularities, where $R' - R D'/D = 0$. The former is typically associated with genuine gravitational collapse leading to a spacetime singularity, while the latter is interpreted as a breakdown of the fluid description --- arising when neighbouring matter shells intersect, a phenomenon the model cannot capture. In the following sections, we will clarify how these singularities relate to the classical dynamical characterisation of the end states of gravitational collapse.

Here, Eq.~\eqref{eq:SkFriedman} furthermore represents the analogue, for each $r = const$ shell, of the Friedman equation and thus can be formally integrated to give
\begin{equation}\label{eq:t}
    t - t_\mathrm{B}(r) = \int_0^R \frac{\dd \tilde{R}}{\sqrt{\frac{2GM}{R} - k + \frac{1}{3}\Lambda R^2}} \, ,
\end{equation}
where $t_\mathrm{B}(r)$ is an arbitrary function representing the Big Bang time across the spacetime. In this work, we consistently set $t_\mathrm{B}(r) = 0$, thereby neglecting the decaying mode of perturbations when re-casting the Szekeres solutions as nonlinear exact deviations from FLRW spacetimes. With this choice, the Szekeres model we consider is fully determined, a priori, by specifying five radial functions: \,$S(r),\, P(r),\,\ Q(r)$, and two between $M(r),\, k(r),\,\, \mathrm{and} \, \,R(t_*,r)$ (i.e., a spatial data for the areal radius on a given time slice $t = t_*)$ ), together with a cosmological constant $\Lambda$. However, since we are interested in modelling BH formation from the collapse of initial perturbations during the matter-dominated era, we may safely set $\Lambda = 0$ in the modelling of the perturbation collapse (although it will be maintained in the background dynamics).

In modelling BH formation, and the study of singularities, it is also meaningful to introduce the kinematic and curvature quantities discussed in Sec.~\ref{sec:1+3} in terms of metrics components. One finds (see e.g.,~\cite{MN.Celerier_2024})
\begin{align}
    & \Theta = 2\frac{\dot{R}}{R} + \frac{\dot{R}}{R}\left[\frac{\dot{R}'/\dot{R} - D'/D}{R'/R - D'/D}\right] \; , \label{eq:theta}\\
    & \sigma_+ = -\frac{1}{3}\frac{\dot{R}}{R}\left[\frac{\dot{R}'/\dot{R} - R'/R}{R'/R - D'/D}\right]\; , \label{eq:sigma+}\\
    & E_+ =-\frac{G}{6}\frac{M' - 3M(R'/R)}{R^2\left(R' - RD'/D\right)} \; , \label{eq:E+}\\ 
    &\Rsp =  \frac{6k}{R^2} + \frac{2k\left(k'/k -2R'/R\right)}{R\left(R'-RD'/D\right)}\; .\label{eq:R3}
\end{align}
\subsection{The Goode--Wainwright formalism}
An alternative formalism, which makes the interpretation of the Szekeres solutions as exact deviations of FLRW more transparent --- and simultaneously allows for a more streamlined study of the nature of their late-time singularities --- was developed by Goode and Wainwright\footnote{See App.~\ref{app:mapping} for the complete mapping between the Hellaby and Goode--Wainwright formalisms.}~\cite{S.W.Goode_J.Wainwright_1982}, and largely employed in structure formation studies within Szekeres cosmologies~\cite{N.Meures_M.Bruni_2011,N.Meures_M.Bruni_2012,M.Ishak_A.Peel_2012,A.Peel_etal_2012}. In this approach, the quasispherical Szekeres models are expressed in comoving and synchronous gauge, with the line element given by
\begin{equation}
    \dd s^2 = -\dd t^2 + S^2\left[e^{2\nu}\left(\dd \varsigma^2 +\dd \varkappa^2\right) + Z^2W^2\dd \xi^2\right] \, , \label{eq:GW_linelement}
\end{equation}
where $S = S(t,\xi)$, $W = W(\xi)$ and $Z =Z(t,\varsigma,\varkappa,\xi)$ are all positive functions. In addition, we have
\begin{align}
    Z(t,\varsigma,\varkappa,\xi) &= A(\varsigma,\varkappa,\xi)  - F(t,\xi) \, ,
\end{align}
where $A$ ultimately encodes the dipole structure of the Szekeres solution and 
\begin{equation}
    F(t,\xi) = \beta_+(\xi)f_+(t,\xi) + \beta_-(\xi)f_-(t,\xi) \, .
\end{equation}
Furthermore, we can write (for quasispherical models)
\begin{equation}
    W(\xi)^2 = \left(1 - \tilde{k}f(\xi)^2\right)^{-1} \, ,
\end{equation}
where $\tilde{k} =\pm 1, 0$ represents the sign of the curvature of the $\xi = const$ shells, and $f(\xi)$ is a free function. We also get
\begin{equation}
    A(\varsigma,\varkappa,\xi) = f(\xi)\nu_{,\xi}(\varsigma,\varkappa,\xi) -\tilde{k}\beta_+(\xi)\, ,
\end{equation}
where 
\begin{equation}
        e^{\nu} = f \left[a\left(\varsigma^2+\varkappa^2\right)+2b\varsigma +2c\varkappa + d\right]^{-1} \, , 
\end{equation}
with $a,~b,~c,~\mathrm{and}~d$ functions only of the radial $\xi$ coordinate, and satisfying the one constraint
\begin{equation}
    ad - b^2 - c^{2} = \frac{1}{4} \, . \label{eq:constraintGW}
\end{equation}
Within this formalism, the Einstein's equations for a pure dust source --- i.e., by already considering $\Lambda = 0$ --- reduce to the following two:
\begin{align}
    \left(\frac{\dot{S}}{S}\right)^2 &= \frac{2G\tilde{M}}{S^3}-\frac{\tilde{k}}{S^2} \, ,\label{eq:GwFriedman}\\
     4\pi G\rho &= \frac{3\tilde{M}}{S^3}\left(1+\frac{F}{Z}\right)\, ,\label{eq:rhoGW}
\end{align}
where $\tilde{M} = \tilde{M}(r)$ gives a measure of the active gravitational mass within the shell defined by $\xi = const$.

Remarkably, within this formalism the $F$ function satisfies the evolution equation
\begin{equation}\label{eq:ODESk}
    \ddot{F} + 2\frac{\dot{S}}{S}\dot{F} -\left(\frac{3\tilde{M}}{S^3}\right)F = 0 \, ,
\end{equation}
i.e., formally the same equation satisfied by the linear dust density contrast perturbation on an FLRW background. Hence, as in linear perturbation theory, the two independent solutions of Eq.~\eqref{eq:ODESk} can be directly interpreted as the growing and decaying modes of the dust density with respect to the local FLRW background defined within the shell of $z = \text{const}$. Remarkably, this identification remains valid at the nonlinear, exact level, not only when the density contrast is small. In addition, we can write the amplitudes of the growing and decaying mode, i.e., $\beta_\pm$, as
\begin{align}
    &\beta_+ = -f\tilde{k}\frac{\tilde{M}_{,\xi}}{3\tilde{M}}\; \; ; \; \; \beta_- = \frac{fT_\mathrm{B, \xi}}{6\tilde{M}} \, , \label{eq:beta+}
\end{align}
where we have identified as the bang time in the GW parametrisation the free function $T_\mathrm{B}(\xi)$. Therefore, we find that, although the growing and decaying modes are defined a priori with respect to a local FLRW background, the vanishing of the global decaying mode --- namely, setting a synchronous big bang time --- corresponds to setting all local decaying modes to zero, i.e., $\beta_- = 0$.

Finally, we can express within this formalism the kinematic and curvature variables to find (see e.g.,~\cite{S.W.Goode_J.Wainwright_1982})
\begin{align}
    & \Theta = 3\frac{\dot{S}}{S} - \frac{\beta_+\dot{f}_+}{A -\beta_+{f}_+} \, , \label{eq:thetaGW}\\
    & \sigma_+ = \frac{1}{3}\frac{\beta_+\dot{f}_+}{A -\beta_+{f}_+} \, , \label{eq:sigma+GW}\\
    &  E_+ = -\frac{G\tilde{M}}{S^3}\frac{\beta_+{f}_+}{A -\beta_+{f}_+}\, , \label{eq:E+GW}\\
    & \Rsp = \frac{6}{S^2}\left[\tilde{k}+\frac{2\beta_+}{3Z}(1+\tilde{k}f_+)\right]\, ,
\end{align}
where in expressing these quantities we have already set the condition of vanishing decaying modes. 
\subsection{LTB Models}
The quasispherical Szekeres solutions contain the spherically symmetric LTB models~\cite{G.Lemaitre_1933,R.C.Tolman_1934,H.Bondi_1947}, which are recovered by requiring a vanishing dipole, i.e., $D' = 0$ in the Hellaby representation. These spherical models --- representing a natural first generalisation of the TH profile in the context of gravitational collapse --- are of great interest both in the study of inhomogeneous cosmology~\cite{F.C.Mena_R.Tavakol_1999,A.Krasinski_C.Hellaby_2001,A.Krasinski_C.Hellaby_2004,MN.Celerier_etal_2010,V.Marra_etal_2022,M.Sarma_C.Marinoni_B.Kalbouneh_C.Clarkson_R.Maartens_2025}, as well as gravitaional collapse and BH formation~\cite{C.W.Misner_D.H.Sharp_1964,P.Yodzis_H.J.Seifert_H.MullerzumHagen_1973,D.M.Eardley_L.Smarr_1979,A.G.Polnarev_M.Y.Khlopov_1981,D.Christodoulou_1984,C.Hellaby_K.Lake_1985,V.Gorini_G.Grillo_M.Pelizza_1990,P.S.Joshi_I.H.Dwivedi_1993,T.P.Singh_P.S.Joshi_1996,S.Jhingan_P.S.Joshi_T.P.Singh_1996,T.Harada_C.Goymer_B.J.Carr_2002,M.Kopp_etal_2011,P.S.Joshi_D.Malafarina_2015,T.Kokubu_etal_2018,K.Uehara_etal_2025}. In particular, due to their simplicity, they provide an ideal toy model for investigating nonlinear general relativistic effects in the strong-gravity regime of the collapse. The reader can find a comprehensive and detailed exposition of the LTB solution in Pleb\'{a}nski and Kras\'{i}nski~[Sec.~$18$]~\cite{J.Plebanski_A.Krasinski_2006}, Bolejko et al.~[Sec.~$2.1$]~\cite{K.Bolejko_etal_2009}, and Vittorio~[Sec.~$3.1$]~\cite{N.Vittorio_2018}.

The LTB line element, in synchronous-comoving spherical coordinates is conventionally written as 
\begin{equation}
    \dd s^2 = - \mathrm{d}t^2 + \frac{R'^2}{1-k}\dd r^2+ R^2\dd^2\Omega \, , 
    \label{eq:metricLTB}
\end{equation}
where, the areal radius $R$ satisfies Eq.~\eqref{eq:SkFriedman}, and, following Eq.~\eqref{eq:SkDensity} for a dust only spacetime, the matter density is given by 
\begin{equation}
    4\pi \rho =  \frac{M'}{R^2R'}\, .
\end{equation}
In addition, the kinematic and curvature variables can be directly expressed by setting the term $D'/D = 0$ in Eqs.~\eqref{eq:theta}--\eqref{eq:R3}. We find
\begin{align}
    & \Theta = 2\frac{\dot{R}}{R} + \frac{\dot{R}'}{R'} \; , \label{eq:thetaLTB}\\
    & \sigma_+ = -\frac{1}{3}\left(\frac{\dot{R}'}{R'} - \frac{\dot{R}}{R}\right)\; , \label{eq:sigma+LTB}\\
    & E_+ =-\frac{G}{6}\left(\frac{M'}{R^2R'} - \frac{3M}{R^3}\right) \; , \label{eq:E+LTB}\\ 
    &\Rsp =  \frac{2k}{R^2} +\frac{2k'}{RR'}\; .\label{eq:R3LTB}
\end{align}
\section{\label{sec:Collapse} Exact Gravitational Collapse}

The collapse of overdensities is described by models in which radial shells have a positive curvature parameter, $k(r) > 0$. In particular, we focus on configurations where either $\lim_{r \to \infty} k(r) = 0$ or, alternatively, such that $k(r)$ smoothly matches a flat region at a finite coordinate radius whilst maintaining a central positive core, $k(r)>0$. Then, these models naturally describe localised \textit{compensated} overdensities, namely overdense regions surrounded by an underdense environment and embedded in a spatially flat background, so that the average density contrast within these models ultimately vanishes\footnote{Additionally, we note that the matching of such models to regions with negative curvature parameter would unavoidably introduce shell-crossing singularities at finite time, which are unrelated to the physical collapse of the overdense core~\cite{T.Harada_S.Jhingan_2015}. Since these shell-crossings obscure the analysis of genuine gravitational collapse without introducing qualitatively new behaviour, such Szekeres and LTB regions are excluded in the present work.}.

For these positively curved models, by neglcting $\Lambda$, we can write standard exact parametric solutions for Eq.~\eqref{eq:SkFriedman}, namely 
\begin{align}
    &t = \frac{ GM(r)}{k(r)^{3/2}}\left( \eta - \sin\eta  \right)\, ,\label{eq:t+} \\
    &R = \frac{GM(r)}{k(r)}\left(1-\cos\eta\right)\, . \label{eq:R+}
\end{align}
In addition, for $k(r) = 0$ --- since we have selected $t_\mathrm{B}(r) = 0$ --- the resulting Szekeres (or LTB) model is equivalent to a flat FLRW spacetime written in a non-standard coordinate system. Hence, its dynamics simply follows that of either an EdS or $\Lambda$CDM background.

For $k(r) > 0$ we can instead employ this set of equations to express the matter, kinematic, and curvature variables of the 1+3 approach in terms of $M(r)$, $k(r)$ and either the time coordinate or the $\eta$ parameter. These are reported in App.~\ref{app:param} for completeness.

\subsection{Relevant collapse times}

By focusing on the collapsing shells with positive curvature parameter and employing Eqs.~\eqref{eq:SkFriedman}, \eqref{eq:t+} and \eqref{eq:R+}, we can immediately extract the turnaround and collapse time of a shell, i.e.,
\begin{align}
    & t_{\mathrm{ta}} (r) = \frac{\pi G M(r)}{k(r)^{3/2}} \, , \label{eq:tta}\\
    & t_{\mathrm{col}} (r) = \frac{2\pi G M(r)}{k(r)^{3/2}} \, . \label{eq:tcol}
\end{align}
We note that, as we would expect from an inhomogeneous collapse, these are functions of the radial coordinate $r$, and not constant as in a conventional TH model. 

In addition, as the gravitational collapse of an overdensity into a BH in a Szekeres or LTB spacetime is identified through the formation of an apparent horizon (AH) we are interested in defining the time at which an AH might form. Following~\cite{T.Harada_S.Jhingan_2015}, we employ the definition of AH introduced by Kras\'{i}nski and Bolejko for Szekeres models~\cite{A.Krasinski_K.Bolejko_2012_AHQS}, and consider as AH the locus of marginally trapped two-surfaces of the form $\{t=\mathrm{const}\,, r=\mathrm{const}\}$, characterised by the vanishing of the expansion scalar of the outgoing (or ingoing) null congruences orthogonal to the surface. In accordance with the terminology of Harada and Jhingan~\cite{T.Harada_S.Jhingan_2015}, we designate an AH with vanishing outgoing (respectively ingoing) null expansion as a future (respectively past) apparent horizon\footnote{
Note that, in this context, a past horizon does not correspond to BH formation. Rather, it should be identified with the generalisation of a cosmological Hubble horizon to Szekeres and LTB models.
}. In what follows, we refer to the former simply as the future horizon FH. 

We can now consider the trapping condition for the selected two-surfaces. Their tangent space is spanned by $\partial_p$ and $\partial_q$, so that any normal 4-vector $n^\mu$ to these must have $n^p = n^q = 0$. Thus, for a congruence of null geodesic with tangent 4-vector $k^\mu$ normal to the ${t=\mathrm{const}\,, r=\mathrm{const}}$ two-surfaces we find
\begin{equation}
    k_\mu k^\mu = 0 \;\; ; \;\; k^\nu \nabla_\nu k^\mu = 0 \, . \label{eq:kconditions}
\end{equation}
By assuming $k^t >0$, then $k^r > 0$ (respectively $k^r <  0$) identifies outgoing (ingoing) null geodesic. Furthermore, we can always normalise, on and only on the two-surface, $k^r = \alpha =\pm 1$ by choice of the affine parameter. We then must consider the expansion scalar $\nabla_\mu k^\mu$, i.e.,
\begin{align}
    \nabla_\mu k^\mu =& \left({k}^t\right)^{\dot{}} + \left({k}^r\right)^{'} +   \left(\sqrt{\gamma_{rr}}\right)^{\dot{}} +\sqrt{\gamma_{rr}}\left[\log(\gamma_{pp})\right]^{\dot{}} 
    \label{eq:kexpnasion} \\& +\alpha\left(\left[\log(\sqrt{\gamma_{rr}})\right]^{'}+\left[\log(\gamma_{pp})\right]^{'}\right)\, ,  \nonumber
\end{align}
where $\gamma_{rr} = \left(R'-RD'/D\right)/{\sqrt{1-k}}$, and $\gamma_{pp} = \gamma_{qq} = R/D$ according to the line element in Eq.~\eqref{eq:Skmetric}.
We can then differentiate the first condition in Eq.~\eqref{eq:kconditions} with respect to the time coordinate and consider the radial component of the second constraint to eliminate the first two terms in Eq.~\eqref{eq:kexpnasion}, to obtain~\cite{T.Harada_S.Jhingan_2015}
\begin{equation}
    \nabla_\mu k^\mu = 2\left(\frac{R'}{R} - \frac{D'}{D}\right)\left[\mathrm{sgn}\left(\dot{R}\right)\sqrt{1 + \frac{2M- R}{R(1-k)}} + \alpha\right] \, .
\end{equation}
Hence --- as we could expect --- we find that a FH ($\alpha = 1$) only forms during the collapse ($\dot{R} < 0$) when the condition $R = 2M$ is satisfied, since the term $\left(R'/R-D'/D\right)$ cannot change sign but at a shell-crossing singularity.

Note that from this result we find that, using Eqs.~\eqref{eq:t+} and~\eqref{eq:R+}, the time at which a future horizon forms on a radial shell is given by 
\begin{equation}
    t_{\mathrm{FH}}(r) = 2 GM(r)\left[\frac{\pi-\arcsin\sqrt{k(r)}}{k(r)^{3/2}} + \frac{\sqrt{1-k(r)}}{k(r)}\right]\, . \label{eq:tFH}
\end{equation}

\subsection{Regularity conditions}\label{subSec:regcond}

\subsubsection{Initial data}
To model BH formation we assume the existence of a regular initial data surface ($t = t_i$). Since $R_i(r)$ represents the areal radius of the $\{t = const, r = const\}$ surfaces, we assume it as a monotonic function of $r$, and we choose the latter such that $R_i(r) = \mathcal{O}(r)$ --- where the subscript $i$ denotes evaluation at the initial time $t_i$. 
Furthermore, for the centre to be locally Minkowski we must require $k(0) = 0$. Then, as we require $\rho_i$ and $\dot{R}_i$ to be bounded for $r \rightarrow 0$, one finds the conditions~\cite{T.Harada_S.Jhingan_2015}
\begin{equation}
    M(r) = \mathcal{O}(r^3) \;\; ; \;\; k(r) = \mathcal{O}(r^2) \, .
\end{equation}
To avoid shell-crossing singularities in the initial data we then must have 
\begin{equation}\label{eq:Icregular1}
    R_i' - R_i D'/D > 0 \Longrightarrow \left(\frac{R_i'}{R_i}\right)^2> \frac{\beta^2}{r^2} \, ,
\end{equation}
for which we have assumed all the dipole functions appearing in the metric to be $C^2$ throughout the coordinate chart, which imply $\beta(r) = \mathcal{O}(r)$. Notice that, to have a positive definite local dust density (see Eq.~\eqref{eq:SkDensity}) we must also require
\begin{equation}
    \frac{M'}{M} > 3\frac{\beta}{r} \, . \label{eq:no-shell1}
\end{equation}

\subsubsection{Shell-crossing}
In~\cite{P.Szekeres_1975b, T.Harada_S.Jhingan_2015, C.Hellaby_A.Krasinski_2002, C.Hellaby_A.Krasinski_2008} the conditions to avoid shell-crossing singularities have been extensively studied. In particular, one finds that collapsing models are free from shell-crossing singularities only if, in addition to the constraint in Eq.~\eqref{eq:no-shell1}, we have
\begin{align}
    &r\left[\ln\left(\frac{M^{2/3}}{k}\right)\right]' > 0\, . \label{eq:no-shell2}
\end{align}
Interestingly, this condition shows no explicit dependence on the anisotropy of the Szekeres solution --- this is fully encoded only in the constraint in Eq.~\eqref{eq:no-shell1}. The condition in Eq.~\eqref{eq:no-shell2} is equivalent to~\cite{D.Vrba_O.Svitek_2014,T.Harada_S.Jhingan_2015}
\begin{equation}
\frac{\langle\rho\rangle_i}{\hat{\rho}_i} > 1 \, ,
\end{equation}
where, following~\cite{T.Harada_S.Jhingan_2015}, $\langle\rho\rangle$ denotes the quasi-local average dust density\footnote{This is defined by the average $\langle\rho\rangle = \int_0^r\dd r\int\rho\mathcal{J}\sqrt{\gamma}\dd\Omega/\int_0^r\dd r\int\mathcal{J}\sqrt{\gamma}\dd\Omega$, where $\mathcal{J}:= \sqrt{1-k}$ represents a measure a curvature weighting on each shell (see e.g.,~\cite{A.R.Sussman_K.Bolejko_2012}).} within a sphere of coordinate radius $r$, and $\hat{\rho}$ is the proper angle-averaged dust density on the bounding shell. Namely, we find that shell-crossing is avoided in the collapse if the average density within a shell is higher than that on the shell surface. Thus, the onset of shell-crossing singularities is directly tied to the \textit{magnitude} of the density concentration, giving a first physical intuition about the meaning of Eq.~\eqref{eq:no-shell2}.

However, we further note that the condition in Eq.~\eqref{eq:no-shell2} is formally equivalent to requiring,
\begin{equation}
t'_{\mathrm{col}} = 2\pi G\left(\frac{M}{k^{3/2}}\right)' > 0 \, ,
\end{equation}
namely, that shells at larger coordinate radius collapse more slowly than the internal ones --- as expected for any model in which no shell-crossing can occur, thus making also the constraint in Eq.~\eqref{eq:no-shell2} straightforward.

Remarkably, this condition can also be recast as a statement about the sign of the initial non-null shear eigenvalue of a given shell (see Eq.~\eqref{eq:sigma+GW}). In particular, a negative initial $\sigma_+$ (i.e., a positive shear in the radial direction, $\sigma_{11}$) does not generate shell–crossing during the collapse, whereas a positive initial $\sigma_+$ (negative $\sigma_{11}$) corresponds to a pancake–like collapse and the onset of shell–crossing behaviour. Namely, the formation of cigar, pancakes or point-like singularities during the collapse is determined already at the level of the initial data for the radial shear.

Furthermore, in~\cite{T.Harada_S.Jhingan_2015} by analysing the behaviour of $R'/R$ it was shown that a shell-crossing singularity on a given shell will form after the FH at the shell only if 
\begin{equation}
    3\left[\ln\left(\frac{M^{2/3}}{k}\right)\right]'X_-(k) + \frac{M'}{M} \geq 3\frac{\beta}{r} \, , \label{eq:no-shell3}
\end{equation}
where $X_-(k)$ is given by 
\begin{equation}
    X_-(k) := 1 -\frac{1}{1+(2/3) \dd\!\ln G_-(k)/\dd \!\ln k } \, ,
\end{equation}
with $G_-(k(r)) = t_{\mathrm{FH}}(r)/2M(r)$. It is thus possible to relax the constraint in Eq.~\eqref{eq:no-shell2} to that of Eq.~\eqref{eq:no-shell3} and still obtain, a priori, BH formation as any shell-crossing appear only after the formation of a FH. 

\subsubsection{Naked singularities}

In a gravitational collapse, the nature of a singularity --- i.e., whether it is covered, locally naked, or globally naked --- determines its physical interpretation. Indeed, only covered and locally naked singularities can be understood as physically meaningful in the context of BH formation. A covered singularity is one entirely hidden within an event horizon (or an AH); a locally naked singularity is visible after the horizon forms only to nearby observers; and a globally naked singularity is visible to distant observers along future-directed null geodesics.

Within Szekeres and LTB models naked singularities are known to be a potential outcome of the gravitational collapse~\cite{P.Yodzis_H.J.Seifert_H.MullerzumHagen_1973,D.M.Eardley_L.Smarr_1979,D.Christodoulou_1984,P.S.Joshi_I.H.Dwivedi_1993,T.P.Singh_P.S.Joshi_1996,S.Jhingan_P.S.Joshi_T.P.Singh_1996,P.S.Joshi_A.Krolak_1996,S.S.Deshingkar_S.Jhingan_P.S.Joshi_1998,B.C.Nolan_U.Debnath_2007}. Thus, for BH formation --- which either avoid shell-crossing altogether or allow it only after the formation of a FH --- one must examine the nature of shell-focusing singularities. From Eqs.~\eqref{eq:tcol} and~\eqref{eq:tFH} we have
\begin{equation}
t_{\mathrm{col}}(r) > t_{\mathrm{FH}}(r) \quad \forall\,  r>0 \, .
\end{equation}
Hence any non-central shell-focusing singularity is necessarily a covered singularity: a FH forms on the shell prior to its collapse and the appearance of the singularity. Consequently, for shell-focusing singularities, only the central singularity can, in principle, be locally or globally naked.

Assuming a smooth $k(r)$ near the origin --- so that $k(r)$ admits an expansion of the type $k(r) = k_2 r^2 + k_4 r^4 + \mathcal{O}(r^6)$ --- it was shown in~\cite{T.Harada_S.Jhingan_2015} that a sufficient condition for the central singularity to be covered is $k_2 > 0$ (necessary for a collapsing region), and $k_4 \geq 0$. Instead, for $k_4 < 0$, the central singularity is naked. In particular, in the context of BH formation, a sufficient requirement for ensuring that the singularity is at most locally naked is
\begin{equation}
t_\mathrm{col}(0) \;\ge\;
\min\limits_{r \leq r_s} t_{\mathrm{FH}}(r) \, , \label{eq:no_nakey}
\end{equation}
where we have introduced $r_s = 2 M_{\mathrm{BH}}$ and $M_{\mathrm{BH}}$ denotes the BH mass.

Although one could, in principle, relax this condition by accounting for the light-travel time between the formation of the singularity and the appearance of the FH at larger $r$, implementing this is ultimately impractical: the null geodesic equations are generally not analytically integrable in either LTB or Szekeres spacetimes. Hence, in this work we adopt the more stringent requirement~\eqref{eq:no_nakey} to identify viable models of BH formation for a given comoving curvature initial perturbation.

\subsection{Causal structure of the collapse}

Here, we concentrate on the causal nature of shell-focusing singularities as the occurrence of shell-crossing singularities, which is a genuinely curvature singularity and can be naked, 
is believed to be crucially subject to the validity of dust fluid assumption in a high-density regime~\cite{P.Yodzis_H.J.Seifert_H.MullerzumHagen_1973}.

\begin{figure*}[t]
    \centering
     \includegraphics[width=0.4\linewidth]{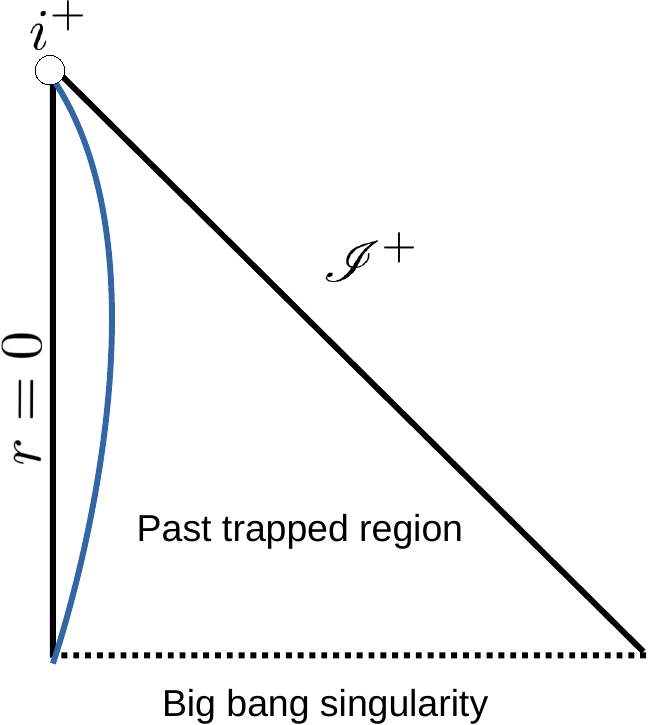}\hfill
     \includegraphics[width=0.49\linewidth]{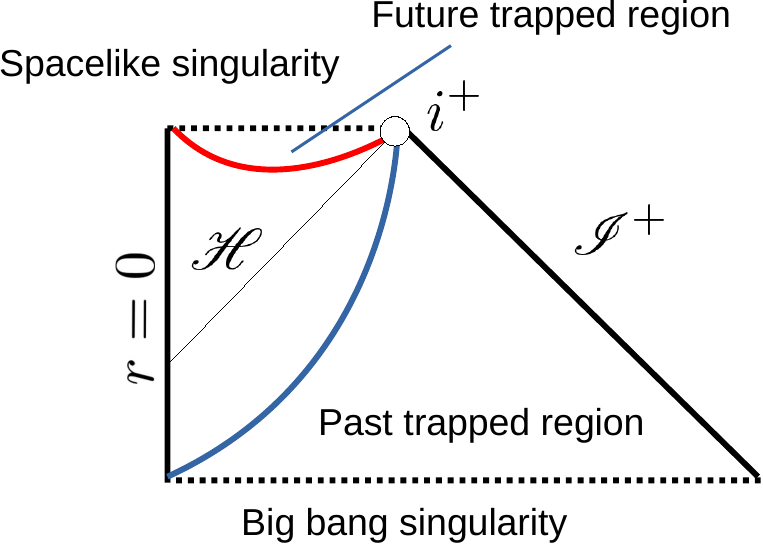}\\
\vspace{0.8cm}
    \includegraphics[width=0.49\linewidth]{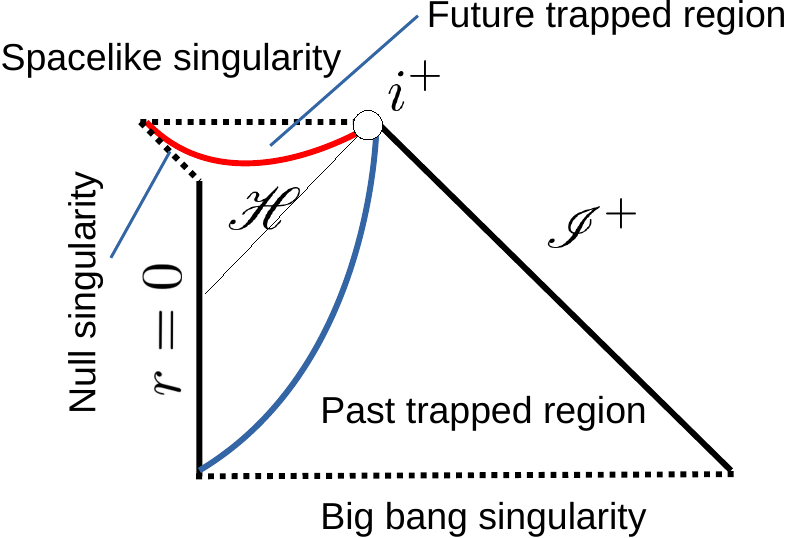}\hfill
     \includegraphics[width=0.49\linewidth]{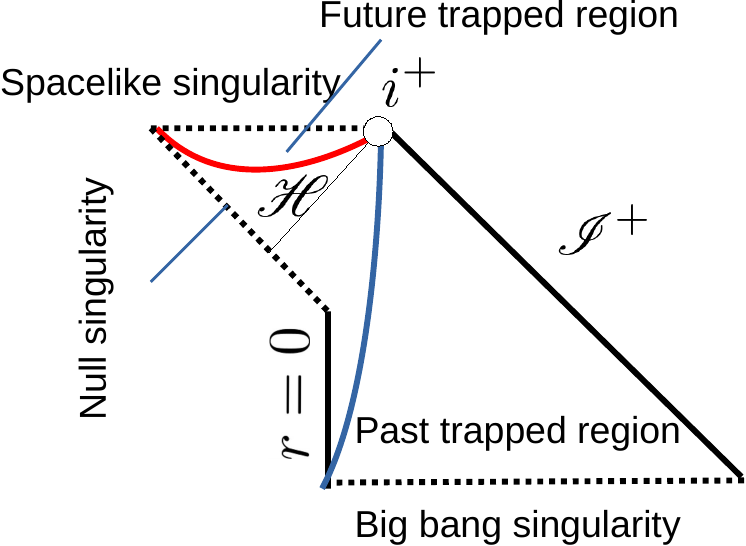}
     \caption{
     Schematic Penrose diagrams illustrating the causal outcomes relevant to the collapse models considered in this work. The upper-left panel shows the unperturbed expanding EdS background, with a spacelike Big Bang singularity and a past trapped region associated with the cosmological expansion, where the blue curve denotes a past trapping horizon, corresponding to a Hubble horizon in this case. The upper-right panel represents a conventional picture of BH formation: a future trapped region forms associated with a future trapping horizon denoted by a red curve followed by the formation of a spacelike singularity. However, it is known that there generically appears a null singularity at the centre in the LTB class of solutions. The bottom-left panel shows a locally naked singularity, from which outgoing null geodesics can emerge only locally from the null singularity but finally terminate at the spacelike singularity. The bottom-right panel displays a globally naked singularity, where an earlier set of outgoing null geodesics from the null singularity reach the future null infinity. In the present work, only the covered cases shown in the top-right and bottom-left panels are retained as viable BH formation channel.
     }
     \label{fig:penrose_collapse}
\end{figure*}

The conditions discussed above determine not only whether a shell-focusing singularity forms, but also its causal character. In particular, the relevant distinction for BH formation is whether the singularity is hidden by a future apparent horizon, or whether future-directed null geodesics can escape from it. The possible causal outcomes are schematically summarised in Fig.~\ref{fig:penrose_collapse}.

The first case is the unperturbed expanding dust background, represented here by the EdS spacetime. This diagram provides the asymptotic cosmological reference structure for the collapse models considered in this work. The spacetime originates from a spacelike Big Bang singularity and contains a past trapped region associated with the cosmological expansion, but no future trapped region associated with BH formation. The cosmological expansion is accompanied  with a Hubble horizon, which corresponds to a past trapping horizon
denoted by a blue curve in the top-left panel.

Once an initially expanding overdensity turns around and collapses, a future trapped region may form. If the future apparent horizon forms before the shell-focusing singularity forms, the collapse leads to a genuine BH spacetime as no distant observer can see the singularity. In the LTB and Szekeres models considered here, as discussed above, every non-central shell-focusing singularity is covered. The only possible obstruction to BH formation is therefore the causal
character of the central singularity. The conventional picture of BH formation involves the formation of a spacelike singularity and the resultant Penrose diagram is provided in the top-right panel. This diagram is the combination of the celebrated Oppenheimer-Snyder model~\cite{Oppenheimer:1939ue} with the EdS spacetime. Nevertheless, in the LTB class of solutions, it is well known that the formation of a central null shell-focusing singularity is also generic~\cite{D.M.Eardley_L.Smarr_1979,D.Christodoulou_1984,P.S.Joshi_I.H.Dwivedi_1993,T.P.Singh_P.S.Joshi_1996,S.Jhingan_P.S.Joshi_T.P.Singh_1996,T.Kokubu_etal_2018,K.Uehara_etal_2025}.

If future-directed outgoing null geodesics emerge from the central singularity but none of them reaches future null infinity, the singularity is locally naked. A possible Penrose diagram in this case is displayed in the bottom-left panel. In this case the singularity can be visible only to observers in a finite neighbourhood of the centre. By contrast, if outgoing null geodesics from the central shell-focusing singularity reach the future null infinity, the singularity is globally naked. This is depicted in the bottom-right panel. This corresponds to a genuine failure of BH formation in the present setting: distant observers can receive signals from the singularity, and the collapse cannot be interpreted as producing a physically acceptable BH seed. Initial comoving curvature profiles leading to either these causal structure are therefore excluded from the viable formation channels considered below. In fact, the exact threshold between the locally and globally naked cases can only be determined through the numerical integration of radial null geodesics in the LTB class~\cite{D.M.Eardley_L.Smarr_1979,T.Kokubu_etal_2018} and more general null geodesics in the Szekeres~\cite{S.S.Deshingkar_S.Jhingan_P.S.Joshi_1998,B.C.Nolan_U.Debnath_2007}. Nevertheless, the criterion that the central shell-focusing forms before the future trapping horizon forms acts as a useful sufficient condition for the viable BH formation~\cite{T.Harada_S.Jhingan_2015}.

\subsection{Types of kinematic singularities}

In characterising the gravitational collapse of Szekeres and LTB models, it is of central interest to determine which kinematic types of singularities are admissible. Here, we thus begin by following the original treatment of Goode and Wainwright~\cite{S.W.Goode_J.Wainwright_1982}, as such an investigation is informative in assessing the role played by the initial anisotropic shear in determining BH formation from CDM perturbations.  

Following the discussion of Sec.~\ref{sec:1+3}, and using Eqs.~\eqref{eq:thetaGW} and~\eqref{eq:sigma+GW}, the expansion tensor eigenvalues in LTB and Szekeres models may be written explicitly in terms of the metric functions in the GW formalism as (see also~\cite{S.W.Goode_J.Wainwright_1982})
\begin{align}
    \Theta_{(1)}  &= \frac{\dot{S}}{S} - \frac{\beta_+\dot{f}_+}{A-\beta_+f_+} \,,
    \label{eq:Theta1} \\
    \Theta_{(2)} &= \Theta_{(3)}  = \frac{\dot{S}}{S}\,, 
    \label{eq:Theta23}
\end{align}
where the index $(1)$ labels the principal direction associated with the anisotropic mode (i.e., the radial direction), while $(2)~\mathrm{and}~(3)$ correspond to the degenerate transverse directions (namely, the angular directions).

From Eqs.~\eqref{eq:Theta1}--\eqref{eq:Theta23}, the associated principal length scales along the eigen-directions follow as
\begin{align}
    \ell_1 &= S\left(A - \beta_+f_+\right)\,,
    \label{eq:l1}\\
    \ell_2 &= \ell_3 = S\,.
    \label{eq:l2l3}
\end{align}
Eqs.~\eqref{eq:l1}--\eqref{eq:l2l3} imply that a vanishing of $\ell_1$ at finite $S$ (equivalently $R$, up to a time-independent rescaling, see App.~\ref{app:mapping}) corresponds to a shell-crossing singularity, while $S \to 0$ signals shell-focusing collapse. Consequently, \textit{pancake-like singularities are necessarily associated with shell-crossing}, whereas \textit{point-like and cigar-like singularities can only arise from shell-focusing}. 

Furthermore, since the growing mode $f_+$ diverges more rapidly than the background scale factor $S$ vanishes during the collapse phase, and since one may always choose $A>0$ without loss of generality~\cite{S.W.Goode_J.Wainwright_1982}, it follows that the late–time behaviour of the collapse is controlled by the sign of $\beta_+$. In particular, the condition $\beta_+=0$ yields a point-like collapse, whereas $\beta_+<0$ leads to a cigar-like singularity and $\beta_+>0$ to a pancake-like collapse.

It is then instructive to recast the condition linking the sign of $\beta_+$ to the final singularity type within the Hellaby formalism. From Eqs.~\eqref{eq:beta+} and~\eqref{eq:App_Useful} (see App.~\ref{app:mapping}), one finds
\begin{equation}
    \beta_+ = -\frac{1}{3}\sqrt{k}\left(\frac{M'}{M}-\frac{3}{2}\frac{k'}{k}\right) \propto -t'_{\mathrm{col}} \propto 1-\frac{\langle\rho\rangle_i}{\hat{\rho}_i} \, ,
\end{equation}
linking $\beta_+$ to the gradient of $t_\mathrm{col}$, and to the ratio between the initial sphere and shell averaged densities, $\langle\rho\rangle_i/\hat{\rho}_i$.

In particular, we see that a positive gradient of the collapse time, $t'_{\mathrm{col}}>0$, implies that inner shells collapse earlier than outer ones and therefore favours shell–focusing collapse without shell–crossing. This corresponds to $\beta+<0$, for which the growing anisotropic mode decreases contraction along the principal direction relative to the transverse ones, yielding a cigar-like singularity. Conversely, $t'_{\mathrm{col}}<0$ implies that outer shells overtake inner ones, inevitably producing shell–crossing prior to global focusing; this corresponds to $\beta+>0$ and results in a pancake-like singularity. The marginal case $t'_{\mathrm{col}}=0$, or equivalently $\beta+=0$, corresponds to a simultaneous collapse of neighbouring shells and hence to an isotropic, point-like singularity. An equivalent interpretation follows then directly in terms of the initial averaged density profiles.

A further characterisation of the dynamical behaviour close to the singularities can be obtained by a dynamical system analysis of the Szekeres and LTB evolution equations, firstly applied to silent universe in~\cite{M.Bruni_etal_1995_Jun,M.Bruni_etal_1995_Jul,M.Bruni_1996,J.Wainwright_G.F.R.Ellis_1997}. It is convenient to introduce a novel set of dynamical variables normalised with respect to the expansion scalar, i.e.,
\begin{equation}
    \Omega := \frac{12\pi G\rho}{\Theta^2}\, , \, \, \, \, \, \,\Sigma_+ := \frac{\sigma_+}{\Theta} \, , \, \, \, \, \, \,\mathcal{E}_+ := \frac{E_+}{\Theta^2}\, .
\end{equation}
We also introduce a new time variable $\tau = \pm \int \Theta \dd t \, ,$ where the $\pm$ is given by the sign of the expansion scalar. We can then cast the ODEs system of Eqs.~\eqref{eq:Sk1}--\eqref{eq:E++} in the form (for $\Theta <0$)
\begin{align}
    & \partial_\tau\ln|\Theta| = \left[\frac{1}{3} + 6\Sigma_+^2 + \frac{\Omega}{6}\right] \, , \\
    & \partial_\tau\Omega = -\frac{\Omega}{3}\left[36\Sigma_+^2 -1 +\Omega\right]\, ,\\
    &\partial_\tau \Sigma_+= \Sigma_+\left[\frac{1}{3}-6\Sigma_+^2 - \Sigma_+ -\frac{\Omega}{6}\right] + \mathcal{E}_+ \, ,\\
    &\partial_\tau \mathcal{E}_+= \mathcal{E}_+\left[\frac{1}{3}-12\Sigma_+^2 + 3\Sigma_+ -\frac{\Omega}{3}\right] +  \frac{1}{6}\Sigma_+\Omega\, .
\end{align}
To move to the case $\Theta >0$, a simple sign change in the r.h.s.\ is performed. It is now possible --- due to the decoupling of the equations for $\Omega,\, \Sigma_+,\, \mathrm{and}~\mathcal{E}_+$ from the one for $\Theta$ --- to search for stationary points of this system and analyse their properties. This has been done in~\cite{M.Bruni_etal_1995_Jun} fully classifying the various types of dynamics close to the singularities in a generic Szekeres-like collapse (see also for a complete review~\cite{J.Wainwright_G.F.R.Ellis_1997}) . Here, we can directly employ Eqs.~\eqref{eq:rhoGW}, and~\eqref{eq:thetaGW}--\eqref{eq:E+GW} to extract $\Omega,\, \Sigma_+,\, \mathrm{and}~\mathcal{E}_+$ in terms of metric functions and analyse the behaviour close to the singularities. We find
\begin{align}
&\Omega =\frac{18A\tilde{M}\bigl(A - \beta_+ f_+\bigr)}{
S\Bigl(S\beta_+ \dot{f}_+- 3\bigl(A - \beta_+ f_+\bigr)\dot{S}\Bigr)^2} \, ,\\
& \Sigma_+ = -\frac{1}{3}\frac{S\beta_+\dot{f}_+
}{S\beta_+\dot{f}_+ - 3\bigl(A - \beta_+ f_+\bigr)\, \dot{S}
}\, , \\
& \mathcal{E}_+ = -\frac{\tilde{M}\beta_+f_+\bigl(A -\beta_+ f_+\bigr)}{S\Bigl(S\beta_+\dot{f}_+- 3\bigl(A - \beta_+ f_+\bigr)\dot{S}\Bigr)^2}\, .
\end{align}
We may then examine the asymptotic values of ${\Omega,\,\Sigma_+,\,\mathcal{E}+}$ as the system approaches the various kinematic singularities. In Tab.~\ref{tab:asymptotic_states} we report the obtained values, as well as their classification as stationary points of the associated autonomous dynamical system, following the notation of~\cite{M.Bruni_etal_1995_Jun}. In particular, the asymptotic states correspond to the stationary points $\mathrm{DI}$ (repeller), $\mathrm{DV}$ (attractor), and $\mathrm{DIV}$ (attractor), respectively. These represent, in turn, a flat FLRW-type collapse, a Kasner-like collapse, and a Minkowski-like fixed point~\cite{M.Bruni_etal_1995_Jun,J.Wainwright_G.F.R.Ellis_1997}.
\begin{table*}[htb!]
\centering
\setlength{\tabcolsep}{12pt}
\renewcommand{\arraystretch}{1.2}
\begin{tabular}{lcccccc}
\hline\hline
Singularity type 
& $\Omega$ 
& $\Sigma_+$ 
& $\mathcal{E}_+$ 
& Type of stationary point 
& Collapse asymptotic behaviour \\
\hline
Point-like 
& $1$ 
& $0$ 
& $0$ 
& $\mathrm{DI}$ (repeller) 
& Flat FLRW \\[6pt]

Cigar-like  
& $0$ 
& $1/3$ 
& $2/9$ 
& $\mathrm{DV}$ (attractor) 
& Kasner-like\\[6pt]

Pancake-like   
& $0$ 
& $-1/3$ 
& $0$ 
& $\mathrm{DIV}$ (attractor) 
& Minkowski-like \\
\hline\hline
\end{tabular}
\caption{Classification of the stationary points of the autonomous Szekeres dynamical system for quasispherical models with positive curvature parameter.}
\label{tab:asymptotic_states}
\end{table*}

The point-like state $\mathrm{DI}$ represents an isotropic, matter-dominated collapse in which both shear and tidal fields are dynamically suppressed, corresponding to a flat FLRW-type singularity. By contrast, the cigar-like state $\mathrm{DV}$ is shear-dominated and effectively vacuum-like, with $\Omega\to0$ while $\Sigma_+$ and $\mathcal{E}_+$ remain finite, signalling a Kasner-like anisotropic collapse driven by Weyl curvature. Remarkably, and contrary to intuitive expectations based on spherical collapse, this solution acts as the late-time attractor for generic collapsing configurations in the LTB/Szekeres class. Therefore, in realistic black-hole–forming scenarios one should expect most collapsing shells to evolve toward a cigar-type geometry rather than an isotropic one.

Finally, although the pancake-like state $\mathrm{DIV}$ also appears as an attractor in the reduced phase space, we emphasise that it corresponds to a shell-crossing configuration. In particular, this fixed point is Minkowski-like, with vanishing Weyl curvature and no genuine gravitational focusing, and is therefore dynamically weak. Its emergence is thus better interpreted as signalling a breakdown of the pressureless-dust description rather than the formation of a true spacetime singularity.

\section{\label{sec:BH}Black Hole formation from CDM perturbations}

\subsection{Initial conditions}
We interpret the Szekeres and LTB solutions as exact, nonlinear perturbation of a global FLRW background. Therefore, we set the Initial Conditions (IC) for the gravitational collapse in the LTB and Szekeres models by matching, at an early time in the matter-dominated era, the spatial metric and expansion tensor predicted by cosmological linear perturbation theory (see Eqs.~\eqref{eq:gammaij_Rcg} and~\eqref{eq:Thetaij_Rc}) and corresponding linearised quantities derived from the exact LTB and Szekeres solutions.

\subsubsection{IC for LTB models}
To linearise the LTB metric in Eq.~\eqref{eq:metricLTB} it can be meaningful, a priori, to employ the parametrisation (see e.g.,~\cite{K.Uehara_etal_2025}) 
\begin{equation}
    R(t,r) = r a(t)e^{\zeta(t,r)}\, ,
\end{equation}
where $a(t)$ is the FLRW background scale-factor, and $\zeta(t,r)$ would thus measure deviations of the LTB spacetime from the background. However, it is more convenient at this stage to exploit the remaining gauge freedom in the LTB models --- as well as in the spherically linearly perturbed FLRW picture --- to fix the radial coordinate on the initial data surface, namely $t =t_i$. Thus, at an exact level, we adopt the gauge choice 
\begin{equation}\label{eq:gauge_choice}
R(t_i,r) = R_i = a_i r \, ,
\end{equation}
where, as discussed above, $a_i = a(t_i)$ is the background scale-factor at the initial time-slice, i.e., such that $a_i = 1/(1+z_i)$, and $r$ is the comoving coordinate that coincides with the background comoving distance at the present time, as we fix $a_0=1$, as usual. This gauge choice --- which fixes the initial areal radius to coincide with that of the background FLRW --- offers several advantages: (i) only the radial metric component is modified with respect to the background; (ii) the total gravitational mass $M(r)$ is directly obtained from the integral of the overdensity $\delta$ over the background comoving volume inside the sphere of radius $r$, together with the background mass distribution, thereby clarifying its physical interpretation; and (iii) the separation between geometric effects (encoded in $k(r)$) and physical inhomogeneities (encoded in $M(r)$) becomes more transparent, simplifying both analytical estimates and comparison with perturbative limits.

Indeed, as noted above we now have 
\begin{align}\label{eq:gaugeM}
    M(r) & = \bar{M}(r) + \deltaM (r)  \\
    &  = 4\pi \int_0^{r} \rho_i\,\bigl[1 + \delta(\tilde r)\bigr]\, a_i^{3}\, \tilde r^{2}\, d\tilde r\,, \nonumber
\end{align}
where $\bar{M}(r)$ is the FLRW background mass contained within the shell, and $\deltaM$ is its perturbation. Furthermore, as the initial perturbation in the spatial curvature is small, we consistently assume $ k(r) \ll 1$. Note, that this does not force the spatial curvature to remain small at all times, as $\Rsp$ is an evolving quantity determined by Eq.~\eqref{eq:R3LTB}.  

From a direct comparison, at the initial time $t_i$, between the LTB metric and that of a linearly perturbed FLRW background in spherical coordinates --- for which we select the same radial gauge, i.e., we implement the coordinate transformation $\bar r = r -r\Rci-\Rci'/(a^{2}\mathcal{F}H^{2})$ --- we find 
\begin{equation}\label{eq:k}
    k(r) =\frac{2r\Rci'}{1+2r\Rci'} \eqone 2r\Rci'\, .
\end{equation}
By now employing the \textit{exact} equations for LTB, in particular the Eqs.~\eqref{eq:t+} and \eqref{eq:R+} we then also fix the gravitational mass function $M(r)$. This can be expanded in the perturbative variable $k(r)$ to give
\begin{widetext}
\begin{align}\label{eq:MTrue}
    GM(r) &= \frac{1}{2}a_i^3 H_i^2 r^3\left[ 1 + \frac{3}{5}\frac{k(r)}{a_i^2H_i^2r^2} + \frac{9}{175}\frac{k(r)^2}{a_i^4 H_i^4 r^4} - \frac{2}{525}\frac{k(r)^3}{a_i^6 H_i^6 r^6} + \mathcal{O}(k^4)\right]\\
    & \eqone \frac{1}{2}a_i^3 H_i^2 r^3 + \frac{3}{5}a_i r^2\Rci' \, ,
\end{align}
\end{widetext}
where in the last leading order equality we have substituted $k(r)$ in terms of the initial comoving curvature perturbation. As expected from linear perturbation theory within the matter era, from Eq.~\eqref{eq:MTrue} we can then identify 
\begin{equation}
    \delta M(r) \eqone \frac{3}{5}\frac{a_i}{G} r^2\Rci'\, .
\end{equation}

Finally, we can find the initial values of the other kinematic variables to be
\allowdisplaybreaks
\begin{align}
    &\Rsp_i =  \frac{4}{a_i^2}\left[\frac{\Rci'' +2\Rci'/r\left(1+r\Rci'\right)}{\left(1+2r\Rci'\right)^2}\right]\; ,\label{eq:R3LTBi}\\
    & \Theta_i \eqone 3H_i - \frac{2}{5a_i^2H_i}\left(\Rci''+\frac{2}{r}\Rci'\right)\; , \label{eq:thetaLTBi}\\
    & \sigma_{+,\,i} \eqone \frac{2}{15a_i^2H_i}\left(\Rci''- \frac{1}{r}\Rci'\right)\; , \label{eq:sigma+LTBi}\\
    & E_{+,\,i}  \eqone -\frac{1}{10a_i^2}\left(\Rci''- \frac{1}{r}\Rci'\right) \; , \label{eq:E+LTBi}\\  
\end{align}
where the intrinsic spatial curvature is exact, whilst the other quantities are reported at leading order, as they involve expression containing (or deriving from) $M(r)$.

\subsubsection{IC for axisymmetric Szekeres models}
To initialise Szekeres collpase via linear perturbation we employ the same approach as for the LTB models. In addition, we linearise the dipole functions according to 
\begin{equation}
    L \eqone 1  + L^{(1)}\;\; ; \;\; P  \eqone P^{(1)} \;\; ; \;\; Q  \eqone Q^{(1)} \, .
\end{equation}
Hence, we have 
\begin{widetext}
\begin{equation}
D\eqone \frac{1}{2}\left[1+p^2+q^2 +L^{(1)}\left(1-p^2-q^2\right) + 2pP^{(1)} +  2qQ^{(1)}\right]\, .
\end{equation}
\end{widetext}
However, as also noted in~\cite{R.A.Sussman_etal_2017}, for non-vanishing $P^{(1)}$ and $Q^{(1)}$ the coordinate system of the underlying FLRW background to which such a perturbation could be ascribed is not easily recovered. It is therefore advantageous to restrict our analysis to the case of an axisymmetric Szekeres model, for which $P^{(1)} = Q^{(1)} = 0$. In this case, the background coordinates $\{p,q\}$ correspond to the standard projective (stereographic) coordinates on the $2$–sphere embedded in a three–dimensional spatial slice of the FLRW background, enabling a clear geometric interpretation of the dipole structure. Importantly, we note that the axisymmetric restriction still retains the key physics distinguishing spherical and anisotropic collapse, and it is thus sufficient to the scope of this work. Hence, we consider as linearised dipole function 
\begin{equation}
   D\eqone\frac{1}{2}\left[1+p^2+q^2 +L^{(1)}\left(1-p^2-q^2\right)\right]\, .
\end{equation}
Then, we can look for which profiles of $\Rc$ would allow for a linearly perturbed metric --- in the projective coordinate system --- respecting the same symmetries of a linearised axisymmetric Szekeres spacetime. A direct calculation leads to the initial profile
\begin{equation}
    \Rci(r,p,q) = \Rci^{[m]}(r) + \Rci^{[d]}(r,p,q)\, ,
\end{equation}
where $\Rci^{[m]}(r)$ is a monopole term, whilst $ \Rci^{[d]}(r,p,q)$ is a pure dipole term along the symmetry axis, defined by
\begin{equation}\label{eq:Rcd}
     \Rci^{[d]}(r,p,q) =  \left[\frac{1-p^2-q^2}{1+p^2+q^2}\right]Br\, ,
\end{equation}
with $B$ as an arbitrary constant, and such that
\begin{equation}
    \frac{D'}{D} = \frac{\Rci^{[d]'}}{1+\Rci^{[d]}}\, .
\end{equation}
We then find the same identification as for the LTB model in Eqs.~\eqref{eq:k} and~\eqref{eq:MTrue} by replacing $\Rci(r) \rightarrow \Rci^{[m]}(r)$ in the formulas, e.g., we find
\begin{equation}
    k(r) = \frac{2r\Rci^{[m]'}(r)}{1+2r\Rci^{[m]'}(r)}\, .
\end{equation}
Note then that the perturbations in the active gravitational mass and the curvature function are fully determined by the monopole term, whereas the dipole contribution manifests solely through the dipole function. In particular, we find that for our solution we have a dipole vector of
\begin{equation}
    \vec{\beta} := (0,0,-Br)\, ,
\end{equation}
namely an increasing dipole with respect to the radial coordinate. Interestingly, we also observe that the linear approximation in this model necessarily breaks down for $r \approx 1/B$, where the assumption of a small $L^{(1)}$ is no longer valid. Therefore, any description within this framework remains necessarily local.

Finally, as for the LTB models, the initial profiles for the kinematic quantities of interest can be derived by employing the identifications in terms of $\Rc$ coupled to the formulas in Eqs.~\eqref{eq:theta}--\eqref{eq:R3}.

\subsection{Exact models for BH formations}

We can now employ our ansätze for axisymmetric Szekeres and LTB models to (i) derive the turn-around, collapse and FH time in terms of local $\Rci$ (and its derivative), and (ii) to constrain the $\Rci$ profiles which allow for BH formation when undergoing gravitational collapse. From Eqs.~\eqref{eq:tta},~\eqref{eq:tcol},~\eqref{eq:tFH},~\eqref{eq:k}and~\eqref{eq:MTrue} --- up to third order in the perturbative expansion --- we find for the LTB models
\allowdisplaybreaks
\begin{widetext}
\begin{align}
    & t_{\mathrm{ta}}(r) = \frac{\pi}{2^{5/2}}H_i^2a_i^3r^{3/2}\left[\Rci^{'-3/2} + 3\left(r^2+\frac{2}{5a_i^2H_i^2}\right)\frac{\Rci^{'-1/2}}{r}  +\mathcal{O}\left(\epsilon^{1/2}\right)\right] \, , \label{eq:ttaRci}\\
    & t_{\mathrm{col}}(r) = \frac{\pi}{2^{3/2}}H_i^2a_i^3r^{3/2}\left[\Rci^{'-3/2} + 3\left(r^2+\frac{2}{5a_i^2H_i^2}\right)\frac{\Rci^{'-1/2}}{r}+\mathcal{O}\left(\epsilon^{1/2}\right)\right] \, , \label{eq:tcolRci}\\
    & t_{\mathrm{FH}}(r) = \frac{\pi}{2^{3/2}}H_i^2a_i^3r^{3/2}\left[\Rci^{'-3/2} + 3\left(r^2+\frac{2}{5a_i^2H_i^2}\right)\frac{\Rci^{'-1/2}}{r} +\frac{2^{3/2}r^{3/2}}{3\pi}+\mathcal{O}\left(\epsilon^{1/2}\right)\right] \, , \label{eq:tFHRci}
\end{align}
\end{widetext}
where $\epsilon =r \Rci'$. Here, we stress that while $t_{\mathrm{ta}}$ and $t_{\mathrm{col}}$ share identical asymptotic expansions up to an overall multiplicative factor, the formation time $t_{\mathrm{FH}}$ acquires an additional contribution at lower effective order, reflecting the distinct geometric condition underlying horizon formation compared to complete collapse. We also note that for the axisymmetric Szekeres models it is then sufficient to replace $\Rci \rightarrow \Rci^{[m]}$ to derive analogous results, i.e., it is only the monopole term of the initial perturbation which determines the turn-around, collapse and FH formation time for the radial shells. Furthermore, we see from Eqs.~\eqref{eq:ttaRci}--\eqref{eq:tFHRci} that it is the radial gradient of the initial perturbation, i.e., its steepness, which ultimately determines the BH formation time. 

Additionally, we can also express the regularity conditions listed in Sec.~\ref{subSec:regcond} in terms of the initial comoving curvature profile at leading order. In this regard, we find that Eq.~\eqref{eq:Icregular1} is automatically satisfied for LTB models, whilst for the axisymmetric Szekeres models it becomes
\begin{equation}\label{eq:constraint_weird}
    \Rci^{[d]'}< \frac{1}{r} \, .
\end{equation}
Since $\Rci^{[d]'} = B = {const}$, this further highlights the local nature of such a model. In addition, the constraint in Eq.~\eqref{eq:no-shell1} is automatically satisfied within both an LTB and an axisymmetric Szekeres models describing an overdensity, if Eq.~\eqref{eq:constraint_weird} is also met. Furthermore, at leading order, we find that Eq.~\eqref{eq:no-shell2} becomes for LTB models
\begin{equation}
    \Rci'' <\frac{\Rci'}{r} \, , \label{eq:figata}
\end{equation}
where it is then sufficient to replace $\Rci \rightarrow \Rci^{[m]}$ to derive the analogous result. Here, Eq.~\eqref{eq:figata} thus fully constrain the profile of the initial comoving curvature in order to avoid shell-crossing during the gravitational collapse. Similarly, the more complex condition in Eqs.~\eqref{eq:no-shell3} and~\eqref{eq:no_nakey} can be turned in analytic constraints on $\Rci$. 

In particular, within a cosmological perturbation framework --- i.e., treating $\Rci$ and its derivatives as perturbative quantities --- Eq.~\eqref{eq:no-shell3} reduces at first order to Eq.~\eqref{eq:figata}. Thus, it follows that for \textit{a spherical collapsing structure (including configurations with dipole-like features) seeded by an initial perturbation about an FLRW background, shell-crossing (if present) will generally occur before horizon formation.}

Additionally, let us consider a $\Rci$ profile with a covered central singularity, i.e., such that 
\begin{equation}
    k(r) = k_2r^2 + k_4r^4 + \mathcal{O}\left(r^6\right)\, ,
\end{equation}
with $k_2>0$ and $k_4\geq 0$. This implies
\begin{equation}
    \Rci' = \frac{1}{2}k_2r + \frac{1}{2}k_4r^3 + \mathcal{O}\left(r^5\right)\, ,
\end{equation}
and 
\begin{equation}
    \Rci'' = \frac{1}{2}k_2 + \frac{3}{2}k_4r^2 + \mathcal{O}\left(r^4\right)\, .
\end{equation}
It follows that
\begin{equation}
    \Rci'' - \Rci'/r =  k_4r^2 + \mathcal{O}\left(r^4\right)\, ,
\end{equation}
which is then positive in a neighbourhood of the origin under the current assumptions if $k_4 \neq 0$. This directly contradicts the leading-order shell--crossing avoidance requirement in Eq.~\eqref{eq:figata}. Hence, within our perturbative LTB/Szekeres framework with $k \neq 0$, a locally covered central singularity is in general incompatible with shell-crossing-free evolution: \textit{if shell-crossing is avoided, the central singularity is generally locally naked; conversely, if the central singularity is locally covered, then shell-crossing will generally occur.} Finally, in a shell--crossing-free collapse, Eq.~\eqref{eq:tFHRci} implies that the perturbative $t_{\mathrm{FH}}(r)$ admits no local minima away from the centre; the only minimum occurs at $r=0$, where $t_{\mathrm{FH}}(0) = t_{\mathrm{col}}(0)$. Consequently, trapped surfaces do not form strictly prior to the central collapse, and the earliest apparent-horizon formation is marginal at the centre. As such we also see that within our framework also the practical criterion of Eq.~\eqref{eq:no_nakey} is generally violated in the absence of shell-crossing.

\subsubsection{Top-Hat Collapse}

Before moving to more realistic collpases from initial comoving curvature perturbation profiles, we show that we can directly recover the TH model~\cite{J.E.Gunn_J.R.Gott_1972} as a subcase of the LTB solutions. The TH model describes the evolution of a homogeneous spherical overdensity in the matter-dominated era. This overdense sphere is modeled by a closed (positive spatial curvature) FLRW “separate universe” within an external spatially flat FLRW background. The radius of the TH overdensity expands at a slower rate than the background, gradually slowing down, as it is bound by its positive curvature (see~\cite{N.Vittorio_2018} for a complete review) . It eventually reaches its maximal size, turns around, and then contracts into itself to collapse. Although a relatively simple model, the TH description provides the critical value of the linear density contrast corresponding to collapse, $\delta_C^{(1)} \approx 1.69$, i.e., a crucial benchmark\footnote{This value assumes a negligible cosmological constant, namely that the collapse occurs well before $\Lambda$ becomes relevant in the Friedman equation for the background.} to estimate virialization and for the Press-Schechter mass function and the Sheth-Tormen extension~\cite{W.H.Press_P.Schechter_1974,R.K.Sheth_G.Tormen_1999}. 

We consider for simplicity an EdS background, such that $\mathcal{F} = 5/2$. Then, to obtain a TH collapse, we impose both a homogeneous first-order density contrast $\delta^{(1)}$ and a small curvature perturbation $k \eqone K r^2$. These conditions can be satisfied simultaneously, at the chosen order, only for an initial comoving curvature perturbation of the type 
\begin{equation}
    \Rci (r) = C + \frac{1}{4}K r^2 \, ,
\end{equation}
for which we find
\begin{equation}
    \delta^{(1)}_i = \frac{3K}{5a_i^2H_i^2}\; \; ; \;\; k \eqone K r^2 \, .
\end{equation}
We thus recover --- at the linear level in the initial density contrast --- the standard, radial-independent turn-around and collapse times from Eqs.~\eqref{eq:ttaRci} and \eqref{eq:tcolRci}, i.e.,
\begin{equation}
    t_\mathrm{ta}^{\mathrm{TH}} \eqone \frac{\pi}{2}\frac{H_i^{-1}}{\left(5\delta_i^{(1)}/3\right)^{3/2}} \;\; ; \;\;t_\mathrm{col}^{\mathrm{TH}} \eqone \frac{\pi H_i^{-1}}{\left(5\delta_i^{(1)}/3\right)^{3/2}} \, .
\end{equation}
Finally, note that we have neglected the dynamics of the underdense external shell within the TH model, focusing only the closed FLRW collapse of the overdense core.

\subsubsection{Single-mode collapse}

It is meaningful to consider a \textit{single-mode} collapse, namely a configuration in which the initial curvature profile is characterised by a single scale set by a single wavenumber $\lambda$. Besides providing a minimal test bed for general multi-mode collapse --- a clear baseline against which the impact of additional structure can be assessed --- such single-mode fluctuations have also been widely used as idealised building blocks of the cosmic web: a periodic lattice of over- and under-densities (filament/void elements) whose overdense peaks are approximately spherical in their immediate neighbourhood~\cite{E.Bentivegna_M.Bruni_2016, H.J.Macpherson_etal_2017,W.E.East_etal_2018, J.C.Aurrekoetxea_etal_2020,S.Saga_etal_2021,R.L.Munoz_M.Bruni_2023}. 

Here, we adopt such a single-mode setting in a spherically symmetric form, to isolate the same physics within our collapse geometry. The monopole in the initial comoving curvature profile is given by
\begin{equation}\label{eq:RciSingleMode}
    \Rci^{[m]}(r) = \begin{cases}
        -A\cos\left(\lambda r\right) + A\cos\left(\lambda r*\right)\quad \quad \ r < r_*\, ,\\
        \,0 \quad \quad \quad \quad \quad \quad \quad \quad \quad \quad \quad \quad   \,\,\, \, \, r \ge r_*\, ,
    \end{cases}
\end{equation}
where $A>0$ is the amplitude and $\lambda$ represents the wave-number of the perturbation, $r_* = \pi/\lambda$, and the constant subtraction ensures a smooth matching with the exterior at $r_*$. In addition, here $\Rci^{[d]}$ is either zero (LTB case) or given by Eq.~\eqref{eq:Rcd}.

Interestingly, this curvature profile leads to a shell-crossing-free collapse. Indeed, the condition in Eq.~\eqref{eq:figata} reduces to
\begin{equation}
    A\lambda ^2\cos\left(\lambda r\right)\left[1-\frac{\tan \left(\lambda r\right)}{\lambda r}\right]  < 0 \, ,
\end{equation}
which is always satisfied for $r < r_*$. Hence, a single-mode perturbation of the comoving curvature naturally drives a coherent collapse, even in the presence of an initial shear in the dust's velocity field. 

However, following the discussion in the previous section, and since this choice of $\Rci^{[m]}$ gives a $k_4 <0$ in the curvature parameter expansion, we find that single-mode LTB/Szekeres collapse generates a central naked singularity. As such, we conclude that such single-scale collapse should not be taken as a viable channel for BH formation in the matter-dominated era.

\subsubsection{Gaussian collapse}

It is likewise meaningful to consider a \textit{Gaussian} collapse, as it provides a complementary “single-peak” test bed to the single-mode collapse. We therefore take the monopole of the initial curvature to be\footnote{In this case we are considering a LTB/Szekeres spacetime such that $k(r)$ tends to zero at infinity.}
\begin{equation}\label{eq:RciGaussian}
    \Rci^{[m]}(r)=
        -A\exp\left(-\dfrac{r^2}{2\sigma^2}\right)\, ,
\end{equation}
where $A>0$ is the amplitude, $\sigma$ is the characteristic scale. In addition, $\Rci^{[d]}$ is either zero (LTB case) or given by Eq.~\eqref{eq:Rcd}. Interestingly, this Gaussian profile also leads to shell-crossing–free collapse. Indeed, using the same criterion as in Eq.~\eqref{eq:figata}, we obtain for \eqref{eq:RciGaussian} the exact identity
\begin{equation}
\Rci''(r)-\frac{\Rci'(r)}{r}
=-A\frac{r^2}{\sigma^4}\exp\left(-\frac{r^2}{2\sigma^2}\right)<0\,,
\end{equation}
so the no–shell–crossing condition is automatically satisfied throughout the perturbed region. As such, and given that for this profile it is immediate to check $k_4 >0$, we find that even Gaussian collapse gives rise to a centrally naked singularity. Therefore, even a purely Gaussian collapse should not be considered a BH forming channel within our framework.

\subsubsection{Peak collapse}

Within the context of cosmological perturbation theory, where the initial comoving curvature may be modelled as a (smoothed) Gaussian random field, a prominent role in the study of collapsing structures is played by peak theory, which provides a controlled description of the rare, high-amplitude extrema that seed nonlinear collapse and --- in the early Universe --- is widely used to estimate the formation and abundance of PBH from large curvature peaks~\cite{A.G.Doroshkevich_1970,J.M.Bardeen_J.R.Bond_N.Kaiser_A.S.Szalay_1986,R.J.Adler_1981,C.Germani_T.Prokopec_2017,C.M.Yoo_T.Harada_J.Garriga_K.Kohri_2018,C.Germani_I.Musco_2019,I.Musco_2019}. In peak theory, proto-collapsing regions are associated with rare extrema of the initial comoving curvature field (or of the corresponding density contrast) after smoothing on a characteristic scale. In particular, in the Bardeen--Bond--Kaiser--Szalay (BBKS) peak theory the \textit{conditional mean} profile around a peak is not a pure Gaussian: it is a linear combination of the correlation function and its derivatives, with coefficients set by the peak height and peak curvature parameters (see e.g.,~\cite{J.M.Bardeen_J.R.Bond_N.Kaiser_A.S.Szalay_1986} and subsequent extensions such as~\cite{J.R.Bond_S.T.Myers_1996a,J.R.Bond_S.T.Myers_1996b,J.R.Bond_S.T.Myers_1996c,C.M.Yoo_J.O.Gong_S.Yokoyama_2019,S.Young_M.Musso_2020,N.Kitajima_etal_2021,C.M.Yoo_T.Harada_S.Hirano_K.Kohri_2021,C.Germani_2025}). After Gaussian smoothing, these profiles reduce precisely to a Gaussian envelope times even polynomials --- quadratic in basic BBKS peak theory, and higher order polynomials in general upon imposing local support of the perturbation (or finite-volume compensation) --- which motivates the ansatz we adopt here.

Indeed, we define the monopole of a compact (compensated) peak with a Gaussian envelope,
\begin{widetext}
\begin{equation}\label{eq:RciFamilyA_def}
\Rci^{[m]}(r)= 
\begin{cases} -A\left[
e^{-x^2/2}\Bigl(1+\alpha x^2+\beta x^4\Bigr) - e^{-x_*^2/2}\Bigl(1+\alpha x_*^2+\beta x_*^4\Bigr) \right] & r<r_*,\\[1mm]
0\ & r\ge r_*,
\end{cases}
\qquad x\equiv r/\sigma,
\end{equation}
\end{widetext}
with amplitude $A>0$ and smoothing/size parameter $\sigma>0$. The constant subtraction ensures $\Rci^{[m]}(r_*)=0$ so that the profile matches smoothly to the background in the exterior. In the BBKS framework, the coefficient $\alpha$ controls the relative weight of the peak curvature compared to its height, namely, increasing $\alpha$ corresponds, in BBKS language, to conditioning on a peak with larger curvature parameter at fixed height, i.e., a sharper peak. Additionally, the $\beta$ coefficient --- which captures the leading non-trivial higher-derivative correction beyond the quadratic core --- encodes the peak’s central ``kurtosis'', so that $\beta>0$ corresponds to a locally sharpened core while $\beta<0$ produces a softened (flatter) core\footnote{%
In particular, for a smoothed Gaussian field $\Rci$ with spectral moments $\sigma_j^2=\int\!\frac{d^3k}{(2\pi)^3}k^{2j}P_{\Rci}(k)W^2(kR_f)$, the BBKS peak theory defines the peak height $\nu\equiv \Rci(0)/\sigma_0$, and peak curvature $x\equiv-\nabla^2\Rci(0)/\sigma_2$. For \eqref{eq:RciFamilyA_def} one has $-\nabla^2\Rci(0)/\Rci(0)=3(1-2\alpha)/\sigma^2$, hence $\alpha=\frac12\!\left[1-\frac{\sigma^2}{3}\frac{\sigma_2}{\sigma_0}\frac{x}{\nu}\right]$. Additionally, by including the higher-order correction to the peak introduced by compensation over finite volume $y\equiv\nabla^4\Rci(0)/\sigma_4$ we obtain $\nabla^4\Rci(0)/\Rci(0)=120(\beta-\alpha/2+1/8)/\sigma^4$, i.e., $\beta=\frac{\alpha}{2}-\frac18+\frac{\sigma^4}{120}\frac{\sigma_4}{\sigma_0}\frac{y}{\nu}$.}.
Furthermore, as in the other examples, the full Szekeres/LTB initial curvature may be taken as $\Rci(r,p,q)=\Rci^{[m]}(r)+\Rci^{[d]}(r,p,q)$ with $\Rci^{[d]}$ given by Eq.~\eqref{eq:Rcd} (or set to zero in the purely spherical case). 

Here, the choice \eqref{eq:RciFamilyA_def} yields a curvature function $k(r)$ with coefficients 
\begin{equation}\label{eq:ckFamilyA_coeffs}
k_2=\frac{2A}{\sigma^2}\bigl(1-2\alpha\bigr)\;\; ; \;\; k_4=\frac{A}{\sigma^4}\bigl(4\alpha-8\beta-1\bigr)\, .
\end{equation}
Then --- motivated by the requirement of a collapse with a covered central singularity --- we impose the constraints $k_2 >0$, $k_4 = 0$ on the initial comoving curvature profile, yielding as constraint
\begin{equation}\label{eq:FamilyA_c4zero}
\alpha < \frac{1}{2} \;\; ; \;\; 
\beta=\frac{\alpha}{2}-\frac{1}{8}\, .
\end{equation}
In what follows we will additionally restrict $\alpha$ to ensure compact support with a first derivative zero at the junction with the exterior background. We begin by noting that for $r>0$ we have $\Rci'(r)\propto x\,e^{-x^2/2}\,Q(x^2)$, where
\begin{equation}\label{eq:FamilyA_Qpoly}
Q(y)=\beta y^2+(\alpha-4\beta)y+(1-2\alpha)\, ,\qquad y\equiv x^2\,.
\end{equation}
With \eqref{eq:FamilyA_c4zero}, there is a unique positive root $y_*=x_*^2$ provided
\begin{equation}\label{eq:FamilyA_alpha_range}
0<\alpha<\frac{1}{4},
\end{equation}
and we define the cutoff by choosing $r_*=\sigma x_*$ with
\begin{equation}\label{eq:FamilyA_xstar}
x_*^2 = \frac{2\left[\,1-2\alpha+\sqrt{(1-2\alpha)(3-10\alpha)}\,\right]}{1-4\alpha}.
\end{equation}
This guarantees $\Rci'(r)>0$ for $0<r<r_*$ and $\Rci'(r_*)=0$, hence $k(r_*)=0$,
so the interior patch is smoothly matched to the background FLRW exterior. In addition, we note that by using \eqref{eq:FamilyA_c4zero}, the combination entering the shell-crossing avoidance condition Eq.~\eqref{eq:figata} simplifies to (within $0<x<x_*$)
\begin{equation}\label{eq:FamilyA_shellcross_expr}
-\frac{A}{8\sigma^2}\,x^4\,e^{-x^2/2}\left[\,8(1-3\alpha)-(1-4\alpha)x^2\,\right] < 0 \, . 
\end{equation}
For $0<\alpha<1/4$ the bracket is strictly positive for $0<x<x_*$, and therefore the chosen $\Rci$ profile results in a shell-crossing-free collapse whilst displaying a covered central singularity within an LTB/Szekeres framework. Interestingly, this choice of $\Rci$ selects a sub-set of \textit{moderately curved, broad, and centrally softened} peaks, for which the curvature profile remains single-lobed and monotonic all the way to the compensation radius. In particular, we find that --- within our shell-crossing–free construction --- a \textit{covered} central singularity necessarily requires such a softening of the peak: the leading higher-derivative correction must flatten (rather than sharpen) the core, so that the central evolution is accurately captured by the top-hat dynamics up to the first non-vanishing higher-order (cubic) corrections. Physically, this is tied to the fact that a softened core suppresses the generation of anisotropy at early times: the shear scalar and the electric part of the Weyl tensor decay rapidly toward the centre, making the innermost region effectively close to a locally closed FLRW (quasi-spherical) evolution.

\subsubsection{CDM compensated peaks as massive BH seeds}

Having constructed a family of compensated curvature peaks that (i) satisfies the no--shell-crossing conditions and hence remains single-stream up to focusing (cf.\ Eqs.~\eqref{eq:no-shell1}--\eqref{eq:no-shell2}), and (ii) forms a future-trapped region prior to the central shell-focusing singularity (so that the latter is covered), we now turn to concrete first estimates for matter-era BH formation. The aim of this subsection is phenomenological: to map realistic curvature-profile parameters to the characteristic seed masses and collapse (and trapping) redshifts, and to identify which portion of parameter space yields collapse sufficiently early to act as a viable seeding channel.

\begin{figure}[htb!]
    \centering
    \includegraphics[width=0.49\textwidth]{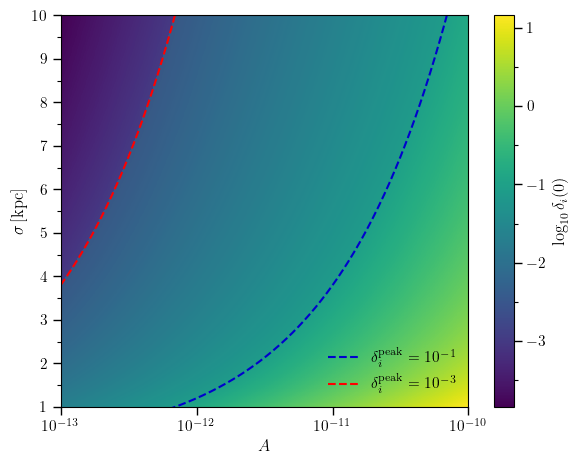}
    \caption{Parameter space of initial curvature perturbations in the $(A,\sigma)$ plane. The dashed lines contour delineates the region satisfying $\delta_i^{\rm peak}\in[10^{-3},10^{-1}]$.}
    \label{fig:phase_space}
\end{figure}

As we are considering direct-collapse BH formation in the matter era, we seed the collapse with perturbations at $z = 300$ and we focus on perturbations describing localised peaks. We therefore select softened profiles with characteristic comoving scales $\sigma \in [1,10]~\mathrm{kpc}$, and fix $\alpha = 0.225$ so that the compensation radius is located at $r_\ast \simeq 5\sigma$ from the centre. The overdense core is then defined operationally as the region enclosed by the first zero-crossing of the initial density contrast, $\delta_i(r_{\rm OD})=0$ with $0<r_{\rm OD}<r_\ast$, and we identify the perturbation relevant for BH formation with $0\le r\le r_{\rm OD}$, rather than with the full compensated configuration out to $r_\ast$. In particular, within the parameter space explored, we find $r_{\rm OD} \simeq 2 \sigma $. Over this family we vary the initial comoving curvature amplitude in the range $A\in[10^{-13},10^{-10}]$\, and select the parameter space area which allows for initial density contrast at the peak still within the linear regime, i.e., $\delta^\mathrm{peak}_i\in [10^{-3},\, 10^{-1}]$. Fig.~\ref{fig:phase_space} shows the allowed parameter space within the range of $A$ and $\sigma$ probed. Furthermore, we restrict our calculation to LTB models for simplicity, as non-trivial dipole contribution to $\Rci$ would still not impact the relevant formation time-scales of total masses of the BH, as previously shown.

\begin{figure}[htb!]
    \centering
    \includegraphics[width=0.49\textwidth]{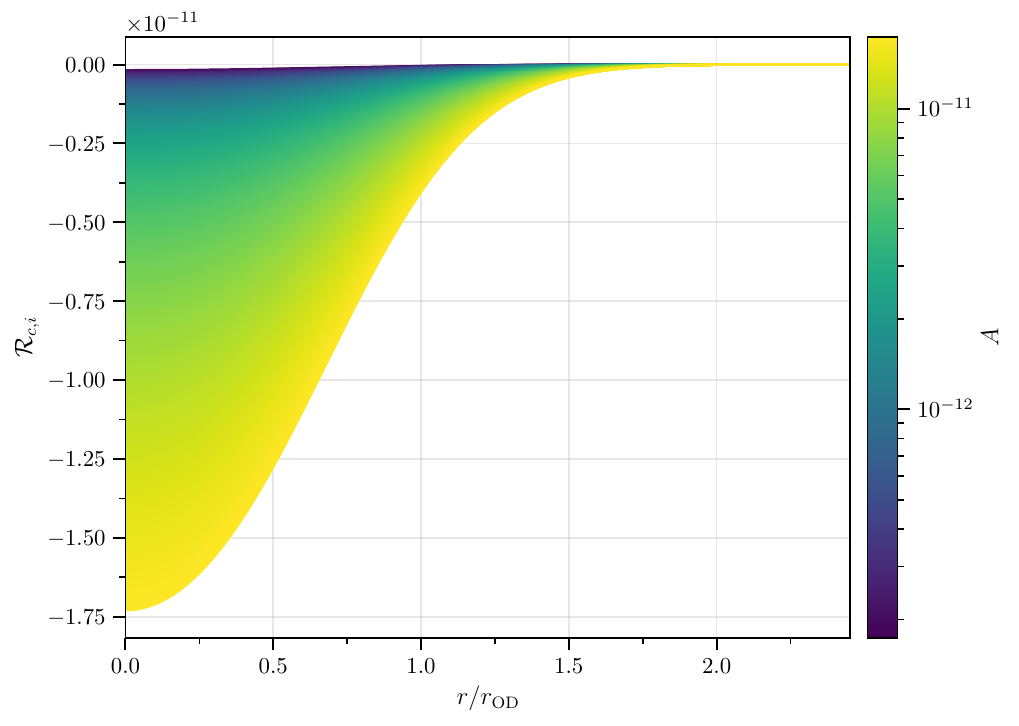}
    \caption{Initial comoving curvature perturbation $\Rci(r)$ for amplitudes $\{A\in[1.7\times10^{-13},\,1.7\times10^{-11}]$, with fixed parameters $\sigma={5}~\mathrm{kpc}$ and $\alpha = 0.225$. For simplicity we adopt a spherically symmetric LTB setup. The initial time slice is given at $z = 300$.}
    \label{fig:1}
\end{figure}

\begin{figure*}[htb!]
    \centering
    \includegraphics[width=0.49\textwidth]{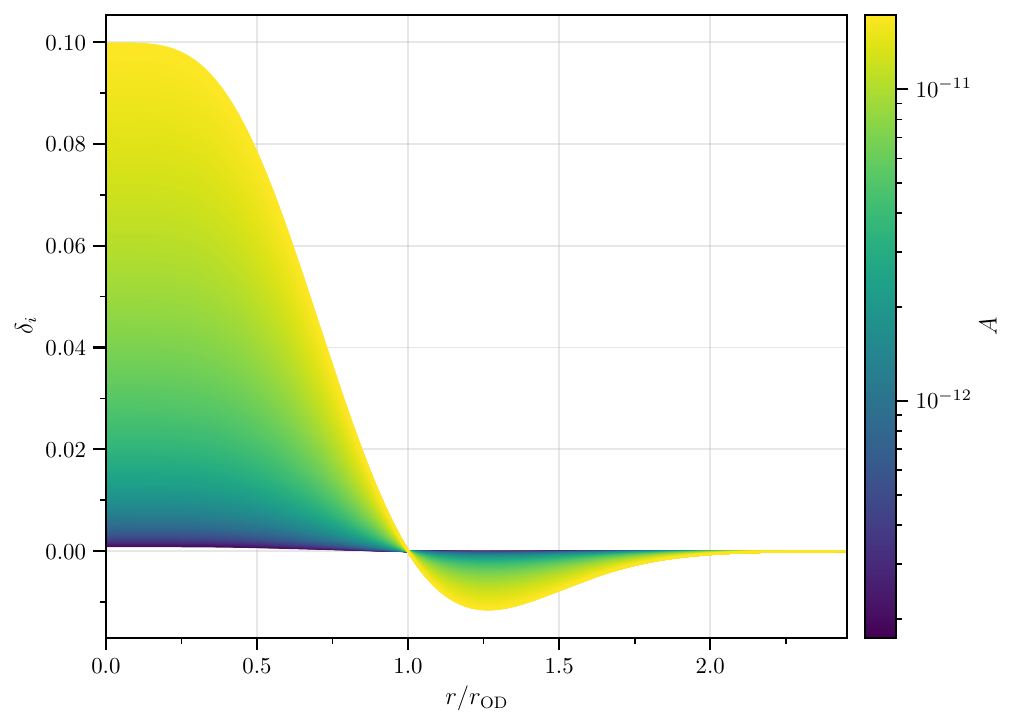}
    \includegraphics[width=0.49\textwidth]{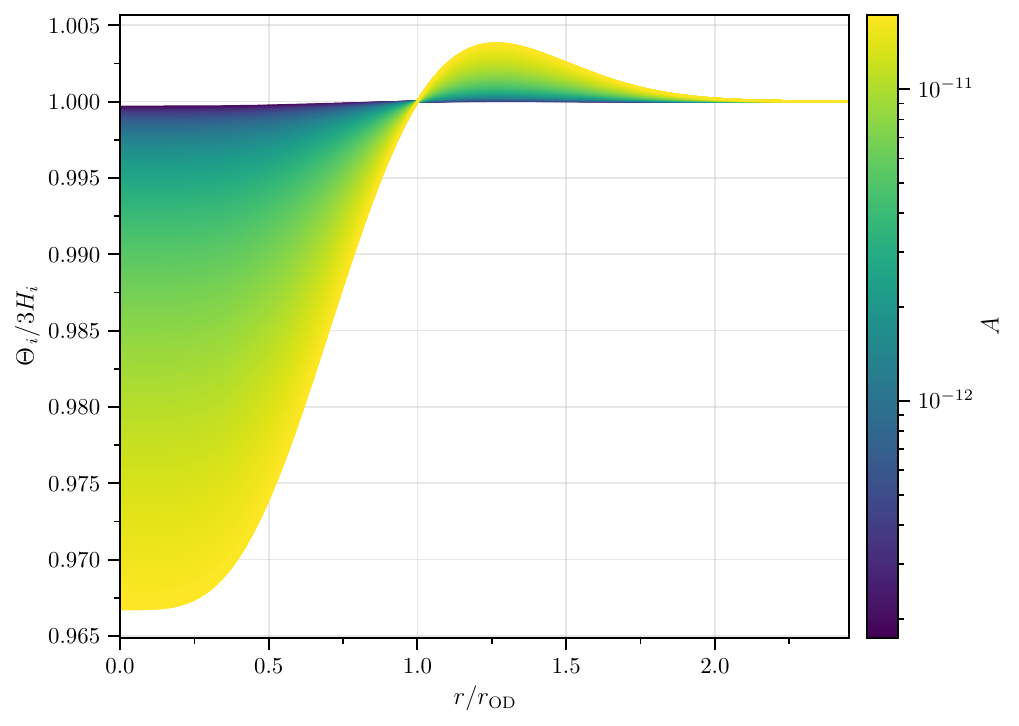}\\
    \includegraphics[width=0.49\textwidth]{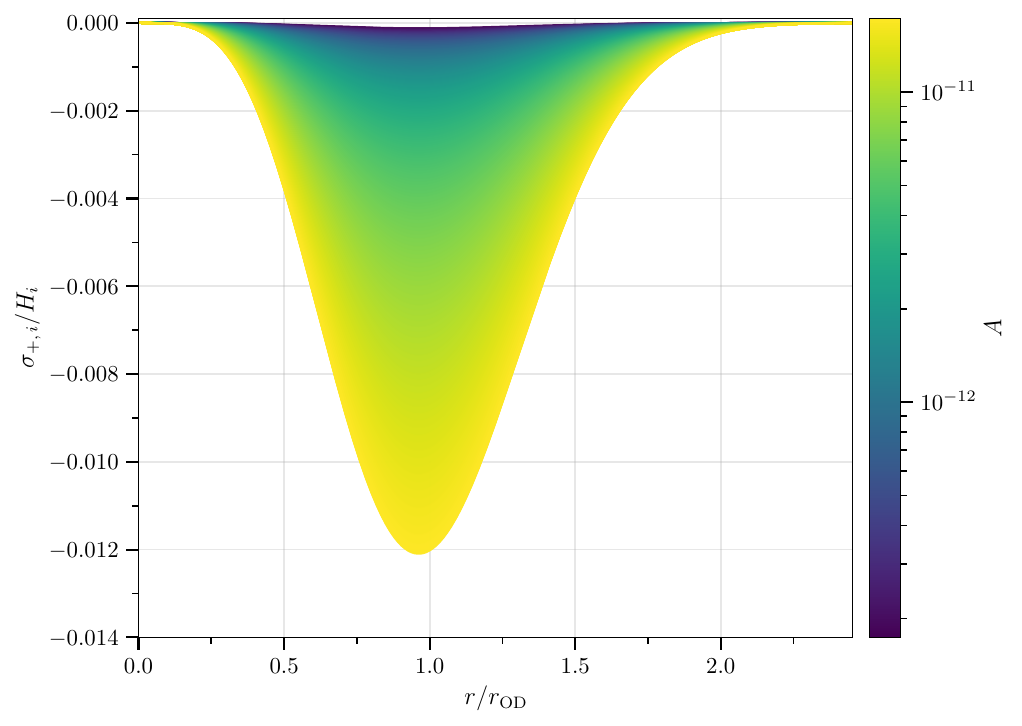}
    \includegraphics[width=0.49\textwidth]{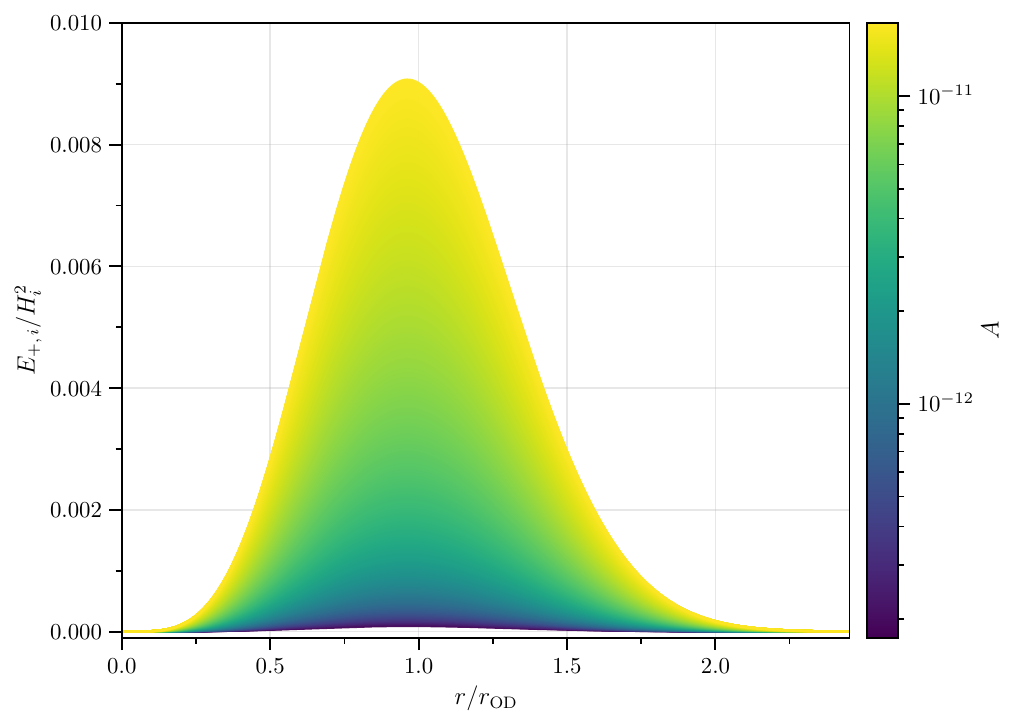}
    \caption{Upper row: initial density contrast $\delta_i$ (left) and initial background-normalised expansion rate $\Theta_i/(3H_i)$ (right). Bottom row: initial background-normalised eigenvalues $\sigma_{+,i}/H_i$ (left) and $E_{+,i}/H_i$ (right) of the shear and electric Weyl tensors, respectively. The profiles shown are obtained starting from the $\Rci$ plotted in Fig.~\ref{fig:1}.}
    \label{fig:2}
\end{figure*}

As an example, in Fig.~\ref{fig:1} we show the initial comoving curvature profile $\Rci$ restricted for visual clarity to $\sigma =5~\mathrm{kpc}$ and the allowed range $A\in[1.7\times10^{-13},\,1.7\times10^{-11}]$. The initial dynamical state associated to these profiles is then summarised in Fig.~\ref{fig:2}, where we plot $\delta_i$ together with the background-normalised expansion $\Theta_i/(3H_i)$, and the eigenvalues of the shear and electric Weyl tensors likewise normalised by $H_i$, i.e., $\sigma_{+,i}/H_i$ and $E_{+,i}/H_i$, respectively. As expected, $\sigma_{+,i}$ and $E_{+,i}$ are suppressed both near the centre and near the matching radius $r_\ast$, while a non-trivial transition region is present between the collapsing overdense core and the compensating underdensity.

\begin{figure*}[htb!]
    \centering
    \includegraphics[width=0.49\textwidth]{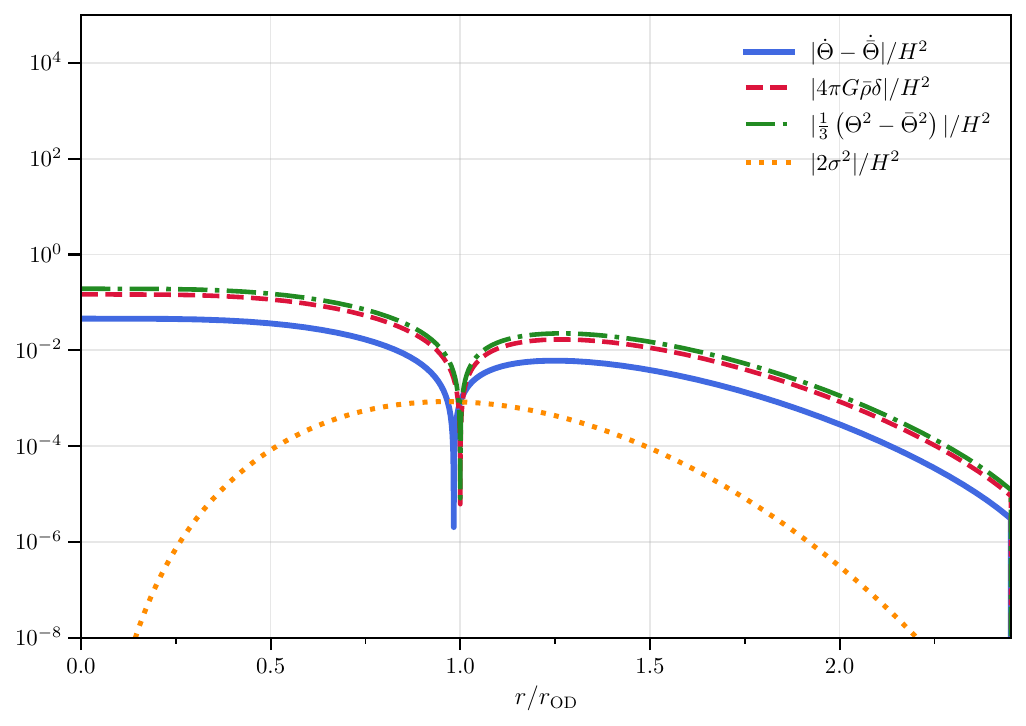}
    \includegraphics[width=0.49\textwidth]{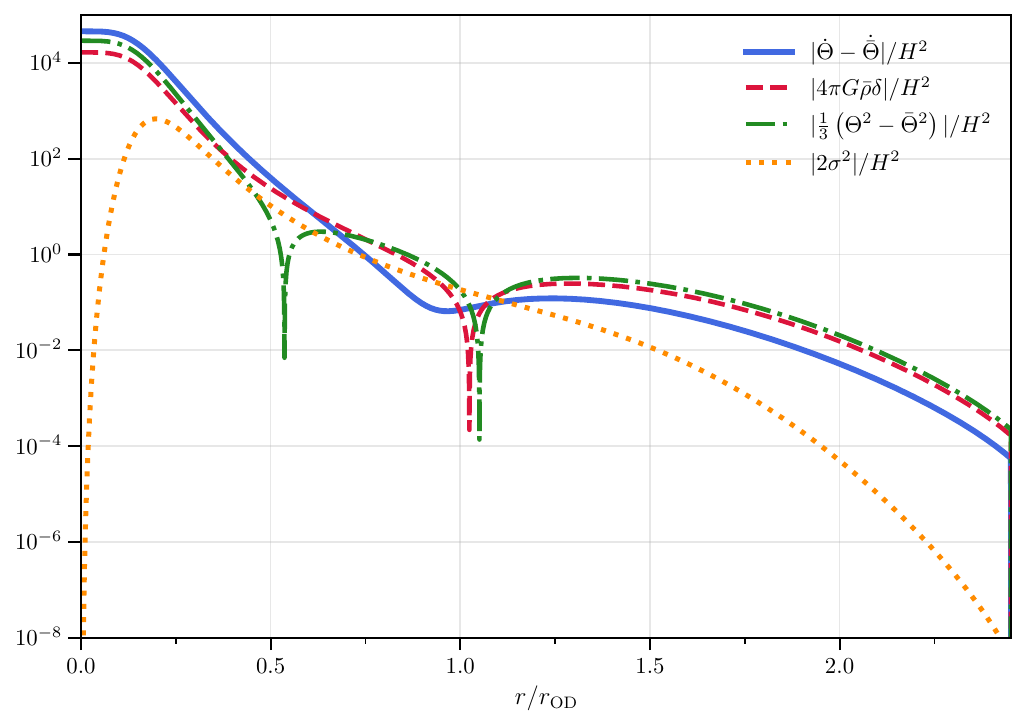} \\
    \includegraphics[width=0.49\textwidth]{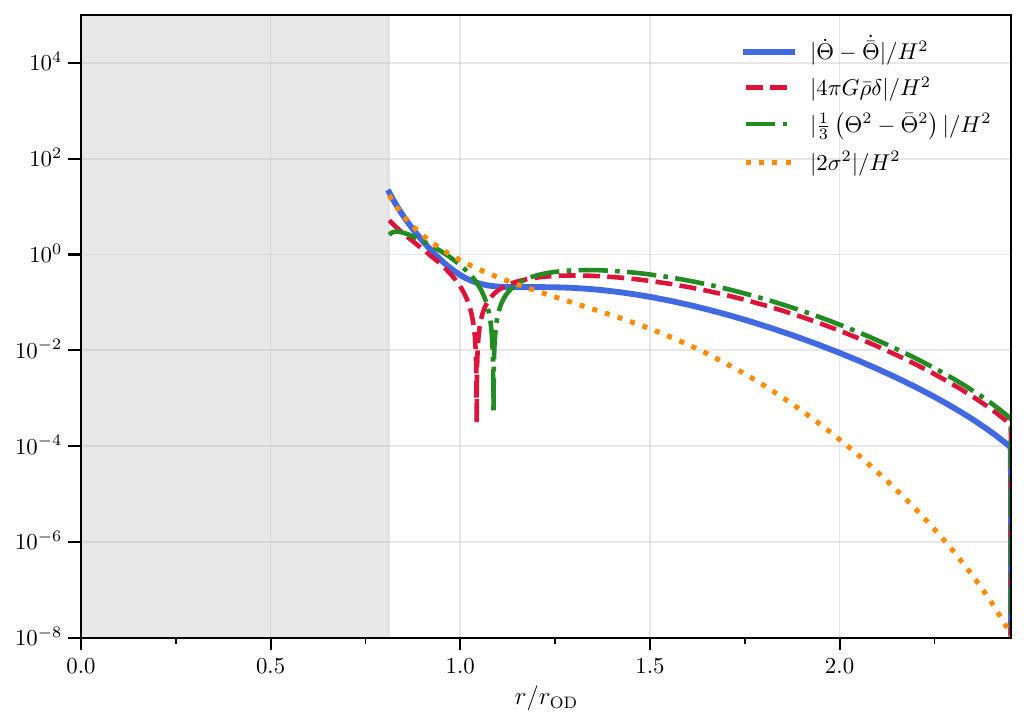}
    \includegraphics[width=0.49\textwidth]{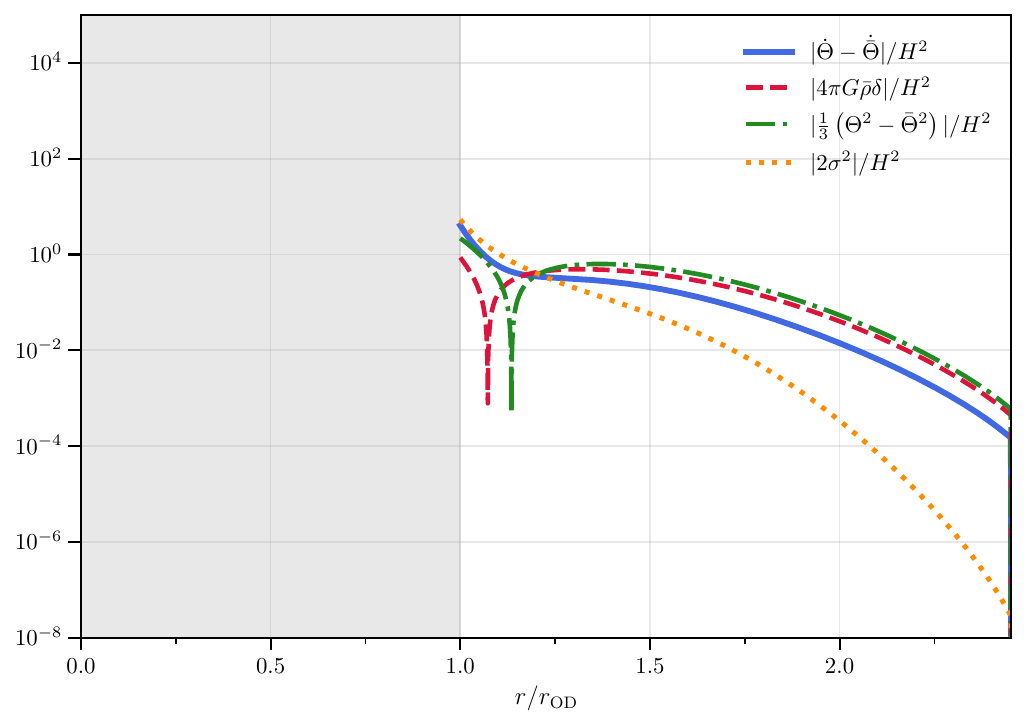}\\
    \caption{Absolute magnitudes of the individual dynamical contributions entering the Raychaudhuri equation normalised by the background Hubble scale $H^2$ for an initial comoving curvature profile with $A=1.7\times10^{-11}$, $\sigma=5~\mathrm{kpc}$ and $\alpha=0.225$. The top-left panel corresponds to the initial slice ($z=300$), the top-right panel to the FH formation time of the central worldline, the bottom right and bottom left panels to the turn-around and  FH formation time of the shell located at $r=r_{\rm OD}$, respectively. While the collapse of the central worldline remains shear-free, the shear contribution grows rapidly away from the centre and progressively dominates the collapse. The shaded gray areas indicate shells which have already undergone collapse}
    \label{fig:Ray}
\end{figure*}

\begin{figure*}[htb!]
    \centering
    \includegraphics[width=0.49\textwidth]{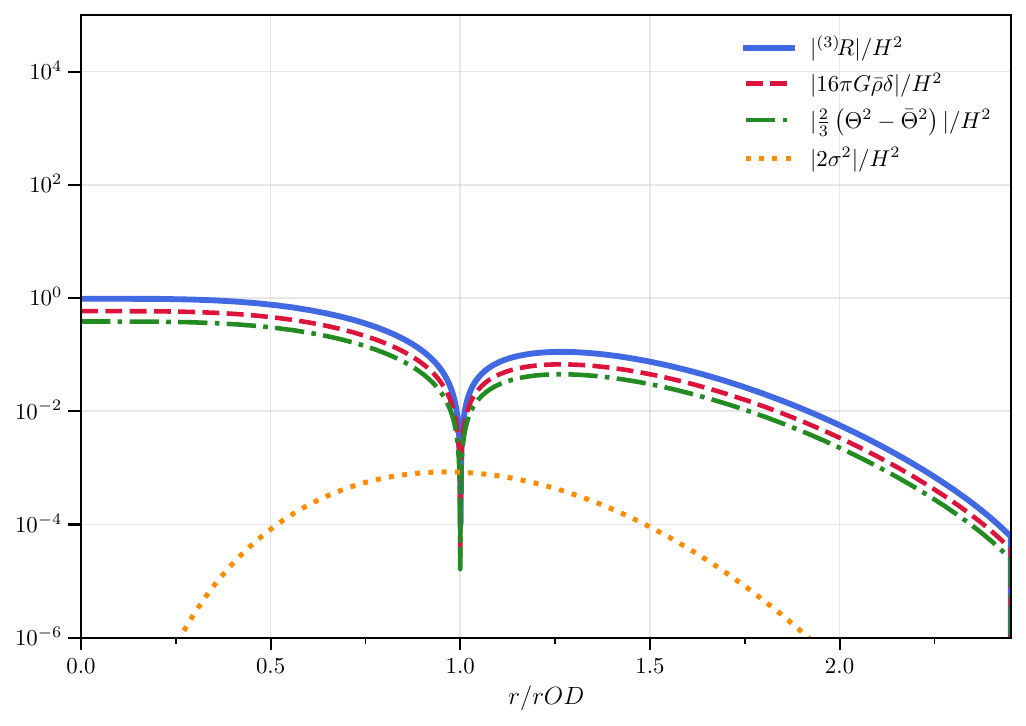}
    \includegraphics[width=0.49\textwidth]{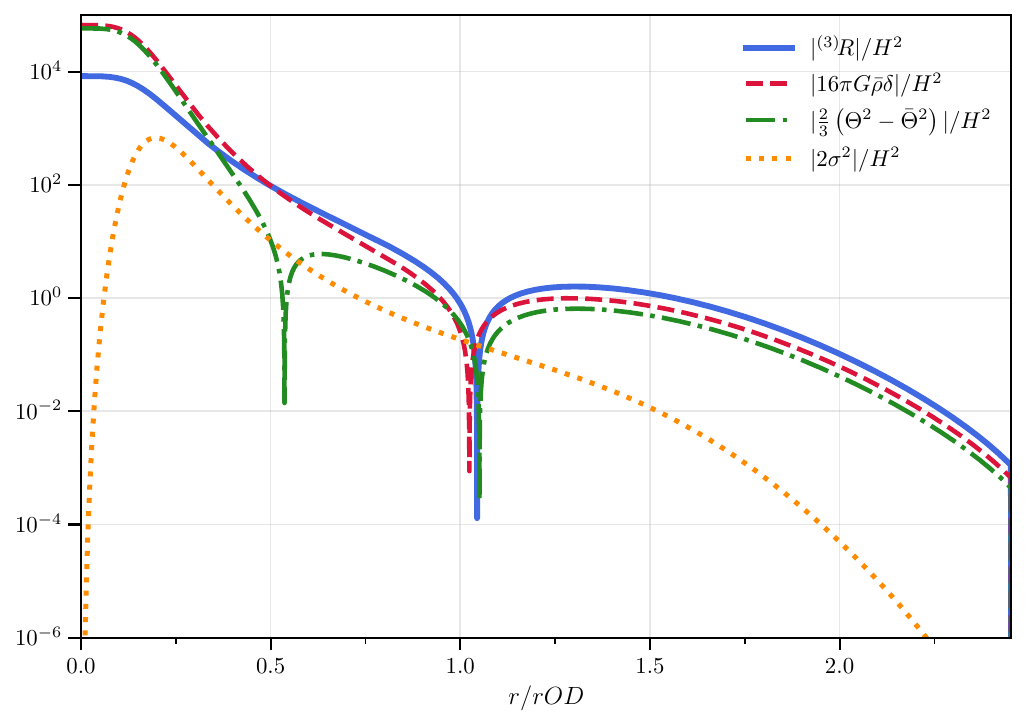}\\
    \includegraphics[width=0.49\textwidth]{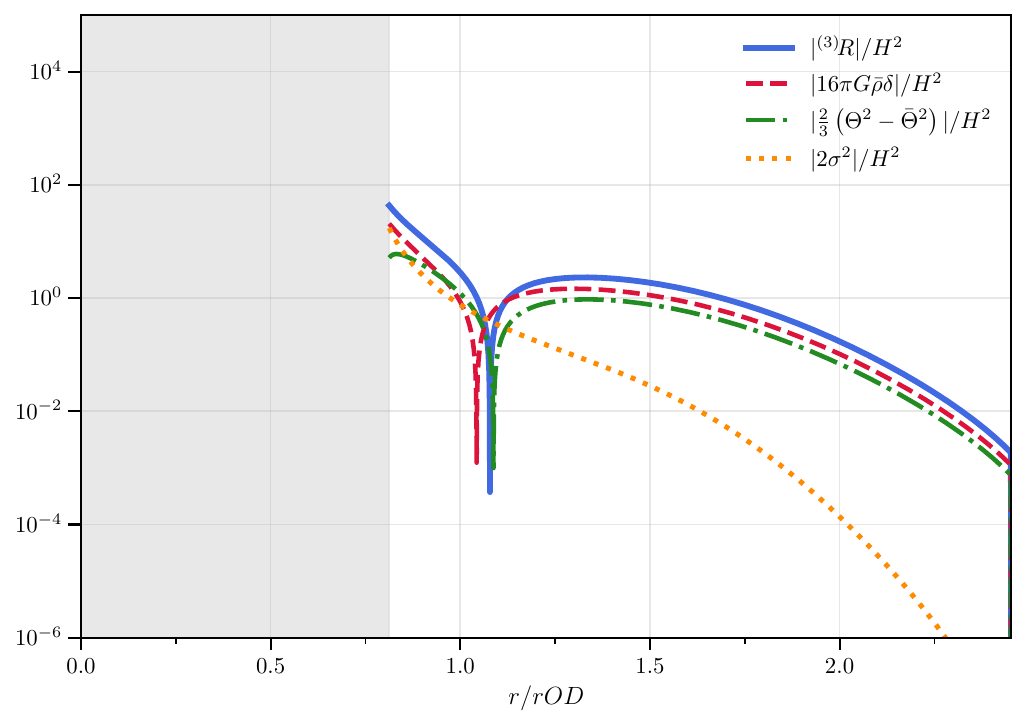}
    \includegraphics[width=0.49\textwidth]{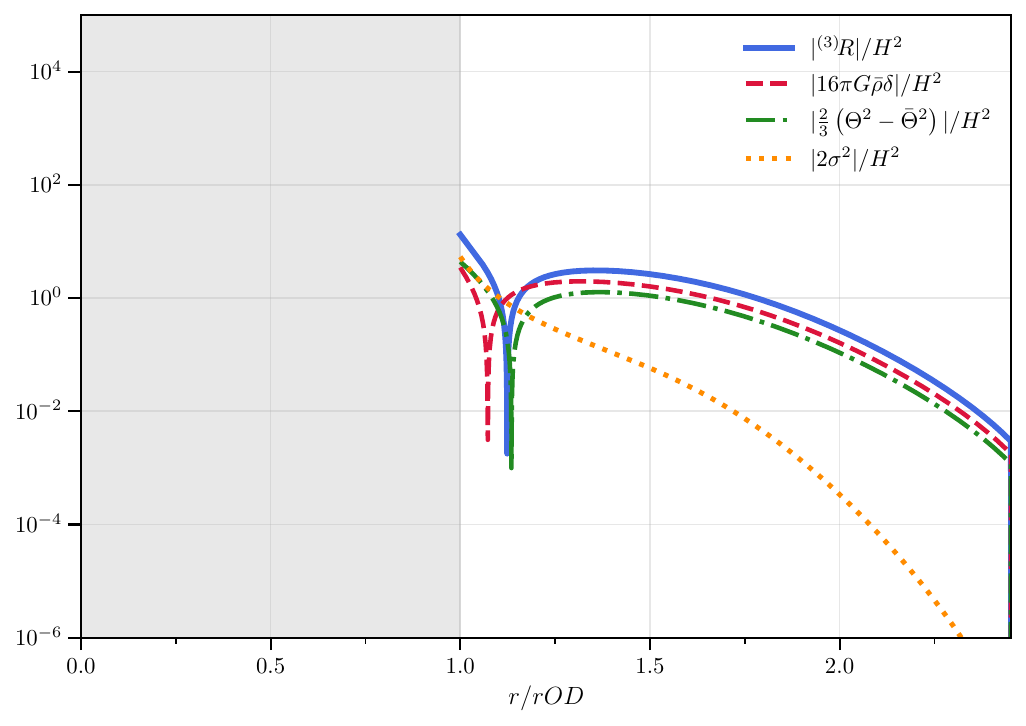}\\
    \caption{Absolute magnitudes of the individual dynamical contributions entering the Hamiltonian constraint, normalised by the background Hubble scale $H^2$  for an initial comoving curvature profile with $A=1.7\times10^{-11}$, $\sigma=5~\mathrm{kpc}$ and $\alpha=0.225$. The top-left panel corresponds to the initial slice ($z=300$), the top-right panel to the FH formation time of the central worldline, the bottom right and bottom left panels to the turn-around and  FH formation time of the shell located at $r=r_{\rm OD}$, respectively. We note that as the collapses progresses the shear overtakes the expansion and the matter as the dominant contribution determining the growth of sptatial curvature. The shaded gray areas indicate shells which have already undergone collapse}
    \label{fig:Hamilton}
\end{figure*}

Additionally, the evolutions of the relative contributions entering the Raychaudhuri equation (see Eq.~\eqref{eq:Raychaudhuri_silent}) and Hamiltonian constraint (see Eq.~\eqref{eq:Raychaudhuri_silent}) are shown in Figs.~\ref{fig:Ray} and~\ref{fig:Hamilton}, respectively. Initially, the dynamics is almost entirely controlled by the density and spatial-curvature terms, while the shear contribution remains strongly suppressed throughout the overdense core (top left panels). This is particularly evident at the centre of the perturbation, where the collapse initially proceeds in an almost isotropic top-hat-like fashion and the shear remains identically zero, consistently with the local spherical symmetry of the central worldline. Nevertheless, already away from the exact centre, anisotropic deformations begin to develop and the shear contribution rapidly grows near the overdensity--underdensity transition region.

At the time at which the central shell reaches FH formation, the overdense core still remains largely dominated by the matter and expansion contributions, although the shear has already become comparable to the remaining terms close to the transition boundary (top right panels). By contrast, by the time the shell located at $r=r_{\rm OD}$ reaches turn-around and then collapses (bottom row), the dynamics has qualitatively changed: the shear term becomes the dominant contribution both in the Raychaudhuri equation, and in the Hamiltonian constraint, over a broad region surrounding the overdensity boundary, thus driving the collapse and singularity formation. This demonstrates that the late-time collapse progressively departs from the local top-hat expectations and becomes increasingly controlled by anisotropic tidal deformations.

Furthermore, we note that Figs.~\ref{fig:Ray} and ~\ref{fig:Hamilton} shows that the location of the overdensity--underdensity interface is itself dynamical. Indeed, as the collapse proceeds, increasingly outer shells approach turn-around and collapse, causing the shear-dominated transition region to move outward.

\begin{figure*}[htb!]
    \centering
    \includegraphics[width=0.49\textwidth]{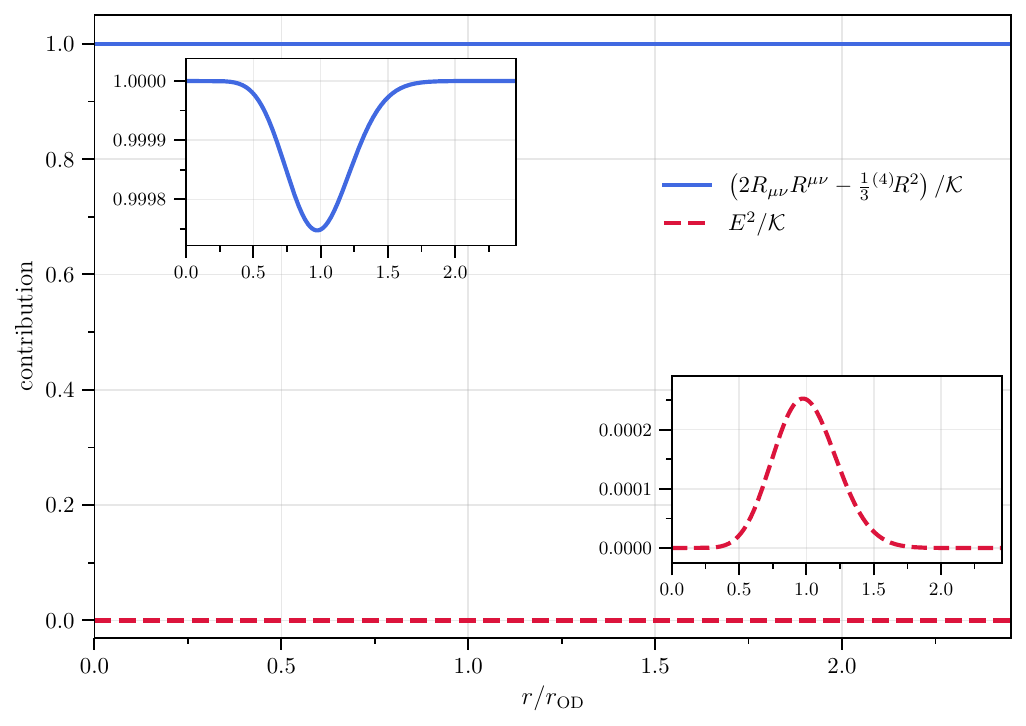}
    \includegraphics[width=0.49\textwidth]{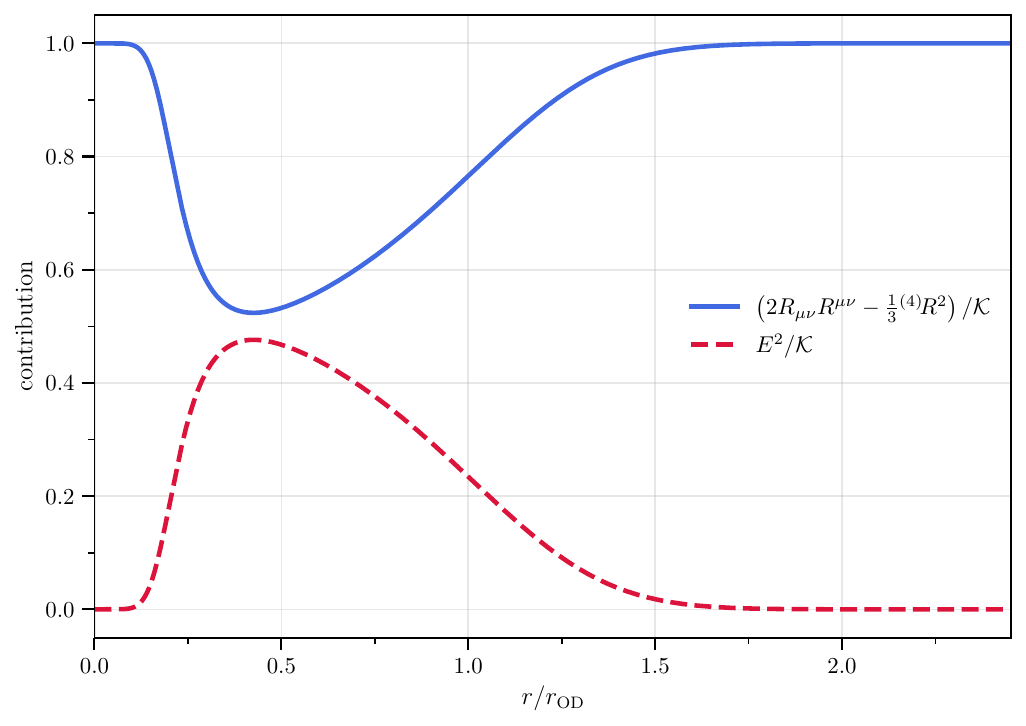}\\

    \includegraphics[width=0.49\textwidth]{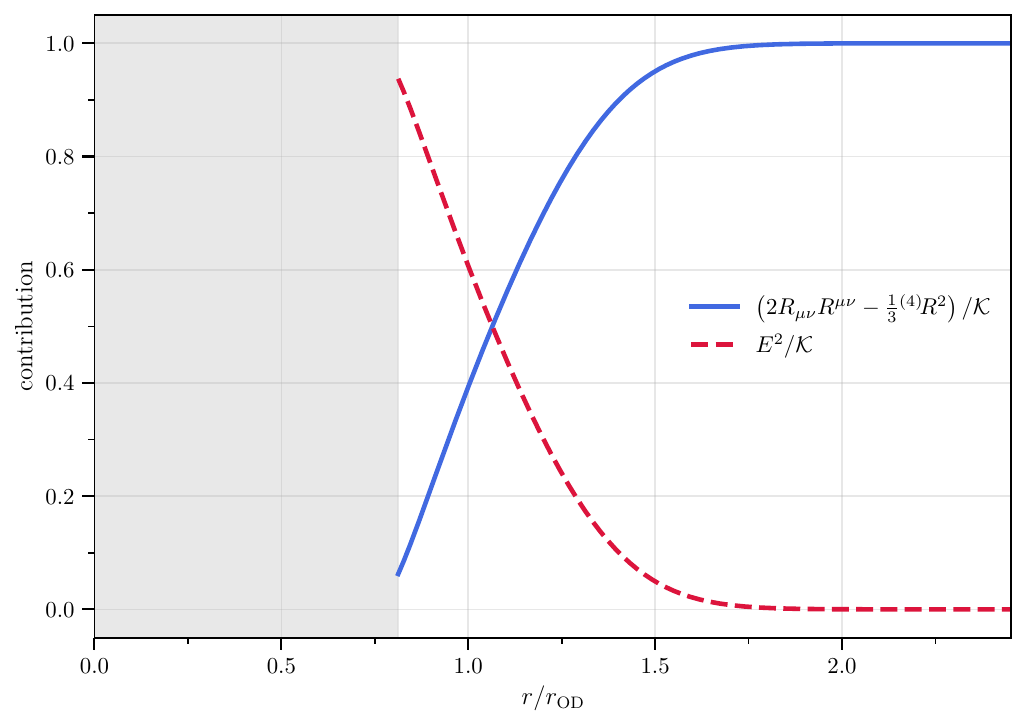}
    \includegraphics[width=0.49\textwidth]{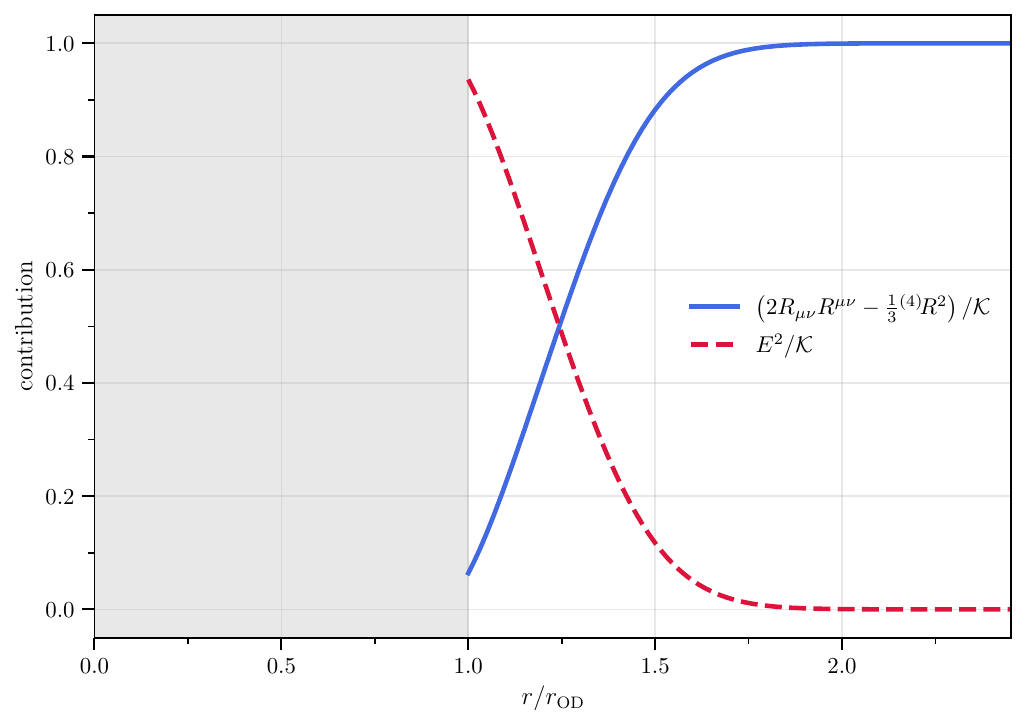}
    \caption{Relative contributions of the Ricci and Weyl curvature sectors to the Kretschmann scalar, evaluated for initial comoving curvature profile with $A=1.7\times10^{-11}$, $\sigma=5~\mathrm{kpc}$ and $\alpha=0.225$. The top-left panel corresponds to the initial slice ($z=300$), the top-right panel to the FH formation time of the central worldline, the bottom right and bottom left panels to the turn-around and  FH formation time of the shell located at $r=r_{\rm OD}$, respectively. The solid blue curve shows the Ricci contribution, $2R_{\mu\nu}R^{\mu\nu}-\frac{1}{3}{^{(4)}\!R}^2$, while the dashed red curve shows the Weyl contribution, $E^2=C_{\mu\nu\rho\sigma}C^{\mu\nu\rho\sigma}$. Initially, the spacetime curvature is almost entirely Ricci dominated, whereas at late stages the Weyl contribution rapidly grows and eventually dominates the Kretschmann scalar throughout the collapsing region.}
    \label{fig:WeylRicci}
\end{figure*}

To further quantify the respective roles of the Electric Weyl and Ricci curvatures within these models, in Fig.~\ref{fig:WeylRicci} we show their respective contribution to the Kretschmann scalar, $\mathcal{K} :=R_{\mu\nu\rho\sigma}R^{\mu\nu\rho\sigma}$, which in four dimensions can be directly decomposed as
\begin{equation}
    \mathcal{K} = C_{\mu\nu\rho\sigma}C^{\mu\nu\rho\sigma} + 2R_{\mu\nu}R^{\mu\nu} - \frac{1}{3} {^{(4)}\!R}^2\, ,
\end{equation}
where for LTB/Szekeres solutions $C_{\mu\nu\rho\sigma}C^{\mu\nu\rho\sigma} = E^2 = 48E_+^2$, and we denoted the standard Ricci curvature as ${^{(4)}\!R}$. In particular, Fig.~\ref{fig:WeylRicci} shows that at the initial time the spacetime curvature is almost entirely dominated by the Ricci contribution, whereas the Weyl contribution remains strongly suppressed, becoming largest only near the transition between overdense and underdense regions. However, as the collapse proceeds, the Weyl contribution rapidly grows and eventually overtakes the Ricci sector, becoming the dominant contribution to the Kretschmann scalar over an increasingly broad radial region. This transition reflects the progressive departure from an almost isotropic matter-dominated collapse toward a strongly tidal and anisotropic regime, where the spacetime curvature becomes primarily sourced by the growth of the electric Weyl tensor rather than by the local Ricci curvature associated with the dust density. In particular, by the time the shell located at $r=r_{\rm OD}$ reaches turn-around, the Kretschmann scalar is almost entirely dominated by the Weyl contribution throughout the collapsing region, signalling the emergence of a genuinely Weyl-curvature-dominated collapse and the approach toward Weyl-dominated singularity formation. 

\begin{figure}[htb!]
    \centering
    \includegraphics[width=0.49\textwidth]{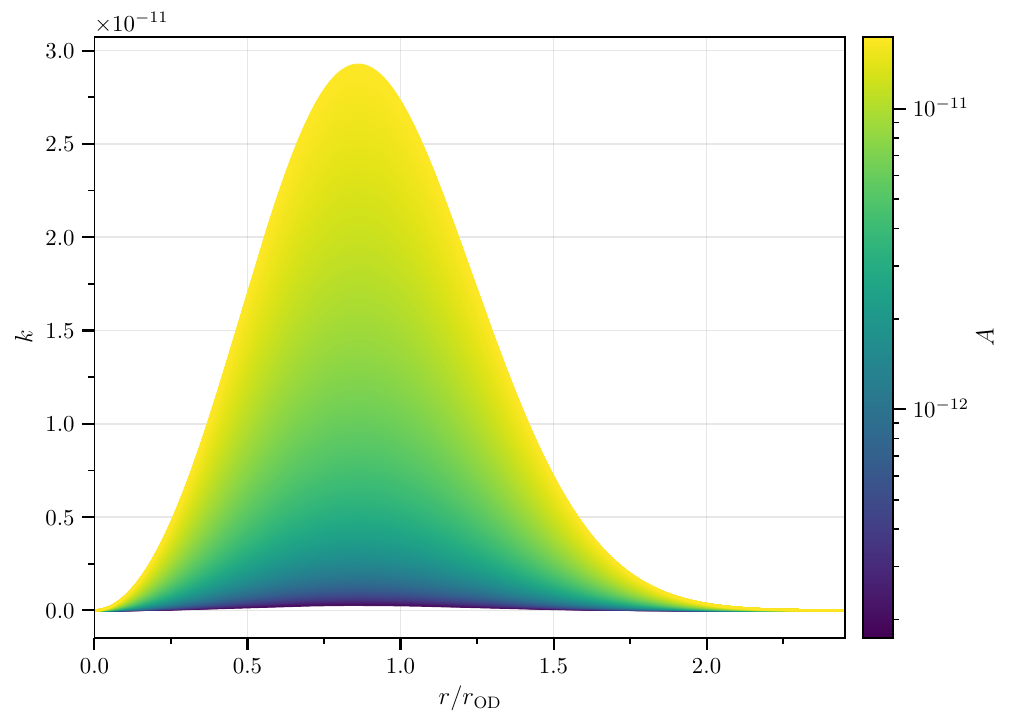}\\
    \includegraphics[width=0.49\textwidth]{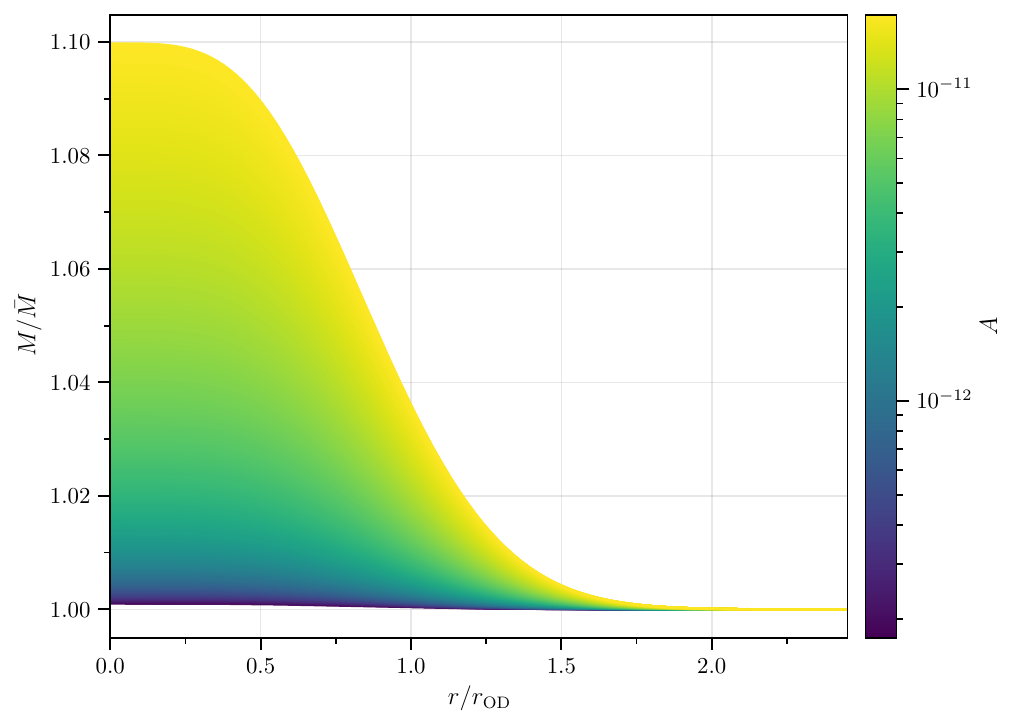}
    \caption{The curvature function $k(r)$ (upper panel) and the enclosed-mass ratio $M(r)/\bar{M}(r)$ (bottom panel) corresponding to the $\Rci$ profiles shown in Fig.~\ref{fig:1}.}
    \label{fig:4}
\end{figure}

The curvature data of Fig.~\ref{fig:1} map uniquely to the LTB free functions, shown in Fig.~\ref{fig:4} as the curvature function $k(r)$ (upper panel) and the enclosed-mass ratio $M(r)/\bar M(r)$ (lower panel). Over the plotted range, the perturbations modify the enclosed mass relative to the background at the $\mathcal{O}(1\%)-\mathcal{O}(10\%)$ level, so that the mass excess is dynamically non-negligible even though the peak density contrast remains perturbative. In particular, comparing Figs.~\ref{fig:2} and~\ref{fig:4} highlights that relatively small values of $k(r)$ can still correspond to appreciable $\delta_i$ once radial gradients are accounted for, and that the largest kinematic anisotropies are localised in the transition region rather than in the softened core.

\begin{figure*}[htb!]
    \centering
    \includegraphics[width=0.49\textwidth]{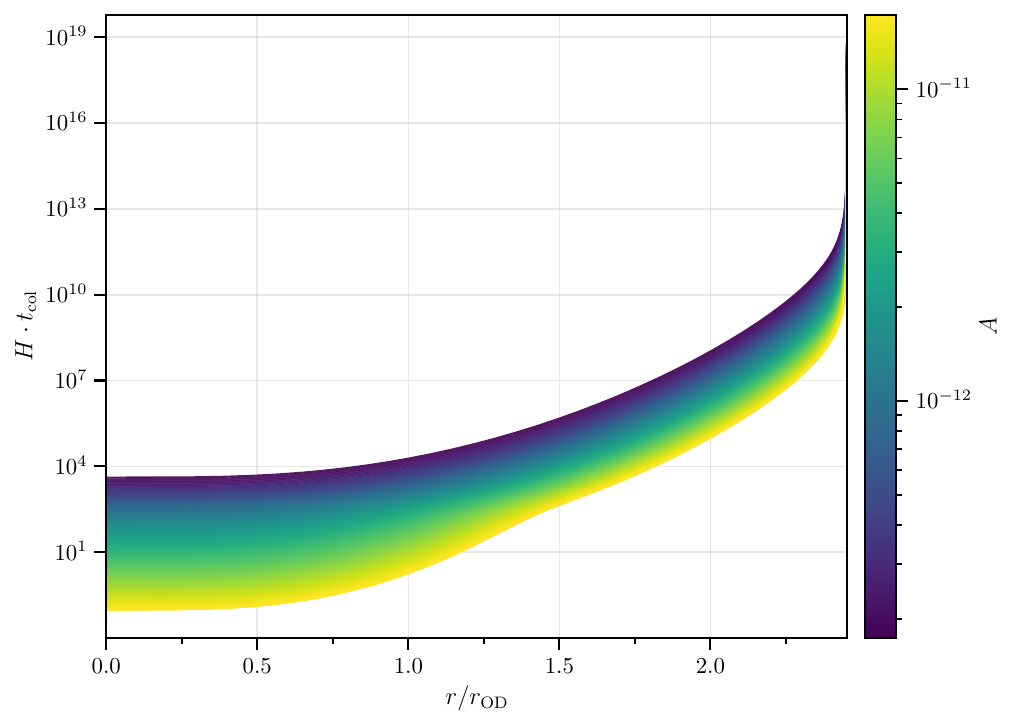}
    \includegraphics[width=0.49\textwidth]{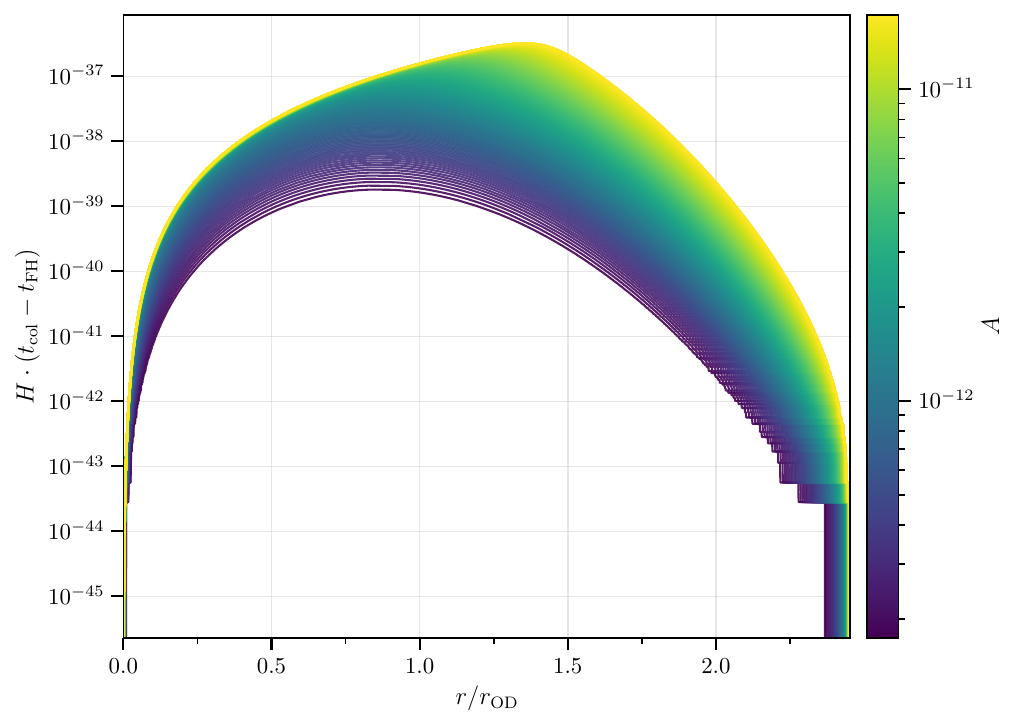}\\
    \includegraphics[width=0.49\textwidth]{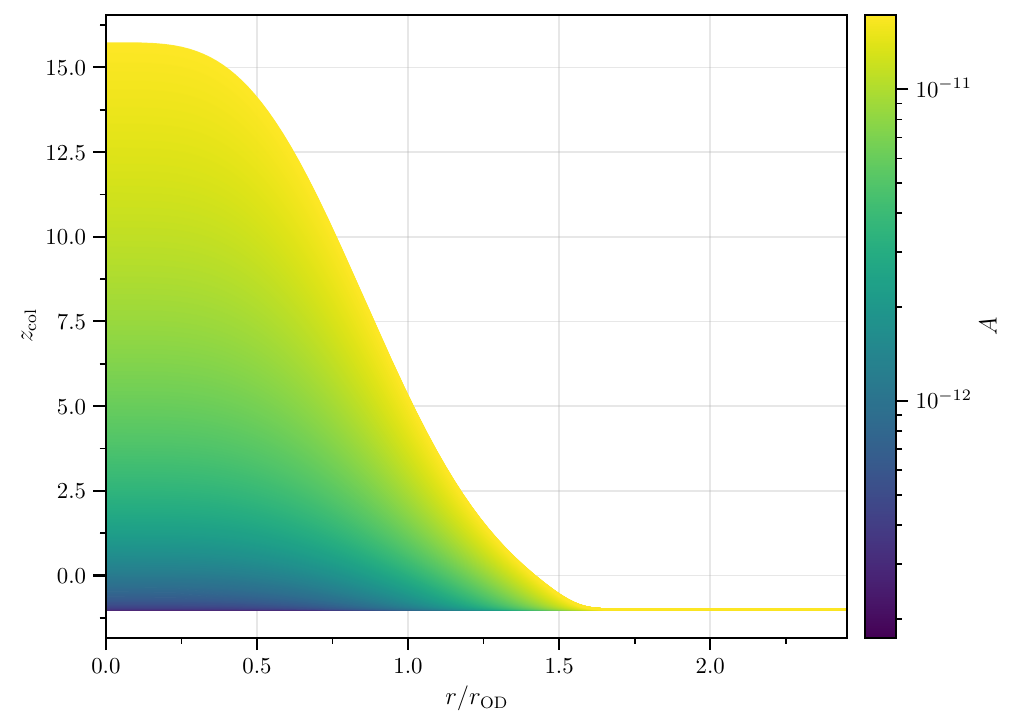}
    \caption{Upper row: the collapse time $t_{\rm col}(r)$ (left) and the fractional delay between collapse and apparent–horizon formation, $(t_{\rm col}-t_{\rm FH})/t_{\rm col}$ (right), both expressed in units of the background Hubble time $H^{-1}$ in a flat $\Lambda$CDM cosmology with Planck 2018 parameters~\cite{Planck_2018}. Bottom row: the corresponding collapse redshift profile $z_{\rm col}(r)$ inferred from $t_{\rm col}(r)$ within the same background model. All curves are computed for the $\Rci$ profiles of Fig.~\ref{fig:1}. The radial range is restricted for clarity, since $t_{\rm col}(r)$ formally diverges as $r\to r_*$.}
    \label{fig:5}
\end{figure*}

Figure~\ref{fig:5} shows --- for the $\Rci$ profiles plotted in Fig.~\ref{fig:1} --- the shell-dependent collapse time normalised with respect to the background Hubble field, $Ht_{\rm col}$ (upper-left), and the corresponding delay between collapse and apparent-horizon formation, $H(t_{\rm col}-t_{\rm FH})$ (upper-right), demonstrating that trapping occurs close to collapse throughout the overdense core. Embedding the solution in a flat $\Lambda$CDM background~\cite{Planck_2018}, the bottom panel of Fig.~\ref{fig:5} reports the corresponding collapse-redshift profile $z_{\rm col}$. In this case, we can see that across the parameter space explored we find that both the central worldline collapse, as well as the redshift by which the entire initial overdense core $0\le r\le r_{\rm OD}$ has been enveloped within a FH spans from $z \sim 16$, for the peaks with the largest gradients, to the present day and beyond, i.e., $z \sim 0$ for the shallowest perturbations. 

\begin{figure}[htb!]
    \centering
    \includegraphics[width=0.49\textwidth]{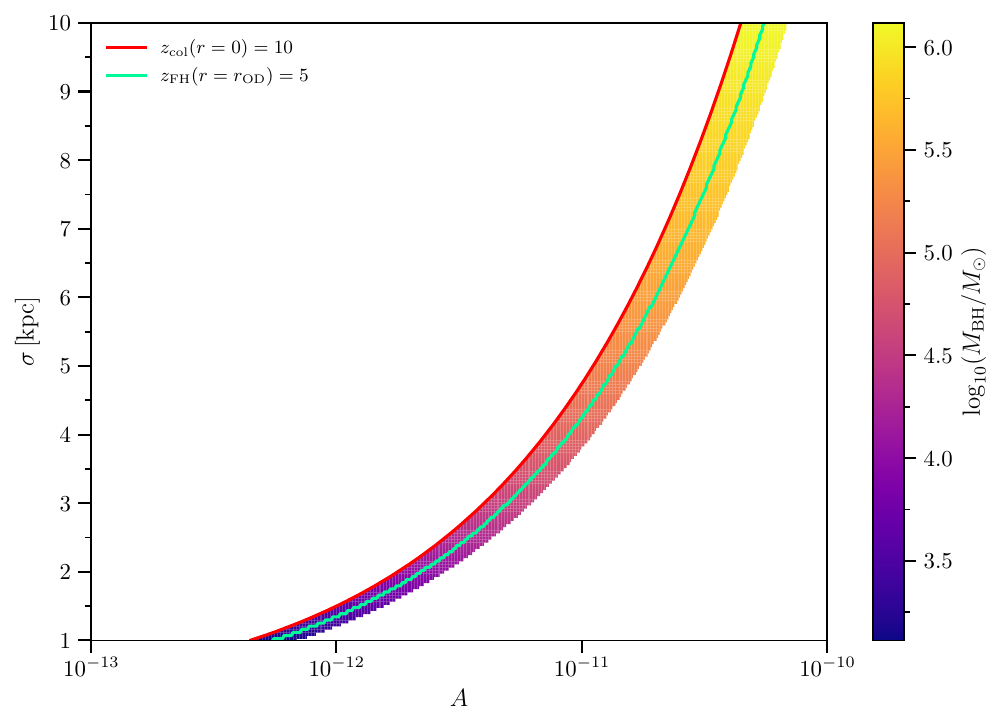}
    \caption{Phase space of collapse outcomes in the $(A,\sigma)$ plane. The red and green curves mark the boundaries $z_{\rm col}(r=0)=10$, and $z_{\rm FH}(r=r_{\rm OD})>5$, respectively. The colour scale shows the resulting black hole mass, $\log_{10}(M_{\rm BH}/M_\odot)$, for models meeting both conditions.}
    \label{fig:BH_phase_space}
\end{figure}

Nonetheless, we find that collapse redshifts over the range $z\simeq 16$ down to $z\simeq 10$, and full BH formations redshifts spanning $z\simeq 7$ down to $z_{\rm col}(r_{\rm OD})\simeq 5$ can still produce seed masses covering the range $M_{\rm BH}\sim 10^{3}$--$10^{6}\mathrm{M}_\odot$. In Fig.~\ref{fig:BH_phase_space}, we show, colour-coded by $M_{\rm BH}$, the region of the $(A,\sigma)$ parameter space for which $z_{\rm col}(r=0) > 10$ and $z_{\rm FH}(r=r_{\rm OD}) > 5$. Here, we note that the mass scale is dominated by the width parameter $\sigma$. On the other hand, the amplitude is found to have a negligible impact on the BH mass, whilst maintaining a significant contribution to the collapse and FH formation times, with broader peaks being able to collapse earlier than more localised one if with larger amplitudes. These results are summerised in Tab.~\ref{tab:seed_range}.

\begin{table*}[]
\centering
\begin{minipage}{0.8\textwidth}
\centering
\hspace*{-1.25cm}
\setlength{\tabcolsep}{11pt}
\begin{tabular}{ccccccc}
\hline\hline 
& $A$ 
& $\sigma~[\mathrm{kpc}]$ 
& $\delta_i^{\rm peak}$ 
& $M_{\rm{BH}}~[\mathrm{M}_\odot]$ 
& $z_{\rm col} (r =0)$ 
& $z_{\rm FH} (r = r_{\rm{OD}})$ 
\\
\hline
& $[10^{-13},\, 7\times10^{-13}]$ 
& $1$
& $[10^{-3},\, 10^{-1}]$ 
& $\sim 10^3$ 
& $[1.5,\, 16]$
& $[-0.3,\, 6]$ 
\\
\hline
& $[1.7\times10^{-13},\, 1.7\times10^{-11}]$ 
& $5$
& $[10^{-3},\, 10^{-1}]$ 
& $\sim 10^5$ 
& $[-0.3,\, 16]$
& $[-1,\, 6]$
\\
\hline
& $[7\times10^{-13},\, 7\times10^{-11}]$ 
& $10$
& $[10^{-3},\, 10^{-1}]$ 
& $\sim 10^6$ 
& $[-0.3,\, 16]$
& $[-1,\, 7]$
\\
\hline\hline
\end{tabular}
\caption{Summary of BH collapse outcomes for initial perturbations set at $z=300$ within a representative subset of the allowed parameter space shown in Fig.~\ref{fig:phase_space}.}
\label{tab:seed_range}
\end{minipage}
\end{table*}

Two further comments are in order. First, the comoving scales considered actually enter the horizon deep in radiation domination; subhorizon linear CDM fluctuations then follow a Mészáros-like evolution prior to equality~\cite{Meszaros_1974,KodamaSasaki_1984}, so that the density contrasts specified at $z=300$ should be interpreted as already processed quantities that have incorporated the pre-equality (and early matter) evolution rather than as direct primordial amplitudes. Second, care must be taken in considering whether the dust approximation remains self-consistent up to trapping. For collisionless CDM the dust limit corresponds to the single-stream regime of the Vlasov dynamics, in which the velocity field is single-valued and the effective pressure and anisotropic stress vanish; this approximation breaks down only at shell-crossing/multi-streaming, when velocities become multivalued and a non-zero velocity-dispersion (hence an effective pressure) is generated (see e.g.,~\cite{S.Pueblas_R.Scoccimarro_2009,O.Erken_2012,I.Sawicki_etal_2013,N.Banik_2015,A.Erschfeld_2019}). Within the family of broad peaks profiles the no--shell-crossing conditions (Eqs.~\eqref{eq:no-shell1}--\eqref{eq:no-shell2}) enforce single-stream evolution up to the FH formation, so that the collapse does not self-generate the velocity dispersion that would spoil the dust closure. Therefore, although representing a strongly restricted class of profiles emerging from peak-theory, their direct collapse does represent a genuine BH formation channel which can be self-consistently described within the CDM dust approximation in the matter era. 

\section{Conclusions}\label{sec:conc}

We have investigated BH formation from the nonlinear growth and collapse of CDM perturbations during the matter-dominated era, modelling CDM as pressureless dust in a fully general-relativistic setting. Interpreting LTB and quasi-spherical Szekeres spacetimes as exact nonlinear deviations from a global FLRW background, we expressed the free functions that determine the collapse --- the active gravitational mass $M(r)$ and curvature function $k(r)$ (and, in the Szekeres case, the dipole function) --- directly in terms of the initial comoving curvature perturbation field, $\Rci$.

We derived an analytic characterisation of the key timescales of the collapse, shell by shell: the turn-around time, the collapse time, and the future apparent-horizon formation time. After mapping to cosmological initial data, these times can be written explicitly in terms of $\mathcal{R}_{c,i}$ and its derivatives, making explicit that it is the \textit{steepness} of the initial profile --- through the radial gradient $\mathcal{R}'_{c,i}$ --- that controls when trapping and collapse occurs on a given shell. In particular, for the axisymmetric quasi-spherical Szekeres subclass, we further showed that these shell-focusing and trapping times are fixed by the monopole $\mathcal{R}^{[m]}_{c,i}$ alone: the dipole affects the anisotropic kinematics, but not the collapse and horizon-formation times of the radial shells.

We expressed previously known regularity conditions for the LTB and Szekeres spacetimes as constraints on the initial comoving curvature profile, such as the conditon for a shell-crossing-free collapse (see Eq.~\eqref{eq:figata}). Within our perturbative framework --- namely, of collapsing proto-structures seeded by initial comoving curvature perturbations about a FLRW background --- we also established that in general whenever shell-crossing occurs, it precedes horizon formation, so that we expect a breakdown of the dust description before a trapped region can form. In other words, this more generic case deserves further studies, beyond the conservative assumption made in this paper in order to establish a secure channel for BHs formation from initial CDM perturbations: trapped regions and BHs can likely also  form if a pancake-like anisotropy arises, but this can probably only be studied using a particle approach, as opposed to the dust fluid used here. Furthermore, we have shown that, provided shell-crossing is avoided throughout the collapse, the ensuing central shell-focusing singularity is generically locally naked, and may in addition be globally naked. We have then showed that single-mode and Gaussian profiles should not be considered as BH forming channels as they determine collapses with a central naked singularity. 

However, within the context of generelised BBKS peak theory \cite{A.G.Doroshkevich_1970,J.M.Bardeen_J.R.Bond_N.Kaiser_A.S.Szalay_1986,R.J.Adler_1981,C.Germani_T.Prokopec_2017,C.Germani_2025} applied to the initial comoving curvature field, and by considering compensated perturbations, we find that $\mathcal{R}_{c,i}$ profiles which allow for shell-crossing-free collapse and a covered central singualrity naturally emerge. These correspond to a sub-set of \textit{softened, broad} peaks: configurations for which the core is sufficiently flattened that its evolution remains extremely close to that of a top-hat collapse. In this regime, kinematic anisotropies sourced by the shear and the electric part of the Weyl tensor are strongly suppressed toward the centre. As a result, the inner region evolves quasi-isotropically up to the onset of focusing, providing a concrete realisation of matter-era collapse that yields a trapped central singularity within a fully relativistic dust model describing CDM --- i.e., a viable BH formation channel in the matter era. In particular, adopting realistic perturbation profiles associated with such localised peaks, we have found that to form BHs with $M_{\rm BH} \in [10^3 \mathrm{M}_\odot,\,10^6\,\mathrm{M}_\odot]$ their cores collapse at redshifts $10 \lesssim z \lesssim 16$, thereby offering a plausible origin for massive BH seeds in early galaxies, a simple channel of formation purely based on CDM and gravity, with no other parameters than those characterizing the amplitude and the profile of the initial perturbation.

Finally, using the covariant $1+3$ formalism and the associated dynamical-systems analysis of quasi-spherical Szekeres/LTB collapse, we characterised the local kinematic end-state of the evolution along each flowline. Interestingly, we find that even in BH–forming solutions, shell-focusing collapse is generically \textit{not} point-like: the late-time attractor for collapsing shells is typically a Kasner-like \textit{cigar} state, so the approach to focusing is intrinsically anisotropic. Moreover, the realised kinematic branch is controlled by the sign of the initial local shear eigenvalue (see Eqs.~\eqref{eq:sigma+GW} and \eqref{eq:Theta1}--\eqref{eq:l2l3}): the shear-free limit corresponds to an isotropic, point-like collapse, a positive shear eigenvalue drives the evolution toward the cigar-like attractor, whilea  negative shear eigenvalue leads toward a pancake-like configuration, which must be associated with shell-crossing and thus signals, in our interpretation, a breakdown of the dust approximation rather than a genuine spacetime singularity.

To conclude, we note that several extensions are natural, and will be investigated in future papers. Indeed, extensions beyond the dust approximation --- e.g., by introducing a radiation component, or moving to a Vlasov description for the matter content --- would further probe the link between cosmological initial conditions and the  formation process of super-massiv BHs, as well as allowing for beyond-shell-crossing descriptions. Additionally, considering non-Gaussianity in the comoving curvature field could lead to further BH formation channels. Nonetheless, the analytic mapping derived in this work provides a first, compact and physically transparent bridge between initial comoving curvature profiles and the timescales, regularity, and kinematic end-states of relativistic CDM collapse. 
\begin{acknowledgments}
The authors are grateful to Francesco Bennetti and Sanjay Jhingan for insightful discussions. MG acknowledges support by the Marsden Fund grant M1271 administered by the Royal Society of New Zealand, Te~Ap\=arangi. The work of TH was partially supported by JSPS KAKENHI Grants No. JP24K07027. For the purpose of open access, we have applied a Creative Commons Attribution (CC
BY) licence to any Author Accepted Manuscript version arising. Supporting research data
are available on reasonable request from the authors.
\end{acknowledgments}


\bibliography{references}

@PREAMBLE{
 "\providecommand{\noopsort}[1]{}" 
 # "\providecommand{\singleletter}[1]{#1}%" 
}

@article{Oppenheimer:1939ue,
    author = "Oppenheimer, J. R. and Snyder, H.",
    title = "{On Continued gravitational contraction}",
    doi = "10.1103/PhysRev.56.455",
    journal = "Phys. Rev.",
    volume = "56",
    pages = "455--459",
    year = "1939"
}

@article{S.Pueblas_R.Scoccimarro_2009,
  author        = {Pueblas, S. and Scoccimarro, R.},
  title         = {{Generation of vorticity and velocity dispersion by orbit crossing}},
  journal       = {\prd},
  volume        = {80},
  pages         = {043504},
  year          = {2009},
  doi           = {10.1103/PhysRevD.80.043504},
}

@article{I.Sawicki_etal_2013,
  title = {{Seeding supermassive black holes with a non-vortical dark-matter subcomponent}},
  author = {Sawicki, I. and Marra, V. and Valkenburg, W.},
  journal = {\prd},
  volume = {88},
  pages = {083520},
  year = {2013},
  doi = {10.1103/PhysRevD.88.083520},
}

@article{A.Erschfeld_2019,
  title = {{Evolution of dark matter velocity dispersion}},
  author = {Erschfeld, A. and Floerchinger, S.},
  journal = {JCAP},
  volume = {06},
  pages = {039},
  year = {2019},
  doi = {10.1088/1475-7516/2019/06/039},
}

@article{O.Erken_2012,
  title = {{Cosmic axion thermalization}},
  author = {Erken, O. and Sikivie, P. and Tam, H. and Yang, Q.},
  journal = {\prd},
  volume = {85},
  pages = {063520},
  year = {2012},
  doi = {10.1103/PhysRevD.85.063520},
}

@article{N.Banik_2015,
  title = {{Linear Newtonian perturbation theory from the Schr\"odinger-Poisson equations}},
  author = {Banik, N. and Christopherson, A. J. and Sikivie, P. and Todarello, E. M.},
  journal = {\prd},
  volume = {91},
  pages = {123540},
  year = {2015},
  doi = {10.1103/PhysRevD.91.123540},
}

@ARTICLE{KodamaSasaki_1984,
   author = {{Kodama}, H. and {Sasaki}, M.},
    title = "{Cosmological Perturbation Theory}",
  journal = {Prog.\ Th.\ Phys.\ Supp.},
     year = 1984,
   volume = 78,
    pages = {1},
      doi = {10.1143/PTPS.78.1}
}

@ARTICLE{Meszaros_1974,
       author = {{M\'esz\'aros}, P.},
        title = "{The behaviour of point masses in an expanding cosmological substratum}",
      journal = {\aap},
         year = 1974,
       volume = {37},
        pages = {225-228},
}

@article{L.Kofman_D.Pogosian_1995,
    title     = {Equations of gravitational instability are nonlocal},
    author    = {Kofman, L. and Pogosian, D.},
    journal   = {\apj},
    volume    = {442},
    number    = {},
    pages     = {30-38},
    year      = {1995},
    month     = {3},
    doi       = {10.1086/175419},
}

@article{E.Bentivegna_M.Bruni_2016,
    title     = {{Effects of Nonlinear Inhomogeneity on the Cosmic Expansion with Numerical Relativity}},
    author    = {Bentivegna, E. and Bruni, M.},
    journal   = {\prl},
    volume    = {116},
    number    = {25},
    pages     = {251302},
    year      = {2016},
    month     = {6},
    doi       = {10.1103/PhysRevLett.116.251302},
}

@article{H.J.Macpherson_etal_2017,
  author        = {Macpherson, Hayley J. and Lasky, Paul D. and Price, Daniel J.},
  title         = {{Inhomogeneous Cosmology with Numerical Relativity}},
  journal       = {\prd},
  volume        = {95},
  number        = {6},
  pages         = {064028},
  year          = {2017},
  doi           = {10.1103/PhysRevD.95.064028},
}

@article{W.E.East_etal_2018,
  author        = {East, William E. and Wojtak, Rados{\l}aw and Abel, Tom},
  title         = {{Comparing Fully General Relativistic and Newtonian Calculations of Structure Formation}},
  journal       = {\prd},
  volume        = {97},
  number        = {4},
  pages         = {043509},
  year          = {2018},
  doi           = {10.1103/PhysRevD.97.043509},
}

@article{J.C.Aurrekoetxea_etal_2020,
    title     = {{The Effects of Potential Shape on Inhomogeneous Inflation}},
    author    = {Aurrekoetxea, J. C. and Clough, K. and Flauger, R. and Lim, E. A.},
    journal   = {\jcap},
    publisher = {{IOP} Publishing},
    volume    = {2020},
    number    = {5},
    pages     = {30-30},
    year      = {2020},
    month     = {5},
    doi       = {10.1088/1475-7516/2020/05/030},
}

@article{S.Saga_etal_2021,
    title     = {{Cold dark matter protohalo structure around collapse: Lagrangian cosmological perturbation theory versus Vlasov simulations}},
    author    = {Saga, S. and Taruya, A. and Colombi, S.},
    journal   = {\aap},
    publisher = {{EDP} Sciences},
    volume    = {664},
    number    = {},
    pages     = {A3},
    year      = {2022},
    month     = {8},
    doi       = {10.1051/0004-6361/202142756}
}

@article{J.E.Gunn_J.R.Gott_1972,
    title     = {{On the Infall of Matter Into Clusters of Galaxies and Some Effects on Their Evolution}},
    author    = {Gunn, J. E. and Gott, J. R. III},
    journal   = {\apj},
    publisher = {{IOP} Publishing},
    volume    = {176},
    number    = {},
    pages     = {1},
    year      = {1972},
    month     = {8},
    doi       = {10.1086/151605},
    archivePrefix = {arXiv},
    eprint    = {}}

@article{W.H.Press_P.Schechter_1974,
    title     = {{Formation of Galaxies and Clusters of Galaxies by Self-Similar Gravitational Condensation}},
    author    = {Press, W. H. and Schechter, P.},
    journal   = {\apj},
    publisher = {{IOP} Publishing},
    volume    = {187},
    number    = {},
    pages     = {425-438},
    year      = {1974},
    month     = {2},
    doi       = {10.1086/152650},
    archivePrefix = {arXiv},
    eprint    = {}}

@article{R.K.Sheth_G.Tormen_1999,
    title     = {Large-scale bias and the peak background split},
    author    = {Sheth, R. K. and Tormen, G.},
    journal   = {\mnras},
    publisher = {Oxford University Press},
    volume    = {308},
    number    = {1},
    pages     = {119-126},
    year      = {1999},
    month     = {9},
    doi       = {10.1046/j.1365-8711.1999.02692.x},
    archivePrefix = {arXiv},
    eprint    = {astro-ph/9901122}}

@article{Y.B.Zeldovich_1970,
    title     = {{Gravitational instability: An approximate theory for large density perturbations.}},
    author    = {Zel'dovich, Y. B.},
    journal   = {\aap},
    publisher = {{EDP} Sciences},
    volume    = {5},
    number    = {},
    pages     = {84-89},
    year      = {1970},
    month     = {3},
    doi       = {},
    archivePrefix = {arXiv},
    eprint    = {}}

@article{R.Arnowitt_S.Deser_C.W.Misner_1959a,
    author = "Arnowitt, Richard L. and Deser, Stanley and Misner, Charles W.",
    title = "{Dynamical Structure and Definition of Energy in General Relativity}",
    doi = "10.1103/PhysRev.116.1322",
    journal = "Phys.\ Rev.",
    volume = "116",
    pages = "1322--1330",
    year = "1959"
}

@article{R.Arnowitt_S.Deser_C.W.Misner_1959b,
  author  = {Arnowitt, Richard and Deser, Stanley},
  title   = {{Quantum Theory of Gravitation: General Formulation and Linearized Theory}},
  journal = {Phys.\ Rev.},
  year    = {1959},
  volume  = {113},
  number  = {3},
  pages   = {745--750},
  doi     = {10.1103/PhysRev.113.745}
}

@article{R.Arnowitt_S.Deser_C.W.Misner_1960,
  author  = {Arnowitt, Richard and Deser, Stanley and Misner, Charles W.},
  title   = {{Canonical Variables for General Relativity}},
  journal = {Phys.\ Rev.},
  year    = {1960},
  volume  = {117},
  number  = {6},
  pages   = {1595--1602},
  doi     = {10.1103/PhysRev.117.1595}
}

@article{R.Arnowitt_S.Deser_C.W.Misner_1961,
  author       = {Arnowitt, Richard and Deser, Stanley and Misner, Charles W.},
  title        = {{Coordinate Invariance and Energy Expressions in General Relativity}},
  journal      = {Phys.\ Rev.},
  volume       = {122},
  number       = {3},
  pages        = {997--1006},
  year         = {1961},
  month        = may,
  doi          = {10.1103/PhysRev.122.997},
}

@article{P.A.M.Dirac_1958,
  author  = {Dirac, P. A. M.},
  title   = {{The Theory of Gravitation in Hamiltonian Form}},
  journal = {Proc.\ R.\ Soc.\ Lond.\ A},
  year    = {1958},
  volume  = {246},
  number  = {1246},
  pages   = {333--343},
  doi     = {10.1098/rspa.1958.0142}
}

@article{P.A.M.Dirac_1959,
  author  = {Dirac, P. A. M.},
  title   = {{Fixation of Coordinates in the Hamiltonian Theory of Gravitation}},
  journal = {Phys.\ Rev.},
  year    = {1959},
  volume  = {114},
  number  = {3},
  pages   = {924--930},
  month   = may,
  doi     = {10.1103/PhysRev.114.924}
}

@article{S.A.Hojman_K.Kuchar_C.Teitelboim_1976,
  author  = {Hojman, Sergio A. and Kucha{\v r}, Karel and Teitelboim, Claudio},
  title   = {{Geometrodynamics Regained}},
  journal = {Annals of Physics},
  year    = {1976},
  volume  = {96},
  number  = {1},
  pages   = {88--135},
  month   = jan,
  doi     = {10.1016/0003-4916(76)90112-3}
}

@article{Y.ChoquetBruhat_1952,
  author  = {Four{\`e}s-Bruhat, Yvonne},
  title   = {{Th{\'e}or{\`e}me d'existence pour certains syst{\`e}mes d'{\'e}quations aux d{\'e}riv{\'e}es partielles non lin{\'e}aires}},
  journal = {Acta Math.},
  year    = {1952},
  volume  = {88},
  pages   = {141--225},
  doi     = {10.1007/BF02392131}
}

@article{Y.ChoquetBruhat_R.Geroch_1969,
  author  = {Choquet-Bruhat, Yvonne and Geroch, Robert},
  title   = {{Global Aspects of the Cauchy Problem in General Relativity}},
  journal = {Commun.\ Math.\ Phys.},
  year    = {1969},
  volume  = {14},
  pages   = {329--335},
  doi     = {10.1007/BF01645389}
}

@article{York1973_ConformalDecomposition,
  author  = {York, James W., Jr.},
  title   = {{Conformally Invariant Orthogonal Decomposition of Symmetric Tensors on Riemannian Manifolds and the Initial-Value Problem of General Relativity}},
  journal = {J.\ Math.\ Phys.},
  year    = {1973},
  volume  = {14},
  number  = {4},
  pages   = {456--464},
  doi     = {10.1063/1.1666338}
}

@article{R.L.Munoz_M.Bruni_2023,
    author = "Munoz, Robyn L. and Bruni, Marco",
    title = "{Structure formation and quasispherical collapse from initial curvature perturbations with numerical relativity simulations}",
    doi = "10.1103/PhysRevD.107.123536",
    journal = "\prd",
    volume = "107",
    number = "12",
    pages = "123536",
    year = "2023"
}

@article{K.A.Malik_D.Wands_2008,
    title     = {Cosmological perturbations},
    author    = {Malik, K. A. and Wands, D.},
    journal   = {Phys.\ Rep.},
    publisher = {Elsevier},
    volume    = {475},
    number    = {1},
    pages     = {1-51},
    month     = {5},
    year      = {2009},
    doi       = {10.1016/j.physrep.2009.03.001},
}

@article{Bruni:2002xk,
    author = "Bruni, Marco and Maartens, Roy and Tsagas, Christos G.",
    title = "{Magnetic field amplification in CDM anisotropic collapse}",
    eprint = "astro-ph/0208126",
    archivePrefix = "arXiv",
    doi = "10.1046/j.1365-8711.2003.06095.x",
    journal = "Mon. Not. Roy. Astron. Soc.",
    volume = "338",
    pages = "785",
    year = "2003"
}

@article{B.Kalbouneh_etal_2025,
  author  = {Kalbouneh, Basheer and Santiago, Jessica and Marinoni, Christian and Maartens, Roy and Clarkson, Chris and Sarma, Maharshi},
  title   = {Expanding covariant cosmography of the local Universe: incorporating the snap and axial symmetry},
  journal = {\jcap},
  year    = {2025},
  pages   = {02 076},
  doi     = {10.1088/1475-7516/2025/02/076},
}

@misc{M.Sarma_C.Marinoni_B.Kalbouneh_C.Clarkson_R.Maartens_2025,
  author  = {Sarma, Maharshi and Marinoni, Christian and Kalbouneh, Basheer and Clarkson, Chris and Maartens, Roy},
  title   = {Covariant cosmography in the presence of local structures: comparing exact solutions and perturbation theory},
  year    = {2025},
  eprint  = {2510.03517},
  archivePrefix = {arXiv},
  primaryClass  = {astro-ph.CO},
  doi = {10.48550/arXiv.2510.03517}
}

@article{J.Ehlers_1961,
  author    = {Ehlers, J.},
  title     = {{Beitr{\"a}ge zur relativistischen Mechanik kontinuierlicher Medien}},
  journal = {Akad.\ Wiss.\ Mainz.\ Abh.\ Math.-Nat.\ Kl.},
  volume       = {11},
  pages        = {792--837},
  year         = {1961},
  note      = {[Republished in {\protect\textit{Gen.\ Relativ.\ Grav.}} {\textbf25} 1225--1266 (1993)]}
}

@article{S.W.Hawking_1966,
  author  = "Hawking, Stephen W.",
  title   = "{Perturbations of an expanding universe}",
  journal = "\apj",
  volume  = "145",
  pages   = "544--554",
  year    = "1966", 
  doi     = " 10.1086/148793"
}

@incollection{G.F.R._Ellis1971,
  author    = "Ellis, George F. R.",
  title     = "{Relativistic Cosmology}",
  booktitle = "General Relativity and Cosmology (Proc.\ Int.\ School of Physics ``Enrico Fermi'', Course 47)",
  editor    = "Sachs, R. K.",
  pages     = "104--182",
  year      = "1971",
  publisher = "Academic Press",
}

@article{M.Bruni_etal_2014_Mar,
    title     = {{Non-Gaussian initial conditions in $\Lambda$CDM: Newtonian, relativistic, and primordial contributions}},
    author    = {Bruni, M. and Hidalgo, J. C. and Meures, N. and Wands, D.},
    journal   = {The Astrophysical Journal},
    publisher = {{IOP} Publishing},
    volume    = {785},
    number    = {1},
    pages     = {2},
    year      = {2014},
    month     = {3},
    doi       = {10.1088/0004-637X/785/1/2},
}

@article{M.Bruni_etal_2014_Sep,
    title     = {Einstein's signature in cosmological large-scale structure},
    author    = {Bruni, M. and Hidalgo, J. C. and Wands, D.},
    journal   = {\apjl},
    volume    = {794},
    number    = {1},
    pages     = {L11},
    year      = {2014},
    month     = {9},
    doi       = {10.1088/2041-8205/794/1/L11},
}

@article{D.Langlois_F.Vernizzi_2010,
    title     = {A geometrical approach to nonlinear perturbations in relativistic cosmology},
    author    = {D. Langlois and F. Vernizzi},
    journal   = {\cqg},
    publisher = {{IOP} Publishing},
    volume    = {27},
    number    = {12},
    pages     = {124007},
    year      = {2010},
    month     = {5},
    doi       = {10.1088/0264-9381/27/12/124007},
}

@article{D.H.Lyth_1985,
    title     = {Large-scale energy-density perturbations and inflation},
    author    = {Lyth, D. H.},
    journal   = {\prd},
    volume    = {31},
    number    = {8},
    pages     = {1792-1798},
    year      = {1985},
    month     = {4},
    doi       = {10.1103/PhysRevD.31.1792},
}

@article{J.M.Stewart_M.Walker_1974,
    title     = {{Perturbations of Space-Times in General Relativity}},
    author    = {Stewart, J. M. and Walker, M.},
    journal   = {Proc.\ R.\ Soc.\ London. A. Math.\ Phys.\ Sc.},
    publisher = {Royal Society},
    volume    = {341},
    number    = {1624},
    pages     = {49-74},
    year      = {1974},
    month     = {10},
    doi       = {10.1098/rspa.1974.0172},
}

@article{M.Bruni_etal_1992,
    title     = {Cosmological perturbations and the physical meaning of gauge invariant variables},
    author    = {Bruni, M. and Dunsby, P. K. S. and Ellis, G. F. R.},
    journal   = {\apj},
    publisher = {{IOP} Publishing},
    volume    = {395},
    number    = {Aug},
    pages     = {34-53},
    year      = {1992},
    month     = {8},
    doi       = {10.1086/171629},
}

@article{M.Bruni_etal_1995_Jun,
    title     = {Dynamics of silent universes},
    author    = {Bruni, M. and Matarrese, S. and Pantano, O.},
    journal   = {\apj},
    publisher = {{IOP} Publishing},
    volume    = {445},
    number    = {},
    pages     = {958-977},
    year      = {1995},
    month     = {6},
    doi       = {10.1086/175755}
}

@article{M.Bruni_etal_1995_Jul,
    author    = {Bruni, M. and Matarrese, S. and Pantano, O.},
    title     = {A Local view of the observable universe},
    journal   = {\prl},
    publisher = {American Physical Society},
    volume    = {74},
    number    = {11},
    pages     = {1916-1919},
    year      = {1995},
    month     = {3},
    doi       = {10.1103/PhysRevLett.74.1916},
}

@article{G.F.R.Ellis_M.Bruni_1989,
    title     = {Covariant and gauge-invariant approach to cosmological density fluctuations},
    author    = {Ellis, G. F. R. and Bruni, M.},
    journal   = {\prd},
    volume    = {40},
    number    = {6},
    pages     = {1804-1818},
    year      = {1989},
    month     = {9},
    doi       = {10.1103/PhysRevD.40.1804},
}

@article{P.K.S.Dunsby_etal_1992,
    title     = {Covariant Perturbations in a multifluid cosmological medium},
    author    = {Dunsby, P. K. S. and Bruni, M. and Ellis, G. F. R.},
    journal   = {\apj},
    publisher = {{IOP} Publishing},
    volume    = {395},
    number    = {},
    pages     = {54-73},
    year      = {1992},
    month     = {8},
    doi       = {10.1086/171630},
}

@article{L.Wang_P.J.Steinhardt_1998,
    title     = {{Cluster Abundance Constraints for Cosmological Models with a Time‐varying, Spatially Inhomogeneous Energy Component with Negative Pressure}},
    author    = {Wang, L. and Steinhardt, P. J.},
    journal   = {\apj},
    publisher = {{IOP} Publishing},
    volume    = {508},
    number    = {2},
    pages     = {483-490},
    year      = {1998},
    month     = {12},
    doi       = {10.1086/306436},
}

@article{P.Szekeres_1975a,
  title ="{A class of inhomogeneous cosmological models}",
  author = {Szekeres, P.},
  journal = {Comm.\ Math.\ Phys.},
  volume = {41},
  pages = {55–64},
  year = {1975},
  doi = {10.1007/BF01608547},
}

@article{P.Szekeres_1975b,
  title = "{Quasispherical gravitational collapse}",
  author = {Szekeres, Peter},
  journal = {\prd},
  volume = {12},
  pages = {2941--2948},
  numpages = {0},
  year = {1975},
  doi = {10.1103/PhysRevD.12.2941},
}

@article{W.B.Bonnor_N.Tomimura_1976,
    author = {Bonnor, W. B. and Tomimura, N.},
    title = "{Evolution of Szekeres's Cosmological Models}",
    journal = {\mnras},
    volume = {175},
    pages = {85-93},
    year = {1976},
    doi = {10.1093/mnras/175.1.85},
}

@article{MN.Celerier_2024,
    author = {C{\'e}l{\'e}rier, Marie-No{\"e}lle},
    title = "{Precision cosmology with exact inhomogeneous solutions of general relativity: The Szekeres models}",
    doi = "10.1103/PhysRevD.110.123526",
    journal = "\prd",
    volume = "110",
    pages = "123526",
    year = "2024"
}

@article{K.Bolejko_2009,
  author        = {Bolejko, Krzysztof},
  title         = {The {S}zekeres {S}wiss {C}heese model and the {CMB} observations},
  journal       = {\grg},
  volume        = {41},
  pages         = {1737--1755},
  year          = {2009},
  doi           = {10.1007/s10714-008-0746-x},
}

@article{K.Bolejko_MN.Celerier_2010,
  author        = {Bolejko, Krzysztof and C{\'e}l{\'e}rier, Marie-No{\"e}lle},
  title         = {Szekeres {S}wiss-cheese model and supernova observations},
  journal       = {\prd},
  volume        = {82},
  number        = {10},
  pages         = {103510},
  year          = {2010},
  doi           = {10.1103/PhysRevD.82.103510},
}

@article{K.Bolejko_etal_2016,
   title="{Differential cosmic expansion and the Hubble flow anisotropy}",
   volume={06},
   doi={10.1088/1475-7516/2016/06/035},
   journal={\jcap},
   author={Bolejko, Krzysztof and Nazer, M. Ahsan and Wiltshire, David L.},
   year={2016},
  pages={035} 
}

@article{M.Ishak_A.Peel_2012,
  title = "{Growth of structure in the Szekeres class-II inhomogeneous cosmological models and the matter-dominated era}",
  author = {Ishak, Mustapha and Peel, Austin},
  journal = {Phys. Rev. D},
  volume = {85},
  pages = {083502},
  year = {2012},
  doi = {10.1103/PhysRevD.85.083502},
}

@article{A.Peel_etal_2012,
   title={Large-scale growth evolution in the Szekeres inhomogeneous cosmological models with comparison to growth data},
   volume={86},
   doi={10.1103/physrevd.86.123508},
   number={12},
   journal={\prd},
   author={Peel, Austin and Ishak, Mustapha and Troxel, M. A.},
   year={2012},
}

@article{N.Meures_M.Bruni_2011,
  title = "{Exact nonlinear inhomogeneities in $\ensuremath{\Lambda}\mathrm{CDM}$ cosmology}",
  author = {Meures, Nikolai and Bruni, Marco},
  journal = {\prd},
  volume = {83},
  pages = {123519},
  year = {2011},
  doi = {10.1103/PhysRevD.83.123519},
}

@article{N.Meures_M.Bruni_2012,
   title="{Redshift and distances in a $\ensuremath{\Lambda}\mathrm{CDM}$ cosmology with non-linear inhomogeneities: Redshift and distances in inhomogeneous $\ensuremath{\Lambda}\mathrm{CDM}$}",
   volume={419},
   doi={10.1111/j.1365-2966.2011.19850.x},
   number={3},
   journal={\mnras},
   author={Meures, Nikolai and Bruni, Marco},
   year={2011},
   pages={1937–1950}
}

@article{P.S.Apostolopoulos_2017,
   title={Szekeres models: a covariant approach},
   volume={34},
   doi={10.1088/1361-6382/aa66df},
   journal={\cqg},
   publisher={IOP Publishing},
   author={Apostolopoulos, P. S.},
   year={2017},
   pages={095013},
}

@article{I.DelgadoGasparBuchert_T.Buchert_2021,
 author = {Delgado Gaspar, I. and Buchert, T.},
 doi = {10.1103/PhysRevD.103.023513},
 journal = {\prd},
 pages = {023513},
 title = {{Lagrangian theory of structure formation in relativistic cosmology. VI. Comparison with Szekeres exact solutions}},
 volume = {103},
 year = {2021}
}

@article{S.Najera_R.A.Sussman_2020,
   title="{Pancakes as opposed to Swiss cheese}",
   volume={38},
   doi={10.1088/1361-6382/abcaec},
   journal={\cqg},
   author={Nájera, S and Sussman, R A},
   year={2020},
   pages={015016} 
}

@article{S.Najera_R.A.Sussman_2021,
    author = "N{\'a}jera, Sebasti{\'a}n and Sussman, Roberto A.",
    title = "{Non-comoving cold dark matter in a $\Lambda $CDM background}",
    doi = "10.1140/epjc/s10052-021-09154-0",
    journal = "Eur.\ Phys.\ J.\ C",
    volume = "81",
    number = "4",
    pages = "374",
    year = "2021"
}

@article{F.A.Pizana_etal_2024,
    author = "Piza{\~n}a, Fernando A. and Hidalgo, Juan Carlos and Delgado Gaspar, Ismael and Sussman, Roberto A.",
    title = "{Growth rate of spherical voids with non-comoving dark matter and baryons}",
    doi = "10.1088/1361-6382/ad0f4e",
    journal = "\cqg",
    volume = "41",
    number = "1",
    pages = "015013",
    year = "2024"
}

@article{S.W.Goode_J.Wainwright_1982,
  title = "{Singularities and evolution of the Szekeres cosmological models}",
  author = {Goode, S. W. and Wainwright, J.},
  journal = {\prd},
  volume = {26},
  pages = {3315--3326},
  year = {1982},
  doi = {10.1103/PhysRevD.26.3315},
}

@article{D.A.Szafron_1977,
  author  = {Szafron, D. A.},
  title   = {{Inhomogeneous cosmologies: New exact solutions and their evolution}},
  journal = {J.\ Math.\ Phys.},
  volume  = {18},
  number  = {8},
  pages   = {1673--1677},
  year    = {1977},
  doi     = {10.1063/1.523468}
}

@article{R.A.Sussman_I.DelgadoGaspar_2015,
  title = "{Multiple nonspherical structures from the extrema of Szekeres scalars}",
  author = {Sussman, Roberto A. and Delgado Gaspar, I.},
  journal = {\prd},
  volume = {92},
  pages = {083533},
  year = {2015},
  doi = {10.1103/PhysRevD.92.083533},
}

@article{R.A.Sussman_etal_2016,
    author = {{Sussman}, Roberto A. and {Delgado Gaspar}, I. and {Hidalgo}, Juan Carlos},
    title = "{Coarse-grained description of cosmic structure from Szekeres models}",
    journal = {\jcap},
    year = {2016},
    volume = {03},
    pages = {012},
    doi = {10.1088/1475-7516/2016/03/012},
   note      = {[Erratum in {\protect{JCAP}} {\textbf06} (2016) E03]}
}

@misc{M.Galoppo_etal_2025,
      title="{An effective $\boldsymbol{\Lambda}$-Szekeres modelling of the local Universe with Cosmicflows-4}", 
      author={Marco Galoppo and Leonardo Giani and Morag Hills and Aurélien Valade},
      year={2025},
      eprint={2512.16591},
      archivePrefix={arXiv},
      primaryClass={astro-ph.CO},
      doi={10.48550/arXiv.2512.16591}, 
}

@article{Bruni:2003hm,
    author = "Bruni, Marco and Sopuerta, Carlos F.",
    title = "{Covariant fluid dynamics: A Long wavelength approximation}",
    eprint = "gr-qc/0307059",
    archivePrefix = "arXiv",
    doi = "10.1088/0264-9381/20/24/003",
    journal = "Class. Quant. Grav.",
    volume = "20",
    pages = "5275--5290",
    year = "2003"
}

@book{Landau1975,
    author = "Landau, L. D. and  Lifshitz, E. M.",
    title = "{The Classical Theory of Fields}",
    isbn = "978-0-08-018176-9",
    publisher = "Pergamon Press",
    address = "Oxford",
    series = "Course of Theoretical Physics",
    Volume = "{2}",
    year = "1975"
}

@article{R.A.Sussman_etal_2017,
  title = "{Nonspherical Szekeres models in the language of cosmological perturbations}",
  author = {Sussman, Roberto A. and Hidalgo, Juan Carlos and Delgado Gaspar, Ismael and Germ\'an, Gabriel},
  journal = {\prd},
  volume = {95},
  pages = {064033},
  year = {2017},
  doi = {10.1103/PhysRevD.95.064033},
}

@inproceedings{K.Bolejko_2008,
  author       = {Bolejko, Krzysztof},
  title        = {{Evolution of a void and an adjacent galaxy supercluster in the quasispherical Szekeres model}},
  booktitle    = {{The Eleventh Marcel Grossmann Meeting: On Recent Developments in Theoretical and Experimental General Relativity, Gravitation and Relativistic Field Theories}},
  publisher    = {World Scientific},
  year         = {2008},
  pages        = {1847--1856},
  doi          = {10.1142/9789812834300_0271},
}

@ARTICLE{C.Hellaby_1996,
       author = {{Hellaby}, Charles},
        title = "{The null and KS limits of the Szekeres model}",
      journal = {\cqg},
         year = 1996,
        month = sep,
       volume = {13},
       number = {9},
        pages = {2537-2546},
          doi = {10.1088/0264-9381/13/9/017}
}

@article{C.Hellaby_A.Krasinski_2002,
  author        = {Hellaby, Charles and Krasi{\'n}ski, Andrzej},
  title         = {{You cannot get through Szekeres wormholes: Regularity, topology, and causality in quasispherical Szekeres models}},
  journal       = {\prd},
  volume        = {66},
  number        = {8},
  pages         = {084011},
  year          = {2002},
  doi           = {10.1103/PhysRevD.66.084011},
}

@article{C.Hellaby_A.Krasinski_2008,
  author        = {Hellaby, Charles and Krasi{\'n}ski, Andrzej},
  title         = {{Physical and geometrical interpretation of the {$\epsilon \le 0$} Szekeres models}},
  journal       = {\prd},
  volume        = {77},
  number        = {2},
  pages         = {023529},
  year          = {2008},
  doi           = {10.1103/PhysRevD.77.023529},
}

@article{A.R.Sussman_K.Bolejko_2012,
       author = {{Sussman}, A. Roberto and {Bolejko}, Krzysztof},
        title = "{A novel approach to the dynamics of Szekeres dust models}",
      journal = {\cqg},
         year = 2012,
       volume = {29},
          doi = {10.1088/0264-9381/29/6/065018},
        pages = {065018},
}

@article{A.Krasinski_2008,
  author       = {Krasi{\'n}ski, Andrzej},
  title        = {{Geometry and Topology of the Quasiplane Szekeres Model}},
  journal      = {\prd},
  volume       = {78},
  number       = {6},
  pages        = {064038},
  year         = {2008},
  doi          = {10.1103/PhysRevD.78.064038},
}

@article{A.Krasinski_K.Bolejko_2012,
  author       = {Krasi{\'n}ski, Andrzej and Bolejko, Krzysztof},
  title        = {{Geometry of the Quasihyperbolic Szekeres Models}},
  journal      = {\prd},
  volume       = {86},
  number       = {10},
  pages        = {104036},
  year         = {2012},
  doi          = {10.1103/PhysRevD.86.104036},
}

@article{M.M.deSouza_1985,
  author  = {de~Souza, M. M.},
  title   = {{Hidden Symmetries of Szekeres Quasispherical Solutions}},
  journal = {Rev.\ Bras.\ F{\'i}s\.},
  volume  = {15},
  pages   = {379--386},
  year    = {1985}
}

@article{R.G.Buckley_E.M.Schlegel_2020,
  author       = {Buckley, Robert G. and Schlegel, Eric M.},
  title        = {{Physical Geometry of the Quasispherical Szekeres Models}},
  journal      = {\prd},
  volume       = {101},
  number       = {2},
  pages        = {023511},
  year         = {2020},
  doi          = {10.1103/PhysRevD.101.023511},
}

@article{G.Ira_C.Hellaby_2017,
  author       = {Georg, Ira and Hellaby, Charles},
  title        = {{Symmetry and Equivalence in Szekeres Models}},
  journal      = {\prd},
  volume       = {95},
  number       = {12},
  pages        = {124016},
  year         = {2017},
  doi          = {10.1103/PhysRevD.95.124016},
}

@article{K.Bolejko_2006,
  author       = {Bolejko, Krzysztof},
  title        = {{Structure Formation in the Quasispherical Szekeres Model}},
  journal      = {\prd},
  volume       = {73},
  number       = {12},
  pages        = {123508},
  year         = {2006},
  doi          = {10.1103/PhysRevD.73.123508},
}

@article{K.Bolejko_2007,
  author       = {Bolejko, Krzysztof},
  title        = {{Evolution of Cosmic Structures in Different Environments in the Quasispherical Szekeres Model}},
  journal      = {\prd},
  volume       = {75},
  number       = {4},
  pages        = {043508},
  year         = {2007},
  doi          = {10.1103/PhysRevD.75.043508},
}

@inproceedings{M.Bruni_1996,
  author    = {Bruni, Marco},
  title     = {{Cosmological Collapses of Irrotational Dust}},
  booktitle = {{Mapping, Measuring, and Modelling the Universe}},
  editor    = {Coles, Peter and Mart{\'\i}nez, Vicent J. and Pons-Border{\'\i}a, Maria-Jesus},
  series    = {Astronomical Society of the Pacific Conference Series},
  volume    = {94},
  pages     = {31--36},
  year      = {1996},
  publisher = {Astronomical Society of the Pacific}
}

@article{J.Wainwright_L.Hsu_1989,
  author  = {Wainwright, John and Hsu, L.},
  title   = {{A dynamical systems approach to Bianchi cosmologies: orthogonal models of class A}},
  journal = {\cqg},
  volume  = {6},
  number  = {10},
  pages   = {1409--1431},
  year    = {1989},
  doi     = {10.1088/0264-9381/6/10/011}
}

@article{A.Barnes_R.R.Rowlingson_1989,
  author  = {Barnes, A. and Rowlingson, R. R.},
  title   = {{Irrotational perfect fluids with a purely electric Weyl tensor}},
  journal = {\cqg},
  volume  = {6},
  number  = {7},
  pages   = {949--960},
  year    = {1989},
  doi     = {10.1088/0264-9381/6/7/003}
}

@article{Matarrese:1993zf,
    author = "Matarrese, Sabino and Pantano, Ornella and Saez, Diego",
    title = "{General relativistic dynamics of irrotational dust: Cosmological implications}",
    eprint = "astro-ph/9310036",
    archivePrefix = "arXiv",
    reportNumber = "DFPD-93-A-67",
    doi = "10.1103/PhysRevLett.72.320",
    journal = "Phys. Rev. Lett.",
    volume = "72",
    pages = "320--323",
    year = "1994"
}

@article{S.Matarrese_etal_1993,
  author  = {Matarrese, Sabino and Pantano, Ornella and Saez, Diego},
  title   = {General-relativistic approach to the nonlinear evolution of collisionless matter},
  journal = {\prd},
  volume  = {47},
  number  = {4},
  pages   = {1311--1323},
  year    = {1993},
  doi     = {10.1103/PhysRevD.47.1311}
}

@article{A.Gierzkiewicz_A.Z.Golda,
  author  = {Gierzkiewicz, Anna and Golda, Zdzis{\l}aw A.},
  title   = {A complete set of integrals and solutions to the {S}zekeres system},
  journal = {Phys.\ Lett.\ A},
  volume  = {382},
  number  = {32},
  pages   = {2085--2091},
  year    = {2018},
  doi     = {10.1016/j.physleta.2018.05.038}
}

@article{A.Krasinski_K.Bolejko_2012_AHQS,
  author       = {Krasi{\'n}ski, Andrzej and Bolejko, Krzysztof},
  title        = {Apparent horizons in the quasi-spherical Szekeres models},
  journal      = {\prd},
  volume       = {85},
  pages        = {124016},
  year         = {2012},
  doi          = {10.1103/PhysRevD.85.124016},
}

@article{D.Vrba_O.Svitek_2014,
  author        = {Vrba, David and Sv{\'i}tek, Otakar},
  title         = {Modelling inhomogeneity in {Szekeres} spacetime},
  journal       = {\grg},
  volume        = {46},
  number        = {10},
  pages         = {1808},
  year          = {2014},
  doi           = {10.1007/s10714-014-1808-x},
}

@article{T.Harada_C.Goymer_B.J.Carr_2002,
  author        = {Harada, T. and Goymer, C. and Carr, B. J.},
  title         = {{Tolman--Bondi} collapse in scalar--tensor theories as a probe of gravitational memory},
  journal       = {\prd},
  volume        = {66},
  pages         = {104023},
  year          = {2002},
  doi           = {10.1103/PhysRevD.66.104023},
}

@article{P.Yodzis_H.J.Seifert_H.MullerzumHagen_1973,
  author  = {Yodzis, P. and Seifert, H.-J. and M{\"u}ller zum Hagen, H.},
  title   = {On the occurrence of naked singularities in general relativity},
  journal = {Commun\. Math.\ Phys.},
  volume  = {34},
  number  = {2},
  pages   = {135--148},
  year    = {1973},
  doi     = {10.1007/BF01646443}
}

@article{D.M.Eardley_L.Smarr_1979,
  author  = {Eardley, Douglas M. and Smarr, Larry},
  title   = {{Time functions in numerical relativity: Marginally bound dust collapse}},
  journal = {\prd},
  volume  = {19},
  pages   = {2239--2259},
  year    = {1979},
  doi     = {10.1103/PhysRevD.19.2239}
}

@article{D.Christodoulou_1984,
  author  = {Christodoulou, Demetrios},
  title   = {Violation of cosmic censorship in the gravitational collapse of a dust cloud},
  journal = {Commun.\ Math.\ Phys.},
  volume  = {93},
  number  = {2},
  pages   = {171--195},
  year    = {1984},
  doi     = {10.1007/BF01223743}
}

@article{P.S.Joshi_I.H.Dwivedi_1993,
  author        = {Joshi, P. S. and Dwivedi, I. H.},
  title         = {Naked singularities in spherically symmetric inhomogeneous {Tolman--Bondi} dust cloud collapse},
  journal       = {\prd},
  volume        = {47},
  number        = {12},
  pages         = {5357--5369},
  year          = {1993},
  doi           = {10.1103/PhysRevD.47.5357},
}

@article{T.P.Singh_P.S.Joshi_1996,
  author        = {Singh, T. P. and Joshi, P. S.},
  title         = {The final fate of spherical inhomogeneous dust collapse},
  journal       = {\cqg},
  volume        = {13},
  number        = {3},
  pages         = {559--572},
  year          = {1996},
  doi           = {10.1088/0264-9381/13/3/019},
}

@article{S.Jhingan_P.S.Joshi_T.P.Singh_1996,
  author        = {Jhingan, Sanjay and Joshi, P. S. and Singh, T. P.},
  title         = {The final fate of spherical inhomogeneous dust collapse {II}: {I}nitial data and causal structure of singularity},
  journal       = {\cqg},
  volume        = {13},
  number        = {11},
  pages         = {3057--3068},
  year          = {1996},
  doi           = {10.1088/0264-9381/13/11/019},
}

@article{P.S.Joshi_A.Krolak_1996,
  author        = {Joshi, P. S. and Kr{\'o}lak, Andrzej},
  title         = {Naked strong curvature singularities in {Szekeres} space-times},
  journal       = {\cqg},
  volume        = {13},
  number        = {11},
  pages         = {3069--3074},
  year          = {1996},
  doi           = {10.1088/0264-9381/13/11/020},
}

@article{S.S.Deshingkar_S.Jhingan_P.S.Joshi_1998,
  author        = {Deshingkar, S. S. and Jhingan, S. and Joshi, P. S.},
  title         = {On the global visibility of singularity in quasi-spherical collapse},
  journal       = {\grg},
  volume        = {30},
  pages         = {1477--1499},
  year          = {1998},
  doi           = {10.1023/A:1018813108516},
}

@article{B.C.Nolan_U.Debnath_2007,
  author        = {Nolan, Brien C. and Debnath, Ujjal},
  title         = {Is the shell-focusing singularity of {Szekeres} space-time visible?},
  journal       = {\prd},
  volume        = {76},
  pages         = {104046},
  year          = {2007},
  doi           = {10.1103/PhysRevD.76.104046},
  eprint        = {0709.3152},
}

@article{G.Lemaitre_1933,
  author  = {Lema{\^{i}}tre, Georges},
  title   = {{L'Univers en expansion}},
  journal = {Ann.\ Soc\. Sci.\ Brux.},
  volume  = {53A},
  pages   = {51--85},
  year    = {1933}
}

@article{R.C.Tolman_1934,
  author  = {Tolman, Richard C.},
  title   = {{Effect of Inhomogeneity on Cosmological Models}},
  journal = {Proc.\ Natl.\ Acad.\ Sci.\ U.S.A.},
  volume  = {20},
  number  = {3},
  pages   = {169--176},
  year    = {1934},
  doi     = {10.1073/pnas.20.3.169}
}

@article{H.Bondi_1947,
  author  = {Bondi, Hermann},
  title   = {{Spherically Symmetrical Models in General Relativity}},
  journal = {\mnras},
  volume  = {107},
  number  = {5-6},
  pages   = {410--425},
  year    = {1947},
  doi     = {10.1093/mnras/107.5-6.410},
  note    = {[Republished in {\protect\textit{Gen.\ Relativ.\ Grav.}} {\textbf31} 1783 (1999)]}
}

@article{C.W.Misner_D.H.Sharp_1964,
  author       = {Misner, Charles W. and Sharp, David H.},
  title        = {Relativistic equations for adiabatic, spherically symmetric gravitational collapse},
  journal      = {Phys.\ Rev.},
  volume       = {136},
  number       = {2B},
  pages        = {B571--B576},
  year         = {1964},
  doi          = {10.1103/PhysRev.136.B571}
}

@article{C.Hellaby_K.Lake_1985,
  author       = {Hellaby, Charles and Lake, Kayll},
  title        = {{Shell Crossings and the Tolman Model}},
  journal      = {\apj},
  volume       = {290},
  pages        = {381--387},
  year         = {1985},
  doi          = {10.1086/162995},
  note         = {[Erratum: Apj \textbf{300} (1986) 461]}
}

@article{P.S.Joshi_D.Malafarina_2015,
  author       = {Joshi, Pankaj S. and Malafarina, Daniele},
  title        = {{All black holes in Lema{\^{\i}}tre--Tolman--Bondi inhomogeneous dust collapse}},
  journal      = {\cqg},
  volume       = {32},
  number       = {14},
  pages        = {145004},
  year         = {2015},
  doi          = {10.1088/0264-9381/32/14/145004},
}

@article{V.Gorini_G.Grillo_M.Pelizza_1990,
  author       = {Gorini, Vittorio and Grillo, Gabriele and Pelizza, Mauro},
  title        = {{Black holes in Tolman--Bondi spacetimes}},
  journal      = {Mod.\ Phys.\ Lett.\ A},
  volume       = {5},
  number       = {10},
  pages        = {719--723},
  year         = {1990},
  doi          = {10.1142/S0217732390000810}
}

@article{MN.Celerier_etal_2010,
  author  = {C{\'e}l{\'e}rier, Marie-No{\"e}lle and Bolejko, Krzysztof and Krasi{\'n}ski, Andrzej},
  title   = {{A (Giant) Void Is Not Mandatory to Explain Away Dark Energy with a Lema{\^\i}tre--Tolman Model}},
  journal = {\aap},
  volume  = {518},
  pages   = {A21},
  year    = {2010},
  doi     = {10.1051/0004-6361/200913581}
}

@article{F.C.Mena_R.Tavakol_1999,
  author       = {Mena, F. C. and Tavakol, Reza},
  title        = {{Evolution of the Density Contrast in Inhomogeneous Dust Models}},
  journal      = {\cqg},
  volume       = {16},
  number       = {2},
  pages        = {435--452},
  year         = {1999},
  doi          = {10.1088/0264-9381/16/2/009},
}

@article{V.Marra_etal_2022,
   title="{The BEHOMO project: \ensuremath{\Lambda} Lemaître-Tolman-Bondi N-body simulations}",
   volume={664},
   doi={10.1051/0004-6361/202243539},
   journal={A\&A},
   author={Marra, V. and Castro, T. and Camarena, D. and Borgani, S. and Ragagnin, A.},
   year={2022},
   pages={A179}
}

@article{A.Krasinski_C.Hellaby_2001,
  author  = {Krasi{\'n}ski, Andrzej and Hellaby, Charles},
  title   = {{Structure formation in the Lema{\^\i}tre--Tolman model}},
  journal = {Phys.\ Rev.\ D},
  volume  = {65},
  pages   = {023501},
  year    = {2001},
  doi     = {10.1103/PhysRevD.65.023501}
}

@article{A.Krasinski_C.Hellaby_2004,
  author  = {Krasi{\'n}ski, Andrzej and Hellaby, Charles},
  title   = {{Formation of a galaxy with a central black hole in the Lema{\^\i}tre--Tolman model}},
  journal = {Phys.\ Rev.\ D},
  volume  = {69},
  pages   = {043502},
  year    = {2004},
  doi     = {10.1103/PhysRevD.69.043502}
}

@article{I.DelgadoGaspar_etal_2018,
    author = "Delgado Gaspar, Ismael and Hidalgo, Juan Carlos and Sussman, Roberto A. and Quiros, Israel",
    title = "{Black hole formation from the gravitational collapse of a nonspherical network of structures}",
    doi = "10.1103/PhysRevD.97.104029",
    journal = "\prd",
    volume = "97",
    number = "10",
    pages = "104029",
    year = "2018"
}

@article{W.Ye_etal_2025,
    author = "Ye, Weitao and Gong, Yungui and Harada, Tomohiro and Kang, Zhaofeng and Kohri, Kazunori and Saito, Daiki and Yoo, Chul-Moon",
    title = "{Primordial black hole formation and spin in matter domination revisited}",
    doi = "10.1103/qg7j-vg7v",
    journal = "\prd",
    volume = "112",
    number = "10",
    pages = "103524",
    year = "2025"
}

@article{K.Uehara_etal_2025,
    author = "Uehara, Koichiro and Escriv{\`a}, Albert and Harada, Tomohiro and Saito, Daiki and Yoo, Chul-Moon",
    title = "{Primordial black hole formation from a type II perturbation in the absence and presence of pressure}",
    doi = "10.1088/1475-7516/2025/08/042",
    journal = "\jcap",
    volume = "08",
    pages = "042",
    year = "2025"
}

@article{A.G.Polnarev_M.Y.Khlopov_1981,
  author  = {Polnarev, A. G. and Khlopov, M. Yu.},
  title   = {{Primordial Black Holes and the Era of Superheavy Particle Dominance in the Early Universe}},
  journal = {Soviet Astronomy},
  volume  = {25},
  pages   = {406--411},
  year    = {1981},
  note    = {English translation of Astron.\ Zh.\ 58, 706--716 (1981).}
}

@article{M.Kopp_etal_2011,
  author        = {Kopp, Michael and Hofmann, Stefan and Weller, Jochen},
  title         = {{Separate Universes Do Not Constrain Primordial Black Hole Formation}},
  journal       = {\prd},
  volume        = {83},
  number        = {12},
  pages         = {124025},
  year          = {2011},
  doi           = {10.1103/PhysRevD.83.124025},
}

@article{T.Kokubu_etal_2018,
  author        = {Kokubu, Takafumi and Kyutoku, Koutarou and Kohri, Kazunori and Harada, Tomohiro},
  title         = {{Effect of Inhomogeneity on Primordial Black Hole Formation in the Matter Dominated Era}},
  journal       = {\prd},
  volume        = {98},
  number        = {12},
  pages         = {123024},
  year          = {2018},
  doi           = {10.1103/PhysRevD.98.123024},
}

@book{C.Byrnes_etal_2025,
    editor = "Byrnes, Christian and Franciolini, Gabriele and Harada, Tomohiro and Pani, Paolo and Sasaki, Misao",
    title = "{Primordial Black Holes}",
    doi = "10.1007/978-981-97-8887-3",
    publisher = "Springer",
    year = "2025"
}

@article{T.Harada_etal_2023,
    author = "Harada, Tomohiro and Kohri, Kazunori and Sasaki, Misao and Terada, Takahiro and Yoo, Chul-Moon",
    title = "{Threshold of primordial black hole formation against velocity dispersion in matter-dominated era}",
    doi = "10.1088/1475-7516/2023/02/038",
    journal = "\jcap",
    volume = "02",
    pages = "038",
    year = "2023"
}

@article{C.M.Yoo_etal_2018,
    author = "Yoo, Chul-Moon and Harada, Tomohiro and Garriga, Jaume and Kohri, Kazunori",
    title = "{Primordial black hole abundance from random Gaussian curvature perturbations and a local density threshold}",
    doi = "10.1093/ptep/pty120",
    journal = "PTEP",
    volume = "2018",
    number = "12",
    pages = "123E01",
    year = "2018",
    note = "[Erratum: PTEP 2024, 049202 (2024)]"
}

@article{T.Harada_etal_2016,
    author = "Harada, Tomohiro and Yoo, Chul-Moon and Kohri, Kazunori and Nakao, Ken-ichi and Jhingan, Sanjay",
    title = "{Primordial black hole formation in the matter-dominated phase of the Universe}",
    doi = "10.3847/1538-4357/833/1/61",
    journal = "\apj",
    volume = "833",
    number = "1",
    pages = "61",
    year = "2016"
}

@article{T.Harada_S.Jhingan_2015,
    author = "Harada, Tomohiro and Jhingan, Sanjay",
    title = "{Spherical and nonspherical models of primordial black hole formation: exact solutions}",
    doi = "10.1093/ptep/ptw123",
    journal = "PTEP",
    volume = "2016",
    number = "9",
    pages = "093E04",
    year = "2016"
}

@article{Y.B.Zeldovich_I.D.Novikov_1967,
  author  = {Zel'dovich, Ya. B. and Novikov, I. D.},
  title   = {{The Hypothesis of Cores Retarded during Expansion and the Hot Cosmological Model}},
  journal = {Sov.\ Astron.},
  volume  = {10},
  number  = {4},
  pages   = {602--603},
  year    = {1967},
  note    = {English translation of Astron.\ Zh.\ \textbf{43}, 758 (1966).}
}

@article{S.W.Hawking_1971,
  author  = {Hawking, Stephen W.},
  title   = {{Gravitationally Collapsed Objects of Very Low Mass}},
  journal = {\mnras},
  volume  = {152},
  number  = {1},
  pages   = {75--78},
  year    = {1971},
  doi     = {10.1093/mnras/152.1.75}
}

@article{B.J.Carr_S.W.Hawking_1974,
  author  = {Carr, Bernard J. and Hawking, Stephen W.},
  title   = {{Black Holes in the Early Universe}},
  journal = {\mnras},
  volume  = {168},
  number  = {2},
  pages   = {399--415},
  year    = {1974},
  doi     = {10.1093/mnras/168.2.399}
}

@article{B.J.Carr_1975,
  author  = {Carr, Bernard J.},
  title   = {{The Primordial Black Hole Mass Spectrum}},
  journal = {\apj},
  volume  = {201},
  pages   = {1--19},
  year    = {1975},
  doi     = {10.1086/153853}
}

@article{J.M.Bardeen_J.R.Bond_N.Kaiser_A.S.Szalay_1986,
  author  = {Bardeen, J. M. and Bond, J. R. and Kaiser, N. and Szalay, A. S.},
  title   = {{The Statistics of Peaks of Gaussian Random Fields}},
  journal = {\apj},
  volume  = {304},
  pages   = {15--61},
  year    = {1986},
  doi     = {10.1086/164143}
}

@article{M.Y.Khlopov_A.G.Polnarev_1980,
  author  = {Khlopov, M. Y. and Polnarev, A. G.},
  title   = {Primordial black holes as a cosmological test of grand unification},
  journal = {Phys.\ Lett.\ B},
  volume  = {97},
  number  = {3--4},
  pages   = {383--387},
  year    = {1980},
  doi     = {10.1016/0370-2693(80)90624-3}
}

@article{A.G.Doroshkevich_1970,
  author  = {Doroshkevich, A. G.},
  title   = {{The Spatial Structure of Perturbations and the Origin of Rotation of Galaxies in the Theory of Fluctuations}},
  journal = {Astrophys.},
  volume  = {6},
  number  = {4},
  pages   = {320--330},
  year    = {1970},
  doi     = {10.1007/BF01001625}
}

@ARTICLE{Planck_2018,
   author = {{Planck Collaboration} and {Aghanim}, N. and others},
    title = "{Planck 2018 results. VI. Cosmological parameters}",
  journal = {\aap},
    pages = {A6},
    volume = {641},
     year = 2020,
    doi = {10.1051/0004-6361/201833910},
}

@article{J.R.Bond_S.T.Myers_1996a,
  author  = {Bond, J. Richard and Myers, Stephen T.},
  title   = {{The Peak-Patch Picture of Cosmic Catalogs. {I}. Algorithms}},
  journal = {ApJ Suppl. Ser.},
  volume  = {103},
  pages   = {1--39},
  year    = {1996},
  doi     = {10.1086/192267}
}

@article{J.R.Bond_S.T.Myers_1996b,
  author  = {Bond, J. Richard and Myers, Stephen T.},
  title   = {{The Peak-Patch Picture of Cosmic Catalogs. {II}. Validation}},
  journal = {ApJ Suppl. Ser.},
  volume  = {103},
  pages   = {41--62},
  year    = {1996},
  doi     = {10.1086/192268}
}

@article{J.R.Bond_S.T.Myers_1996c,
  author  = {Bond, J. Richard and Myers, Stephen T.},
  title   = {{The Peak-Patch Picture of Cosmic Catalogs. {III}. Application to Clusters}},
  journal = {ApJ Suppl. Ser.},
  volume  = {103},
  pages   = {63--79},
  year    = {1996},
  doi     = {10.1086/192269}
}

@book{R.J.Adler_1981,
  author    = {Adler, Robert J.},
  title     = {The Geometry of Random Fields},
  publisher = {Wiley},
  address   = {New York},
  year      = {1981}
}

@article{C.Germani_T.Prokopec_2017,
  author  = {Germani, Cristiano and Prokopec, Tomislav},
  title   = {{On Primordial Black Holes from an Inflection Point}},
  journal = {Phys.\ Dark Universe},
  volume  = {18},
  pages   = {6--10},
  year    = {2017},
  doi     = {10.1016/j.dark.2017.09.001},
}

@article{C.M.Yoo_T.Harada_J.Garriga_K.Kohri_2018,
  author  = {Yoo, Chul-Moon and Harada, Tomohiro and Garriga, Jaume and Kohri, Kazunori},
  title   = {{{PBH} Abundance from Random Gaussian Curvature Perturbations and a Local Density Threshold}},
  journal = {Prog.\ Theor.\ Exp.\ Phys.},
  volume  = {2018},
  number  = {12},
  pages   = {123E01},
  year    = {2018},
  doi     = {10.1093/ptep/pty120},
}

@article{C.Germani_I.Musco_2019,
  author  = {Germani, Cristiano and Musco, Ilia},
  title   = {{Abundance of Primordial Black Holes Depends on the Shape of the Inflationary Power Spectrum}},
  journal = {\prl},
  volume  = {122},
  pages   = {141302},
  year    = {2019},
  doi     = {10.1103/PhysRevLett.122.141302},
}

@article{I.Musco_2019,
  author  = {Musco, Ilia},
  title   = {{Threshold for Primordial Black Holes: Dependence on the Shape of the Cosmological Perturbations}},
  journal = {\prd},
  volume  = {100},
  pages   = {123524},
  year    = {2019},
  doi     = {10.1103/PhysRevD.100.123524},
}

@article{C.M.Yoo_J.O.Gong_S.Yokoyama_2019,
  author  = {{Yoo, Chul-Moon and Gong, Jinn-Ouk and Yokoyama, Shuichiro}},
  title   = {Abundance of Primordial Black Holes with Local Non-Gaussianity in Peak Theory},
  journal = {\jcap},
  volume  = {09},
  pages   = {033},
  year    = {2019},
  doi     = {10.1088/1475-7516/2019/09/033},
}

@article{S.Young_M.Musso_2020,
  author  = {Young, Sam and Musso, Marcello},
  title   = {{Application of Peaks Theory to the Abundance of Primordial Black Holes}},
  journal = {\jcap},
  volume  = {11},
  pages   = {022},
  year    = {2020},
  doi     = {10.1088/1475-7516/2020/11/022},
}

@article{N.Kitajima_etal_2021,
   title={{Primordial black holes in peak theory with a non-Gaussian tail}},
   volume={2021},
   DOI={10.1088/1475-7516/2021/10/053},
   number={10},
   journal={\jcap},
   author={Kitajima, Naoya and Tada, Yuichiro and Yokoyama, Shuichiro and Yoo, Chul-Moon},
   year={2021},
   month=oct, 
   pages={053}
}

@article{C.M.Yoo_T.Harada_S.Hirano_K.Kohri_2021,
  author  = {Yoo, Chul-Moon and Harada, Tomohiro and Hirano, Shin'ichi and Kohri, Kazunori},
  title   = {{Abundance of Primordial Black Holes in Peak Theory for an Arbitrary Power Spectrum}},
  journal = {Prog. Theor. Exp. Phys.},
  volume  = {2021},
  number  = {1},
  pages   = {013E02},
  year    = {2021},
  doi     = {10.1093/ptep/ptaa155},
}

@article{C.Germani_2025,
    author = "Germani, Cristiano and Gorji, Mohammad Ali and Uwabo-Niibo, Michiru and Yamaguchi, Masahide",
    title = "{{Peaks sphericity of non-Gaussian random fields}}",
    doi = "10.1088/1475-7516/2025/09/052",
    journal = "\jcap",
    volume = "09",
    pages = "052",
    year = "2025"
}

@book{M.Alcubierre_2008,
    title     = {{Introduction to 3+1 Numerical Relativity}},
    author    = {Alcubierre, M.},
    publisher = {Oxford Science Publications},
    doi       = {10.1093/acprof:oso/9780199205677.001.0001},
    year      = {2008}}

@book{T.W.Baumgarte_S.L.Shapiro_2010,
    title     = {{Numerical Relativity: Solving Einstein's Equations on the Computer}}, 
    author    = {Baumgarte, T. W. and Shapiro, S. L.}, 
    publisher = {Cambridge University Press}, 
    doi       = {10.1017/CBO9781139193344}, 
    year      = {2010}}

@book{M.Shibata_2015,
    title     = {Numerical Relativity},
    author    = {Shibata, M.},
    publisher = {World Scientific Publishing Company},
    doi       = {10.1142/9692},
    year      = {2015}}

@book{N.Vittorio_2018,
    title     = {Cosmology},
    author    = {Vittorio, N.},
    publisher = {CRC Press, Taylor \& Francis Group},
    doi       = {10.1201/b22176},
    year      = {2018}}

@book{J.Wainwright_G.F.R.Ellis_1997, 
    title     = {Dynamical Systems in Cosmology}, 
    author    = {Wainwright, J. and Ellis, G. F. R.},
    publisher = {Cambridge University Press}, 
    doi       = {10.1017/CBO9780511524660}, 
    year      = {1997}}

@book{R.Arnowitt_S.Deser_C.W.Misner_1962,
  author    = {Arnowitt, Richard and Deser, Stanley and Misner, Charles W.},
  title     = {The Dynamics of General Relativity},
  publisher = {John Wiley \& Sons},
  year      = {1962},
}

@book{G.F.R.Ellis_etal_2012,
    title     = {Relativistic Cosmology},
    author    = {Ellis, G. F. R. and Maartens, R. and MacCallum, M. A. H.},
    publisher = {Cambridge University Press},
    doi       = {10.1017/CBO9781139014403},
    year      = {2012}}

@book{P.J.E.Peebles_1980,
    title     = {The large-scale structure of the universe},
    author    = {Peebles, P. J. E.},
    publisher = {Princeton University Press},
    doi       = {10.1038/nature04805},
    year      = {1980}}

@book{S.W.Hawking_G.F.R.Ellis_1973,
  author    = "Hawking, Stephen W. and Ellis, George F. R.",
  title     = "{The Large Scale Structure of Space-Time}",
  publisher = "Cambridge University Press",
  year      = "1973"
}

@book{Krasiński_1997,
place={Cambridge},
title="{Inhomogeneous Cosmological Models}",
publisher={Cambridge University Press}, 
author={Krasiński, Andrzej}, 
year={1997},
}

@book{K.Bolejko_etal_2009,
  author    = {Bolejko, Krzysztof and Krasi{\'n}ski, Andrzej and Hellaby, Charles and C{\'e}l{\'e}rier, Marie-No{\"e}lle},
  title     = {Structures in the Universe by Exact Methods: Formation, Evolution, Interactions},
  publisher = {Cambridge University Press},
  year      = {2009},
}

@book{J.Plebanski_A.Krasinski_2006,
  author    = {Pleba{\'n}ski, Jerzy and Krasi{\'n}ski, Andrzej},
  title     = {An Introduction to General Relativity and Cosmology},
  publisher = {Cambridge University Press},
  year      = {2006},
  doi       = {10.1017/CBO9780511617676},
}

@incollection{M.A.H.MacCallum_1973,
  author    = {MacCallum, M. A. H.},
  title     = {Cosmological Models from a Geometric Point of View},
  booktitle = {Carg{\`e}se Lectures in Physics, Vol.\ 6},
  editor    = {Schatzman, E.},
  publisher = {Gordon and Breach},
  address   = {New York},
  year      = {1973},
  pages     = {61--174},
  note      = {Lectures at the International Summer School of Physics, Carg{\`e}se, Corsica (1971).}
}

@article{R.Maiolino_etal_2024,
  author  = {Roberto Maiolino and Jan Scholtz and Joris Witstok and others},
  title   = {{A Small and Vigorous Black Hole in the Early Universe}},
  journal = {Nature},
  volume  = {627},
  pages   = {59--63},
  year    = {2024},
  doi     = {10.1038/s41586-024-07052-5}
}

@article{R.Larson_etal_2023,
  author  = {Rebecca L. Larson and Steven L. Finkelstein and Dale D. Kocevski and others},
  title   = {{A CEERS Discovery of an Accreting Supermassive Black Hole 570 Myr after the Big Bang: Identifying a Progenitor of Massive $z > 6$ Quasars}},
  journal = {\apjl},
  volume  = {953},
  pages   = {L29},
  year    = {2023},
  doi     = {10.3847/2041-8213/ace619}
}

@article{F.Pacucci_etal_2023,
  author  = {Fabio Pacucci and Bao Nguyen and Stefano Carniani and Roberto Maiolino and Xiaohui Fan},
  title   = {{JWST CEERS and JADES Active Galaxies at $z=4$--$7$ Violate the Local $M_{\bullet}$--$M_{\star}$ Relation at $>3\sigma$: Implications for Low-mass Black Holes and Seeding Models}},
  journal = {\apjl},
  volume  = {957},
  pages   = {L3},
  year    = {2023},
  doi     = {10.3847/2041-8213/ad0158}
}

@article{S.L.Shapiro_S.A.Teukolsky_1991,
  author = {Shapiro, Stuart L. and Teukolsky, Saul A.},
  title = {{Formation of Naked Singularities: The Violation of Cosmic Censorship}},
  journal = {\prl},
  volume = {66},
  pages = {994--997},
  year = {1991},
  doi = {10.1103/PhysRevLett.66.994}
}

\appendix
\section{\label{app:mapping}Mapping between Hellaby and Goode-Wainwright formalism}

This appendix shows the mapping --- for the class I Szekeres models --- between the Hellaby and GW parametrisations. We point the reader to Kras\'{i}nski~Sec.~$2.5$]~\cite{Krasiński_1997}, Pleb\'{a}nski and Kras\'{i}nski~[Sec~$19.8$]~\cite{J.Plebanski_A.Krasinski_2006}, as well as Delgado Gaspar and Buchert~[App.~A]~\cite{I.DelgadoGasparBuchert_T.Buchert_2021} for a complete treatment, including also class II models.

For the class I solutions, the metric element in the Hellaby parametrisation reads
\begin{align}\label{eq:App1}
    \dd s^2 &= -\dd t^2 +\frac{\left(R'-RD'/D\right)^2}{1-k}\dd r^2 + \frac{R^2}{D^2}\left(\dd p^2 + \dd q^2\right) \\
    & = -\dd t^2 +e^{2\alpha}\dd r^2 + e^{2\beta}\left(\dd p^2 + \dd q^2\right)\, ,\nonumber
\end{align}
where $R(t,r)$ satisfies the local Friedman equation
\begin{align}\label{eq:App2}
    \left(\frac{\dot{R}}{R}\right)^2 = \frac{2GM}{R^3} -\frac{k}{R^2} + \frac{1}{3}\Lambda \, ,
\end{align}
and we have introduced $e^\alpha$ and $e^\beta$ to simplify metric components' identification in this appendix. 
\subsection{The case $k \neq 0$}

To obtain the Szekeres solutions in the GW parametrisation, we normalise the curvature function $k(r)$ by introducing an auxiliary function $f(r)$ such that
\begin{equation}
    k(r) = \tilde{k}f(r)^2 \, , \label{eq:App3}
\end{equation}
i.e., such that $f(r) = |k(r)|^{1/2}$, and $\tilde{k} = \pm 1$. We thus consider the rescaling of the areal radius as 
\begin{equation}
    R(t,r) = S(t,r) f(r) \, .\label{eq:App4}
\end{equation}
Similarly, for the active gravitational mass function we consider 
\begin{equation}
    M(r) = \tilde{M}(r) f^3(r) \, . \label{eq:App5}
\end{equation}
We therefore obtain that $S(t,r)$ solves for
\begin{equation} \label{eq:App6}
    \left(\frac{\dot{S}}{S}\right)^2 = \frac{2G\tilde{M}}{S^3} - \frac{\tilde{k}}{S^2} +\frac{1}{3}\Lambda \, 
    ,
\end{equation}
where we see that the latter equation matches the appropriate evolution equation of the GW parametrisation, i.e., Eq.~\eqref{eq:GwFriedman}.

In addition, we introduce the following identifications
\begin{equation} \label{eq:App7}
    a(r) = \frac{1}{2L} \;\; ; \;\;b(r) = -\frac{P}{2L} \;\; ; \;\;c(r) = -\frac{Q}{2L}\;\; ;   
\end{equation}
and
\begin{equation}
    d(r) = \frac{P^2+Q^2+\epsilon L^2}{2L} \, , \label{eq:App8}
\end{equation}
where $\epsilon = \pm1 $ determines the topology of the model. Following Eqs.~\eqref{eq:App7} and~\eqref{eq:App8} it is clear that the free functions $a,\,b,\,c$ and $d$ now satisfy 
\begin{equation}\label{eq:App9}
      ad - b^2 - c^{2} = \frac{\epsilon}{4} \, ,
\end{equation}
i.e., the constraint in Eq.~\eqref{eq:constraintGW}, generalised to arbitrary $\epsilon$. It is then natural --- to match the two parametrisations --- to introduce
\begin{equation}\label{eq:App10}
    e^{-\nu} := \frac{1}{f}\left[a(p^2+q^2) + 2bp + 2cq +d\right]\ = \frac{D}{f}\, .
\end{equation}
Thus, as needed, we obtain
\begin{equation}
    e^\beta = Se^{\nu} \, .
\end{equation}
Furthermore, we introduce the function
\begin{equation}
    W = \left(\epsilon-\tilde{k}f^2\right)^{-1} \, .
\end{equation}
Hence, by employing Eqs.~\eqref{eq:App3}--\eqref{eq:App10} we can write 
\begin{equation}
    e^\alpha = WSZ \, ,
\end{equation}
where we identify, as in the standard GW parametrisation~\cite{S.W.Goode_J.Wainwright_1982}
\begin{equation}\label{eq:AppBoh}
    Z = f\frac{S'}{S} + f\nu'\, ,
\end{equation}
which naturally splits in a time-dependent and a time independent part. We can now consider the formal solution to Eq.~\eqref{eq:App6}, i.e.,
\begin{equation}
    t - T_\mathrm{B}(r) = \int_0^S\frac{\dd\hat{S}}{\sqrt{\frac{2G\tilde{M}}{\hat{S}} - \tilde{k} + \frac{1}{3}\Lambda\hat{S^2}}}\, ,
\end{equation}
and differentiate with respect to the radial coordinate to get
\begin{equation}\label{eq:App_nice}
    \frac{S'}{\dot{S}} - \tilde{M}'\int_0^S\frac{\dd\hat{S}}{S(\dot{\hat{S}})^3} = - T_\mathrm{B}'
\end{equation}
After some algebra the integral con be re-written as~\cite{I.DelgadoGasparBuchert_T.Buchert_2021} 
\begin{equation}
    \int_0^S\frac{\dd\hat{S}}{S(\dot{\hat{S}})^3} = \frac{1}{3\tilde{M}}\left(\tilde{k}\int_0^S\frac{\dd\hat{S}}{(\dot{\hat{S}})^3} + \frac{S}{\dot{S}}\right)\, .
\end{equation}
In turn, this allows us to write 
\begin{equation}\label{eq:Appderivative}
    \frac{S'}{S} = - T_\mathrm{B}'\left(\frac{\dot{S}}{S}\right) + \tilde{k}\frac{\tilde{M'}}{3\tilde{M}}\left(\frac{\dot{S}}{S}\int_0^S\frac{\dd\hat{S}}{(\dot{\hat{S}})^3} \right)+\frac{\tilde{M'}}{3\tilde{M}} \, .
\end{equation}
Here, it is important to note that the two formal independent solutions to Eq.~\eqref{eq:ODESk}, i.e., the equation determining $f_\pm$ are given by
\begin{equation}
    f_\pm \propto \begin{cases}
        (\dot{S}/S)\int_0^S\dot{\hat{S}}^{-3}\dd \hat{S} \quad \quad \mathrm{growing~mode\!:~}f_+ \, ,\\
        \dot{S}/S \quad \quad\quad \quad\quad \quad \quad\,\mathrm{decaying~mode\!:~}f_-\, .
    \end{cases}
\end{equation}
Therefore by denoting 
\begin{align}
    & f_+  = \frac{\dot{S}}{S}\int_0^S\frac{\dd\hat{S}}{(\dot{\hat{S}})^3} \, ,\\
    & f_-  = 6\tilde{M}\frac{\dot{S}}{S} \, ,
\end{align}
we can re-write Eq.~\eqref{eq:Appderivative} as
\begin{equation}
    \frac{S'}{S} = - \frac{T_\mathrm{B}'}{6\tilde{M}}f_- + \tilde{k}\frac{\tilde{M'}}{3\tilde{M}}f_+ +\frac{\tilde{M'}}{3\tilde{M}} \, .
\end{equation}
Hence, Eq.~\eqref{eq:AppBoh} becomes directly
\begin{equation}
    Z = (f\nu' - \tilde{k}\beta_+) - (\beta_+f_+ + \beta_-f_-) \, ,
\end{equation}
where $\beta_\pm$ are defined exactly as in Eq.~\eqref{eq:beta+}, finally proving the validity mapping via the remaning identifications 
\begin{align}
    & A = f\nu' - \tilde{k}\beta_+\, , \\
    & F = \beta_+f_+ + \beta_-f_-\, .
\end{align}. 
In addition, we note that we have the following useful identifications
\begin{align}
    & T_\mathrm{B} = t_\mathrm{B} \, ,\\
    & \beta_- = \frac{k^{2}t_\mathrm{B}'}{6M}\, , \\
    &\beta_+ = -\mathrm{sgn}(k)\frac{\sqrt{k}}{3}\left(\frac{M'}{M}-\frac{3}{2}\frac{k'}{k}\right)\, .\label{eq:App_Useful} 
\end{align}
\medskip
\subsection{The case k = 0}

Things are considerably simpler in the case $(kr)=0$. Indeed, here we can directly choose
\begin{equation}
    f = M^{1/3} \,,
\end{equation}
as Eq.~\eqref{eq:App3} cannot be used to define $f$.
It then follows that
\begin{equation}
   R = {M}^{1/3} S \,,
\end{equation}
and we immediately find that $S$ satisfies Eq.~\eqref{eq:App6} with $k=0$ and $\tilde{M}=1$. Consequently, for $k=0$ we may take $\tilde{M}=\text{const.}$ in the Friedmann-like equations, and subsequently use the arbitrary function $f$ to fix the parametrisation.

Proceeding along these lines, the remainder of the equations remains unchanged with respect to the previous calculations for $k(r) \neq 0$. However, we must emphasise that, as it is known~\cite{S.W.Goode_J.Wainwright_1982}, the solutions for $F$ now only contain a contribution from $\beta_-$, so that we may consistently set $\beta_+=0$. Indeed, for $\tilde{M}=\text{const.}$, $k(r) = 0$, we find that Eq.~\eqref{eq:App_nice} reduces directly to
\begin{equation}
    \frac{S_{,\xi}}{{S}} = - T_{\mathrm{B}}'\frac{\dot{S}}{S}\,.
\end{equation}
Furthermore, from the definition of $Z$ in Eq.~\eqref{eq:AppBoh}, we obtain
\begin{align}
    Z &= f \frac{S'}{S} + f\nu' = f\nu'- \beta_- f_- \,.
    \tag{A37}
\end{align}
Therefore, for $k(r)=0$ we have
\begin{equation}
    A = f\nu'\,, 
    \qquad
    F = \beta_- f_-\,,
    \qquad
    \beta_+ = 0 \,.
\end{equation}

\section{\label{app:param}Parametric form for the 1+3 variables}
In this appendix, we report the parametric form of the matter, kinematic and curvature variables for the Szekeres models employed in this work. We point the reader to Gierzkiewicz and Golda~\cite{A.Gierzkiewicz_A.Z.Golda} for alternative forms based on a complete set of first-integral of the autonomous Szekeres dynamical system. Here, we instead directly employ Eqs.~\eqref{eq:SkDensity},~\eqref{eq:theta}--\eqref{eq:R3}, with Eqs.~\eqref{eq:t+} and~\eqref{eq:R+} to derive these expressions. 

For the fluid density, $\rho$, we find
\begin{widetext}
    \begin{align}
    & 4\pi\rho = \frac{\csc^{6}\left(\frac{\eta}{2}\right)\, k^{3}\left(M' - 3 MD'/D \right)}{
    M^{2}\left[2\csc^{2}\left(\frac{\eta}{2}\right)\left(2\left(2 - \eta \cot\frac{\eta}{2}\right)M'/M-\left(5 + \cos\eta - 3\eta \cot\frac{\eta}{2}\right)k'/k\right)-8D'/D\right]}\, .
\end{align}
\end{widetext}
For the expansion scalar $\Theta$ we have 
\newpage
\begin{widetext}
    \begin{align}
    & \Theta = \frac{ k^{3/2}\csc^{6}\left(\frac{\eta}{2}\right)\left[3\left(2(1-\cos\eta)D'/D\sin\eta+(\eta + 2\eta\cos\eta - (2+\cos\eta)\sin\eta)k'/k\right)-2(\eta + 2\eta\cos\eta-3\sin\eta)M'/M\right]}{8M(1-\cos\eta)^{-1}\left[(5+\cos\eta - 3\eta\cot(\frac{\eta}{2}))k'/k+2\left((1-\cos\eta)D'/D+(\eta\cot(\frac{\eta}{2})-2)M'/M\right)\right]}\, .
\end{align}
\end{widetext}
For the shear eigenvalue, $\sigma_+$, we get
\newpage
\allowdisplaybreaks
\begin{widetext}
\begin{align}
&\sigma_+ = \frac{k^{3/2}\csc^{6}\!\left(\frac{\eta}{2}\right)\bigl(\eta(2+\cos\eta) - 3\sin\eta\bigr)\left(M'/M- (3/2)k'/k\right)}{3M\left[(5+\cos\eta - 3\eta\cot(\frac{\eta}{2}))\csc^{2}\!\left(\frac{\eta}{2}\right)k'/k+4\left(2D'/D+(\eta\cot(\frac{\eta}{2})-2) \csc^{2}\left(\frac{\eta}{2}\right)M'/M\right)\right]} \, .
\end{align}
\end{widetext}
For the electric Weyl eigenvalue, $E_+$, we get
\begin{widetext}
\begin{align}
&E_+ = \frac{ k^{3}\bigl(M'/M-(3/2)k'/k\bigr)(1-\cos\eta)\,\bigl(5 + \cos\eta - 3\eta \cot(\frac{\eta}{2})\bigr)\,\csc^{8}(\frac{\eta}{2})}{48\, M^{2}\left[2(1-\cos\eta)D'/D- \bigl(5 + \cos\eta - 3\eta \cot(\frac{\eta}{2})\bigr)k'/k- 2\bigl(-2 + \eta \cot(\frac{\eta}{2})\bigr)M'/M\right]}\, .
\end{align}
\end{widetext}
Finally, for the spatial curvature we obtain
\newpage
\begin{widetext}
\begin{align}
&\Rsp =\frac{k^{3}\csc^{4}\left(\frac{\eta}{2}\right)\left[3\,\csc^{4}\!\left(\frac{\eta}{2}\right)\sin\eta(\sin\eta - \eta)k'/k + 4\left(6D'/D + (\eta\cot(\frac{\eta}{2})-2)\csc^{2}\left(\frac{\eta}{2}\right)M'/M\right) \right]}{4 M^{2}\left[(5 + \cos\eta - 3\eta\cot(\frac{\eta}{2}))\csc^{2}\!\left(\frac{\eta}{2}\right)k'/k + 2\left(2D'/D +(\eta\cot(\frac{\eta}{2})-2)\,\csc^{2}\!\left(\frac{\eta}{2}\right)M'/M\right)\right]}\, .
\end{align}
\end{widetext}

\end{document}